\documentclass[aps,prd,reprint,showkeys,showpacs,nofootinbib]{revtex4-1}

\usepackage{bm}
\usepackage[dvipdfmx]{graphicx,color} %
\usepackage{amsmath} %
\usepackage{amssymb} %
\usepackage{float} %
\usepackage{color} %
\usepackage{longtable}
\usepackage[utf8]{inputenc}
\usepackage{ulem}

\usepackage[dvipdfmx]{hyperref}
\usepackage{url}
\hypersetup{colorlinks=true, citecolor=cyan}

\def\prd{\rm{Phys.~Rev.~D}}        

\begin{document}

\title{Generalized framework for testing gravity with gravitational-wave propagation. III. Future prospect}

\author {Atsushi Nishizawa}
\email{anishi@resceu.s.u-tokyo.ac.jp}
\affiliation {Research Center for the Early Universe, University of Tokyo, Tokyo 113-0033, Japan}
\affiliation {Kobayashi-Maskawa Institute for the Origin of Particles and the Universe, Nagoya University, Nagoya 464-8602, Japan}

\author {Shun Arai}
\email{arai.shun@a.mbox.nagoya-u.ac.jp}
\affiliation {Department of Physics and Astrophysics, Nagoya University, Nagoya 464-8602, Japan}
\affiliation {Institute for Astronomy, University of Edinburgh,
Royal Observatory, Blackford Hill, Edinburgh, EH9 3HJ, U.K.}

\date{\today}

\begin{abstract}
The properties of gravitational-wave (GW) propagation are modified in alternative theories of gravity and are crucial observables to test gravity at cosmological distance. The propagation speed has already been measured from GW170817 so precisely and pinned down to the speed of light, while other properties of GW propagation have not constrained tightly yet. In this paper, we investigate the measurement precisions of the amplitude damping rate (equivalently, the time variation of the gravitational coupling for GWs) and graviton mass in the generalized framework of GW propagation with the future detectors such as Voyager, Cosmic Explorer, and Einstein Telescope. As a result, we show that the future GW observation can reach 1\% error for the amplitude damping. We also study the time variation of the gravitational couplings in Horndeski theory by performing Monte Carlo-based numerical simulations. From the simulation results, we find that the current accelerating Universe prefers the models with less damping of GWs and that the equivalence principle can be tested at the level of 1\% by the future GW observation.
\end{abstract}


\maketitle

\section{Introduction}
\label{sec:Intro}

To explain the origin of the accelerating expansion of the Universe at present, a possibility to modify gravity theories at cosmological distance has been proposed \cite{Tsujikawa:2010zza,Nojiri:2010wj,Clifton:2011jh,Joyce:2016vqv,Nojiri:2017ncd}. If gravity strength is modified, it can affect the cosmological observables such as the cosmic microwave background and the large-scale structure (see, for instance, \cite{Ade:2015rim,Bellini:2015xja,Renk:2016olm,Kreisch:2017uet,Nersisyan:2018auj,Peirone:2017ywi,Espejo:2018hxa,Frusciante:2018jzw,Abbott:2018xao,Noller:2018wyv,Camera:2013xfa,SpurioMancini:2019rxy} for recent works and \cite{Ishak:2018his} for a review). 
At the same time, the modification of gravity changes the properties of gravitational-wave (GW) propagation as well \cite{Saltas:2014dha,Nishizawa:2017nef}. Therefore, searching for deviations from general relativity during the propagation of a GW is also crucial to test gravity at cosmological scales.

Recently the coincidence detection of GW170817 and GRB170817A \cite{GW170817:detection} brought us the first opportunity to measure the speed of a GW from the arrival time difference and constrained the deviation from the speed of light at the level of $10^{-15}$ \cite{GW170817:GRB}. Consequently, from this constraint, the strong limit on gravity modifications relevant to the current accelerating expansion of the Universe has been obtained ~\cite{Baker:2017hug,Creminelli:2017sry,Sakstein:2017xjx,Ezquiaga:2017ekz,Arai:2017hxj,Gumrukcuoglu:2017ijh,Oost:2018tcv,Boran:2017rdn,Nojiri:2017hai,Nishizawa:2018srh,Casalino:2018tcd}. 

Other than the modification on the propagation speed, the variation of gravitational constants is one of the prominent signatures of modified gravity. In modified gravity theories, there appear in general multiplicate gravitational couplings for the Poisson equation $G_{\rm matter}$, the gravitational lensing equation $G_{\rm light}$, and GWs $G_{\rm gw}$ \cite{Song:2010fg,Johnson:2015aaa}. One of the theoretical frameworks including many specific theories with a single scalar field is Horndeski theory \cite{Horndeski:1974wa,Deffayet:2011gz,Kobayashi:2011nu}. In the Horndeski theory, the gravitational couplings, $G_{\rm matter}$ and $G_{\rm light}$, are given in a quasi-static regime \cite{DeFelice:2011hq, Tsujikawa:2014mba, Gleyzes:2015rua, Peirone:2017ywi} and are different each other due to the non-trivial contribution from scalar field fluctuations. That is, a dynamical scalar field leads to different gravitational couplings, depending on distance scale and time, and violates the equivalence principle. In addition, the gravitational coupling for GW, $G_{\rm gw}$, is different from other two couplings and its time variation affects the amplitude of a GW \cite{Belgacem:2017ihm,Amendola:2017ovw,Nunes:2018zot}. Therefore, measuring the gravitational couplings at different times by multiple tracers and testing the equivalence to the Newton constant are a crucial direction to pin down a correct theory of gravity at cosmological scales. 

This paper is the third one in the series of our study on GW propagation test of gravity. In the first paper \cite{Nishizawa:2017nef}, we have formulated the generalized GW propagation (gGP) framework, which parametrizes almost all the modifications on GW propagation: propagation speed, modified dispersion relation, amplitude damping, massive graviton, and a source term. Within this framework, we derived a parametrized GW waveform in an analytical way and performed a parameter estimation study with the current GW detector network composed of aLIGO at Hanford and Livingston, and aVIRGO. Then in the second paper \cite{Arai:2017hxj}, we have studied the constraint on the model parameters in the gGP framework from GW170817. To be concrete, we considered the Horndeski theory, in which the GW speed and the amplitude damping rate are modified. With a numerical model sampling based on a Monte-Carlo method, we investigated model distributions in the parameter space and selected out the viable parameter region of the Horndeski theory. In this paper, we focus on the sensitivities of the future GW detectors such as Voyager, Cosmic Explorer (CE), and Einstein Telescope (ET) to the amplitude damping of a GW in the gGP framework, limiting the propagation speed of a GW to the speed of light. Then we compute gravitational couplings in the Horndeski theory and discuss the implication for the model space searched by the future measurement of the GW amplitude damping.

This paper is organized as follows. In Sec.~\ref{sec2}, we review the gravitational couplings appearing in the Horndeski theory and the modification of GW propagation. In Sec.~\ref{sec:GW-constraint}, we forecast the measurement errors of modified gravity parameters in the era of the third-generation GW detectors with the Fisher information matrix. Then the future prospect of the GW observations is discussed in detail. In Sec.~\ref{sec:param_Horn}, we introduce the numerical parametrization of the Horndeski theory, extending the previous parametrization to higher redshifts and focusing on the time variation of the gravitational couplings, and show what parameter region of the theory is tested by the future GW observations. Section \ref{sec:discussions} is devoted to discussions about the other measurements of the time variation of the gravitational couplings and the sensitivity comparison. We summarize in Sec.~\ref{sec:conclusions}.

\section{Gravitational couplings in modified gravity}
\label{sec2}

\subsection{Gravity force}
\label{sec:Gs_Horn}

First we show the commonly-used parametrization for gravity strength in modified gravity theories. Let us assume the flat Friedmann-Lema\^{i}tre-Robertson-Walker (FLRW) Universe and take the conformal Newtonian gauge as
\begin{align}
ds^2 = -(1+2\Psi )dt^2 + a^2(t)(1-2\Phi)\delta_{ij}dx^idx^j\,.\label{FRW_metric}
\end{align}
where $\Psi$ is the Newton potential and $\Phi$ is the spatial curvature. If we treat $\Psi$ and $\Phi$ as perturbations in the FLRW background, $\Psi$ and $\Phi$ are given as solutions of the linear-order perturbation equations. In particular, under the quasi-static approximation (QSA), we ignore all dynamical terms in the equations of motion. Then the Poisson and lensing equations are given by \cite{Song:2010fg,Johnson:2015aaa}
\begin{align}
k^2 \Psi \simeq - 4\pi G_{\rm matter}(k,\tau) \delta \rho_{\rm m}\,,\label{Poisson_Gmatter} \\
k^2 (\Psi + \Phi) \simeq -8\pi G_{\rm light}(k,\tau) \delta \rho_{\rm m}\,,\label{Poisson_Glight} 
\end{align}
where $\tau$ is the conformal time, $\delta \rho_{\rm m}$ is the density fluctuation of matter. $G_{\rm matter}$ and $G_{\rm light}$ denote the effective gravitational couplings for matter clustering and gravitational lensing, respectively. Under the QSA the time evolution of $\Psi$ and $\Phi$ at the scales much smaller than the Hubble radius is ignored and $G_{\rm light}$ and $G_{\rm matter}$ stay almost constant. However, $G_{\rm light}$ and $G_{\rm matter}$ vary in time at the cosmological scales so that it is possible to see the variation of the gravitational couplings at different redshifts via matter clustering and gravitational lensing \cite{Ade:2015rim,Bellini:2015xja,Renk:2016olm,Kreisch:2017uet,Nersisyan:2018auj,Peirone:2017ywi,Espejo:2018hxa,Frusciante:2018jzw,Abbott:2018xao,Noller:2018wyv,Camera:2013xfa,SpurioMancini:2019rxy}.

In modified gravity theories, in general $G_{\rm matter} \neq G_{\rm light}$, i.~e., $\Psi$ and $\Phi$ are no longer equivalent each other, leading to the violation of the equivalence principle. Moreover, the values of $G_{\rm matter}$ and $G_{\rm light}$ also can deviate from the Newton constant $G_N$. On the contrary, at the smaller scales such that a system gravitates by itself and is decoupled from the cosmological expansion, the non-linear screening mechanism \cite{Khoury:2003aq, Khoury:2003rn, Vainshtein:1972sx} may work and set the gravitational couplings to $G_N$ uniquely in order to pass the experimental tests in the Solar System.

In this paper we specify a theory to Horndeski theory \cite{Horndeski:1974wa,Deffayet:2011gz,Kobayashi:2011nu} and focus on gravitation at cosmological scales. The Lagrangian density\footnote{$G_2 (\phi,X)$ is often written $K(\phi, X)$ in literature.} of the Horndeski theory after GW170817/GRB170817A \cite{Baker:2017hug,Creminelli:2017sry,Ezquiaga:2017ekz,Sakstein:2017xjx,Arai:2017hxj}, setting GW propagation speed exactly to unity, is
\begin{equation}
{\cal L} = G_2(\phi, X)-G_3(\phi,X)\Box \phi +G_4(\phi)R \;.
\label{action}
\end{equation}
Here $X = -\phi_{;\mu}\phi^{;\mu}/2$, the canonical kinetic energy density of $\phi$. In this theory, a single scalar field exists and determines the gravitational couplings, depending on the mass $M$ and amplitude fluctuations of the scalar field.

When one considers the fluctuations of a scalar field on a given cosmological background, the scalar field acquires the mass $M$. For a canonical field with a potential $V(\phi)$, $M^2$ is nothing but the second derivative of $V$ with respect to $\phi$, $V_{\phi \phi}$. In the theory given by Eq.~(\ref{action}), however, $M$ arises not only from $V$ in $G_2$ but also from $G_3$ and $G_4$ (see Eq.~(35) in \cite{DeFelice:2011hq} for the exact expression of $M$). When $M$ is much larger than the Hubble scale $H$, the scalar field fluctuation does not propagate at cosmological scales. On the contrary, when $M \sim H$, the fluctuation affects the cosmic expansion. Since we are interested in the case when the scalar field fluctuations significantly modify the gravitational force at cosmological scales in accordance with the late-time acceleration of the Universe, we consider the case of $M \sim H_0$, where $H_0$ is the Hubble constant. Hence it is convenient to divide into two cases: super-Compton limit ($k/a \ll M$) and sub-Compton limit ($k/a \gg M$). According to \cite{DeFelice:2011hq,Pogosian:2016pwr}, the couplings at the super-Compton scales become
\begin{align}
G_{\rm matter} = G_{\rm light} = G_N\frac{M^2_{\rm pl}}{M^2_*}\,, \label{supG_aT0}
\end{align}
 while at the sub-Compton scales 
 \begin{align}
 & G_{\rm matter} = G_N\frac{M^2_{\rm pl}}{M^2_*} (1+\beta^2_\xi) \,, \label{subGmatter_aT0}\\
  &G_{\rm light} =  G_N\frac{M^2_{\rm pl}}{M^2_*} \left [1+\beta^2_\xi + \sqrt{\frac{2}{c^2_S D}}\frac{\alpha_M \beta_\xi}{2}\right ]\,,\label{subGlight_aT0}
 \end{align}
 where 
 \begin{align}
 M^2_* &= 2 G_4\,, \\
 \alpha_M &= \frac{1}{H}\frac{d \log M^2_*}{dt} \;. \label{eq:alphaM}
\end{align}
Here and hereafter we use $M_{\rm pl}$ to denote the reduced Planck mass. The other functions, $\beta_{\xi}$, $D$, and $c_s^2$, are given in Appendix~\ref{app:computations}. Particularly, the function $\beta_{\xi}$ plays an important role to distinguish the gravitational couplings. This difference is originated from the fluctuations of a scalar field.

\subsection{GW propagation}

Following the general formulation of GW propagation in an effective field theory \cite{Saltas:2014dha}, tensor perturbations obey the equation of motion
\begin{equation}
h^{\prime\prime}_{ij}+(2+\nu){\cal H} h^{\prime}_{ij} + (c_{\rm T}^2 k^2 + a^2 \mu^2) h_{ij} = a^2 \Gamma \gamma_{ij} \;,
\label{eq:GWeq1}
\end{equation}
where the prime is a derivative with respect to conformal time, $a$ is the scale factor, ${\cal H} \equiv a^{\prime}/a$ is the Hubble parameter in the conformal time, $\nu={\cal H}^{-1} (d \ln M_{*}^2/d\tau)$ is the Planck mass run rate, $c_{\rm T}$ is GW propagation speed, and $\mu$ is graviton mass. The source term $\Gamma \gamma_{ij}$ arises from anisotropic stress. In the absence of the source term ($\Gamma=0$)\footnote{Even with the source term, an analytical solution can be obtained, but its expression is much more complicated \cite{Nishizawa:2017nef}.}, the WKB solution for Eq.~(\ref{eq:GWeq1}) in the gGP framework is obtained \cite{Nishizawa:2017nef}: 
\begin{align}
h &= {\cal{C}}_{\rm MG} h_{\rm GR} \;, 
\label{eq3} \\
{\cal{C}}_{\rm MG} &= e^{- {\cal D}} e^{- i k \Delta T} \;,
\end{align}
with
\begin{align}
{\cal D} &= \frac{1}{2} \int_0^z \frac{\nu}{1+z^{\prime}} dz^{\prime} \;, 
\label{eq:damping-factor}
\end{align}
\begin{align}
\Delta T &= \int_0^z \frac{1}{\cal H} \left( \frac{\delta_g}{1+z^{\prime}} -\frac{\mu^2}{2k^2 (1+z^{\prime})^3} \right) dz^{\prime} \;.
\label{eq:time-delay}
\end{align}
where ${\cal D}$ is the damping factor, and $\Delta T$ is the time delay due to the effective GW speed different from speed of light, and we defined $\delta_g =1-c_{\rm T}$ as a tiny parameter.

In the Horndeski theory, graviton is massless ($\mu=0$) and the correspondence of other two parameters to the $\alpha$ parametrization commonly used in literature is \cite{Saltas:2014dha}
\begin{align}
\nu &=\alpha_{\rm M} \, \label{eq:nu} \\
c_{\rm T}^2 &=1+\alpha_{\rm T} \;.
\end{align}
Substituting Eq.~(\ref{eq:nu}) into Eq.~(\ref{eq:damping-factor}) together with Eq.~(\ref{eq:alphaM}), we can express the amplitude modification in terms of the gravitational constant as
\begin{equation}
e^{- {\cal D}} = \frac{M_*(z)}{M_*(0)} = \sqrt{\frac{G_{\rm gw}(0)}{G_{\rm gw}(z)}} \;.
\label{eq4}
\end{equation}
Here we defined 
\begin{equation}
G_{\rm gw} \equiv \frac{M^2_{\rm pl}}{M^2_*}G_N \,.
\end{equation}
Since the GR waveform is inversely proportional to the luminosity distance, one can interpret the amplitude modification as a correction to the luminosity distance, defining the effective luminosity distance for GWs \cite{Belgacem:2017ihm,Amendola:2017ovw}
\begin{equation}
d_{\rm L}^{\rm gw} \equiv e^{\cal D} d_{\rm L}(z) = \sqrt{\frac{G_{\rm gw}(z)}{G_{\rm gw}(0)}} d_{\rm L}(z) \;,
\label{eq:dL-gw}
\end{equation} 
with
\begin{align}
d_{\rm L}(z) &= (1+z) \chi(z)\;, 
\label{eq:dL-em} \\
\chi(z) &= \int_{0}^{z} \frac{dz^{\prime}}{H(z^{\prime})} \;, \\
H(z) &= H_0 \left\{ \Omega_{\rm m} (1+z)^3 + (1-\Omega_{\rm m} ) \right\}^{1/2} \;, \label{Hz}
\end{align}
where $\chi(z)$ is the comoving distance to redshift $z$ and $H(z) \equiv \dot{a}/a$ is the Hubble parameter in physical time, $\Omega_{\rm m}$ and $H_0$ are the matter energy density at present and the Hubble constant. Note that the relation Eq.~(\ref{eq4}) is valid only at a level of linear perturbations and cannot be applicable to the whole path of GW propagation from a source to the Earth.

\section{Future constraint from GW observations}
\label{sec:GW-constraint}

In this section, we estimate the parameter errors in the future GW observations with the Fisher information matrix in the same way as our previous work \cite{Nishizawa:2017nef} except for setting $\delta_g=0$ and taking the Earth rotation effect into account for binary neutron stars (BNS). For the complete description of the parameter estimation method with the Fisher matrix, see \cite{Nishizawa:2017nef}.

\subsection{Numerical setup}

We consider the simplest waveform in which arbitrary functions $\nu$ and $\mu$ are assumed to be constant and $\delta_g$ and $\Gamma$ are zero. Setting $\delta_g=0$ is motivated by the recent measurement of the GW speed from GW170817/GRB170817A \cite{GW170817:detection, GW170817:GRB, GW170817:multimessenger}. $\Gamma=0$ is just for simplicity, but is true in most gravity theories including the Horndeski theory. While we simply assume that $\nu$ is constant and the effective luminosity distance in the form of Eq.~(\ref{eq:dLeff}). However, for the case of a time-dependent $\nu$, a concrete parametrization for the effective luminosity distance has been suggested in \cite{Belgacem:2018lbp}.

Under the assumptions above, the waveform in Eqs.~(\ref{eq3})-(\ref{eq:time-delay}) is reduced to
\begin{align}
h &= (1+z)^{-\nu/2} e^{- i k \Delta T}  h_{\rm GR} 
\label{eq9} \;, \\
\Delta T &= - \frac{\mu^2}{2k^2} \int_{0}^{z} \frac{dz^{\prime}}{(1+z^{\prime})^3 {\cal H}} \;.
\end{align}
For the GR waveform, $h_{\rm GR}$, we will use the phenomenological waveform (PhenomD) \cite{Khan:2016PRD} (compiled in Appendix of \cite{Nishizawa:2017nef}), which is an up-to-date version of inspiral-merger-ringdown (IMR) waveform for aligned-spinning (nonprecessing) binary black holes (BBH) with mass ratio up to 1:18. While for BH-NS binaries and BNS, we will use the inspiral waveform up to 3.5 PN order in phase, which is an early inspiral part of the PhenomD waveform, in order to avoid ambiguities in tidal deformation and disruption of a NS. The waveform in Eq.~(\ref{eq9}) has in total 13 parameters: the redshifted chirp mass ${\cal{M}}$, the symmetric mass ratio $\eta$, time and phase at coalescence, $t_{\rm c}$ and $\phi_c$, redshift $z$, symmetric and asymmetric spins, $\chi_s$ and $\chi_a$, the angle of orbital angular momentum measured from the line of sight $\iota$, sky direction angles of a source, $\theta_{\rm S}$ and $\phi_{\rm S}$, polarization angle $\psi$, and gravitational modification parameters, $\nu$ and $\mu$. We will assume a flat Lambda-Cold-Dark-Matter ($\Lambda$CDM) model and fix cosmological parameters to those determined by Planck satellite \cite{Planck2015cosmology}. This is justified because the expansion of the Universe is accelerated at low redshifts ($z \lesssim 1$), while it should be consistent with the standard cosmology at higher redshifts ($z \gg 1$). Then the luminosity distance $d_{\rm L}$ is mapped into redshift $z$ by the standard formula of the luminosity distance in Eq.~(\ref{eq:dL-em}). On the other hand, as seen from Eq.~(\ref{eq:dL-gw}), the effective luminosity distance for GWs is 
\begin{equation}
d_{\rm L}^{\rm gw} = (1+z)^{-\nu/2} d_{\rm L}(z) \;.
\label{eq:dLeff}
\end{equation}

The forecast constraints from the planned GW detectors are estimated with the Fisher information matrix \cite{Finn:1992wt,Cutler:1994ys}
\begin{equation}
\Gamma_{ab} = 4 \sum_{I} \, {\rm{Re}} \int_{f_{\rm{min}}}^{f_{\rm{max}}}
 \frac{\partial_{a} \tilde{h}_I^{\ast}(f)\, \partial_{b}
 \tilde{h}_I(f)}{S_{\rm{h}}(f)} df \;,
\end{equation}
where $\partial_a$ denotes a derivative with respect to a parameter $\theta_a$, $\tilde{h}_I$ is the Fourier amplitude of a GW signal from $I$th detector, which is $h$ in Eq.~(\ref{eq9}) multiplied by the geometrical factor \cite{Nishizawa:2017nef}, and $S_{h}$ is the noise power spectral density of a detector. We consider here Voyager, ET, and CE, whose fitting formulas to the noise curves are given in Appendix \ref{sec:AppB} and are shown in Fig.~\ref{fig:noise-curve}. However, the locations of ET and CE have not fixed yet. For the analysis, we assume that they are at the sites of LIGO Hanford and VIRGO for two detector cases and in addition at the site of LIGO Livingston for three detector cases. To implement a Gaussian prior on a source redshift from the follow-up observation of an electromagnetic counterpart or identification of a unique host galaxy, we take a standard deviation of $z$ as $\Delta z=0.001$ and add $1/(\Delta \log z)^2$ to the $(\log z, \log z)$ component of the Fisher matrix. 

\begin{figure}[t]
\begin{center} 
\includegraphics[width=7.5cm]{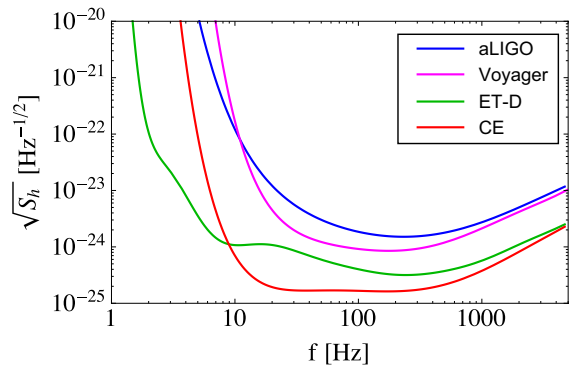}
\caption{Detector noise curves: aLIGO (blue), Voyager (magenta), ET-D (green), CE (red).} 
\label{fig:noise-curve}
\end{center}
\end{figure}

\begin{figure}[t]
\begin{center} 
\includegraphics[width=7.5cm]{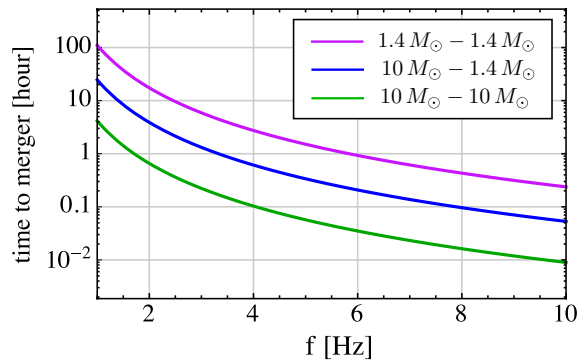}
\caption{Time to merger as a function of frequency for $10M_{\odot}$-$10M_{\odot}$ (green), $10M_{\odot}$-$1.4M_{\odot}$ (blue), $1.4M_{\odot}$-$1.4M_{\odot}$ (purple).} 
\label{fig:time-to-merger}
\end{center}
\end{figure}

As shown in Fig.~\ref{fig:noise-curve}, the third-generation detectors are much more sensitive at lower frequencies and can start observing GWs from compact binaries much earlier than the second-generation detectors. In Fig.~\ref{fig:time-to-merger}, we plot the time to merger at Newtonian order as a function of frequency \cite{Cutler:1994ys}
\begin{equation}
t_{\rm merge}=\frac{5}{256} {\cal M} (\pi {\cal M} f )^{-8/3} \;.
\end{equation}
For BNS, it is $\sim 2$ hours and $\sim 1$ day before merger at $5\,{\rm Hz}$ and $2\,{\rm Hz}$, respectively. Even a BH-NS binary takes several hours before merger below $3\,{\rm Hz}$. This is a merit for detectors because the Earth rotation during observing a signal allows the detector response functions to change their directions and improve the sky localization even with less number of detectors \cite{Zhao:2017cbb}. Therefore, the time evolution of the detector response functions should be taken into account correctly in the Fisher matrix analysis. The longitudes of detector locations is a function of time, $\phi_I(t)=\phi_I(0)+\omega_{\rm E} t$, where $\omega_{\rm E}=2\pi/1{\rm day}$ is the angular frequency of the Earth rotation. Since GW frequency is a function of time for a compact binary, the time $t$ in the Fisher matrix needs to be expressed in terms of frequency as $t(f)= t_{\rm c}-t_{\rm merge}(f)$. We study in Appendix~\ref{sec:AppC} how much the parameter estimation errors are affected by the presence of the Earth rotation effect and find that one-CE, one-ET-D, and two-CE cases need the Earth rotation effect to be considered.

In the following analysis, we will set fiducial parameters, $t_{\rm c}$, $\phi_c$, $\chi_s$, $\chi_a$, $\nu$, and $\mu$ to zero and randomly generate sky locations ($\theta_{\rm S}$, $\phi_{\rm S}$) and other angle parameters ($\iota$, $\psi$) for compact binaries with fixed masses and redshift. In the procedure of the source generation, we set the network signal-to-noise ratio (SNR) threshold for detection to $\rho=8$ and keep only sources with $\rho>8$.

\subsection{Parameter estimation errors: redshift dependence}
\label{sec:err-redshift}

\begin{figure*}[t]
\begin{center} 
\includegraphics[width=15.5cm]{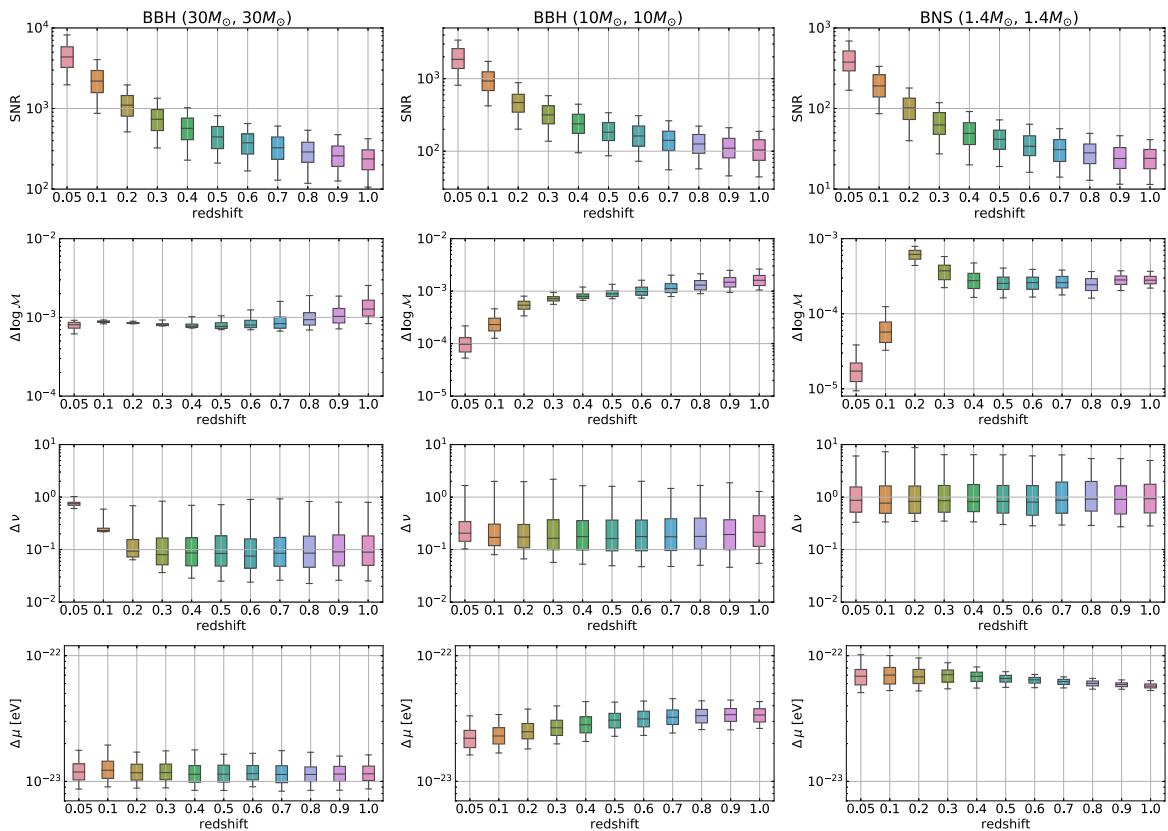}
\caption{Redshift dependences of SNR and parameter estimation errors with 3 CE for BBH with equal masses $30\,M_{\odot}$ (left), BBH with equal masses $10\,M_{\odot}$ (middle), and BNS with equal masses $1.4\,M_{\odot}$ (right). The thick and thin bars show 25-75\% and 5-95\% ranges of the probability distributions}, respectively.
\label{fig:PE-dist-z}
\end{center}
\end{figure*}

The parameter estimation errors are computed from the inverse Fisher matrix. We define the sky localization error as
\begin{equation}
\Delta \Omega_{\rm S} \equiv 2 \pi | \sin \theta_{\rm S}| \sqrt{(\Delta \theta_{\rm S})^2(\Delta \phi_{\rm S})^2 - \langle \delta \theta_{\rm S} \delta \phi_{\rm S} \rangle^2 } \;,
\end{equation}
where $\langle \cdots \rangle$ stands for ensemble average and $\Delta \theta_{\rm S} \equiv \langle (\delta \theta_{\rm S})^2 \rangle^{1/2}$ and $\Delta \phi_{\rm S} \equiv \langle (\delta \phi_{\rm S})^2 \rangle^{1/2}$. 

Figure~\ref{fig:PE-dist-z} shows the redshift dependences of SNR and parameter estimation errors for $30M_{\odot}$ -$30M_{\odot}$ BBH, $10M_{\odot}$ - $10M_{\odot}$ BBH, and $1.4M_{\odot}$ - $1.4M_{\odot}$ BNS observed with three CEs. This choice of the detector network might be too optimistic, but for qualitative understanding about the redshift dependence of the parameter estimation errors we assume the most optimistic detector network. Regarding to other more realistic detector networks, we consider them in the next subsection. 

There are two interesting features in Fig.~\ref{fig:PE-dist-z}: (i) heavier compact binaries give smaller errors in the modification parameters $\nu$ and $\mu$, (ii) the errors in $\nu$ and $\mu$ hardly depend on a redshift or distance except for $30M_{\odot}$ -$30M_{\odot}$ BBH at lower redshifts. The former is merely because of larger SNR from massive binaries. The latter is an accumulation effect during propagation and is explained as follows. SNR is inversely proportional to luminosity distance\footnote{Our choice of the fiducial value of $\nu$ is $\nu=0$, which does not modify the luminosity distance. The same scaling of SNR as in GR holds.}. Particularly at low redshifts, ${\rm SNR} \propto d_{\rm L}^{-1} \propto z^{-1}$. The error of the amplitude modification, forgetting about angular dependences, is at best given by
\begin{align}
\left| \frac{\Delta [(1+z)^{-\nu/2}]}{(1+z)^{-\nu/2}} \right| &= \left| \frac{\nu}{2} \frac{\Delta z}{1+z} +\frac{1}{2} \log (1+z) \Delta \nu \right| \nonumber \\
&\sim \frac{1}{\rm SNR} \;.
\end{align}
Since our choice of the fiducial value is $\nu=0$, the above equation is reduced to
\begin{equation}
\Delta \nu \sim \frac{2}{\log (1+z)\times {\rm SNR}} \sim {\rm constant} \;.
\label{eq:nu-error}
\end{equation}
The second equality holds, particularly at lower redshifts $z \lesssim 1$, at which the dependence of ${\rm SNR}\propto z^{-1}$ is compensated by the factor $\log(1+z)$ due to the distance traveled by GWs. Thus, thanks to the accumulation effect during propagation, the errors of $\nu$ is almost independent of a source redshift. The anomalous deviation from the redshift independence for $30M_{\odot}$ -$30M_{\odot}$ BBH at lower redshifts is attributed to the systematic error arising from parameter degeneracies. The errors of $\log {\cal M}$ for $10M_{\odot}$ - $10M_{\odot}$ BBH and $1.4M_{\odot}$ - $1.4M_{\odot}$ BNS are basically limited by detector sensitivity or SNR (the redshift prior also helps determine the parameters to some extent). However, the $\log {\cal M}$ error for $30M_{\odot}$ -$30M_{\odot}$ BBH is not improved as the source redshift is smaller and SNR increases. This is the parameter degeneracy caused by the shorter GW signal of a heavier compact binary, for which the chirp mass in GW phase is more difficult to be determined\footnote{The same behavior has also been seen in a similar analysis with the Fisher matrix \cite{Takeda:2018uai}.}. Since the chirp mass appears also in GW amplitude, the systematic error prevents $\Delta \nu$ in Eq.~(\ref{eq:nu-error}) from scaling with SNR and worsens $\Delta \nu$ at lower redshifts (shorter propagation distance).

The redshift-independent behavior implies that GW sources of a same kind at different redshifts (similar masses) are almost equivalent for the use to test GW amplitude damping and graviton mass.

\subsection{Parameter estimation errors: source and detector dependences}

\begin{figure*}[t]
\begin{center} 
\includegraphics[width=14cm]{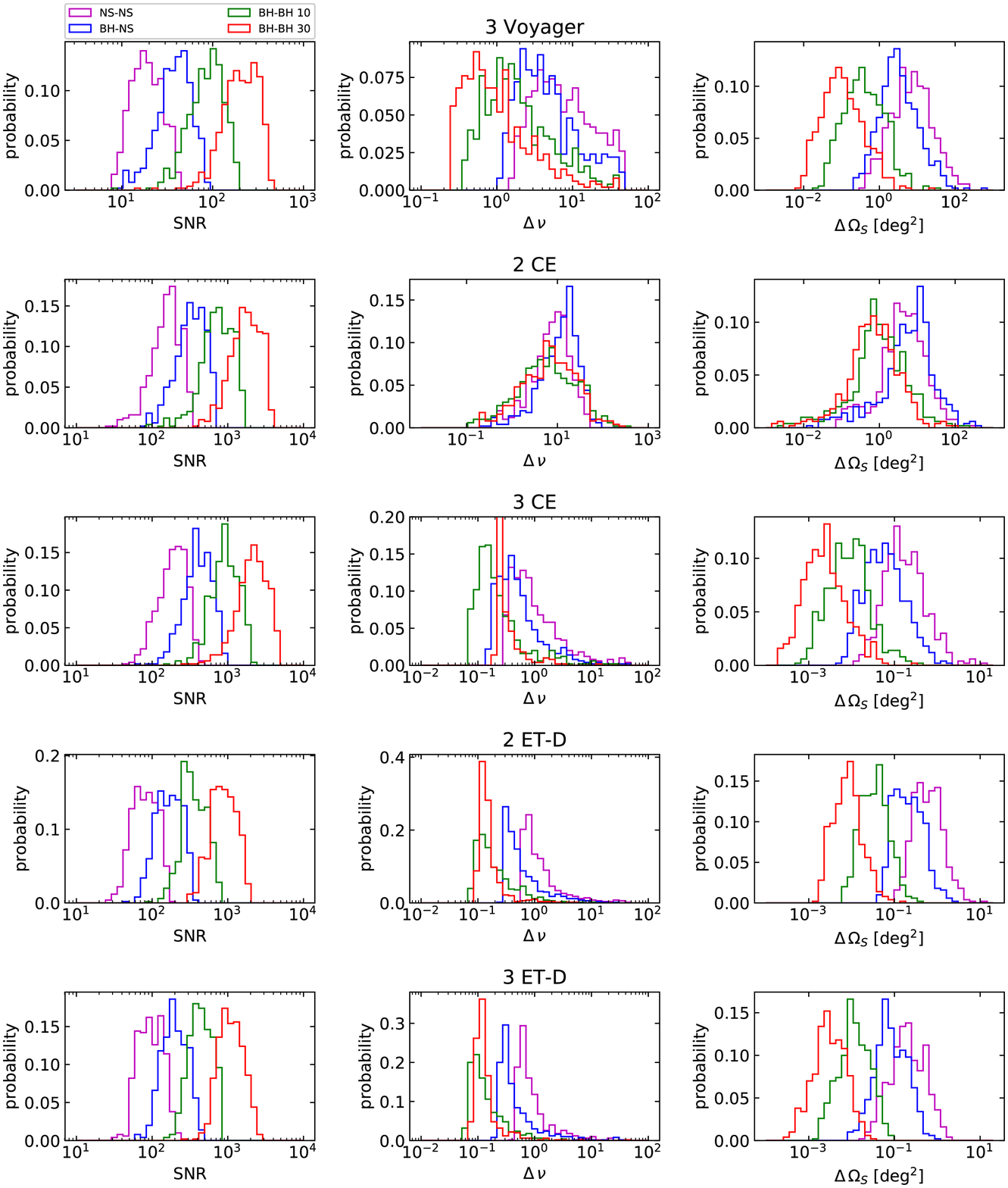}
\caption{Source dependence of SNR and parameter estimation errors for a source at $z=0.1$, showing mass dependence: $30M_{\odot}$-$30M_{\odot}$ (red), $10M_{\odot}$-$10M_{\odot}$ (green), $10M_{\odot}$-$1.4M_{\odot}$ (blue), $1.4M_{\odot}$-$1.4M_{\odot}$ (purple).} 
\label{fig:PE-dist-m-z01}
\end{center}
\end{figure*}

\begin{figure*}[t]
\begin{center} 
\includegraphics[width=14cm]{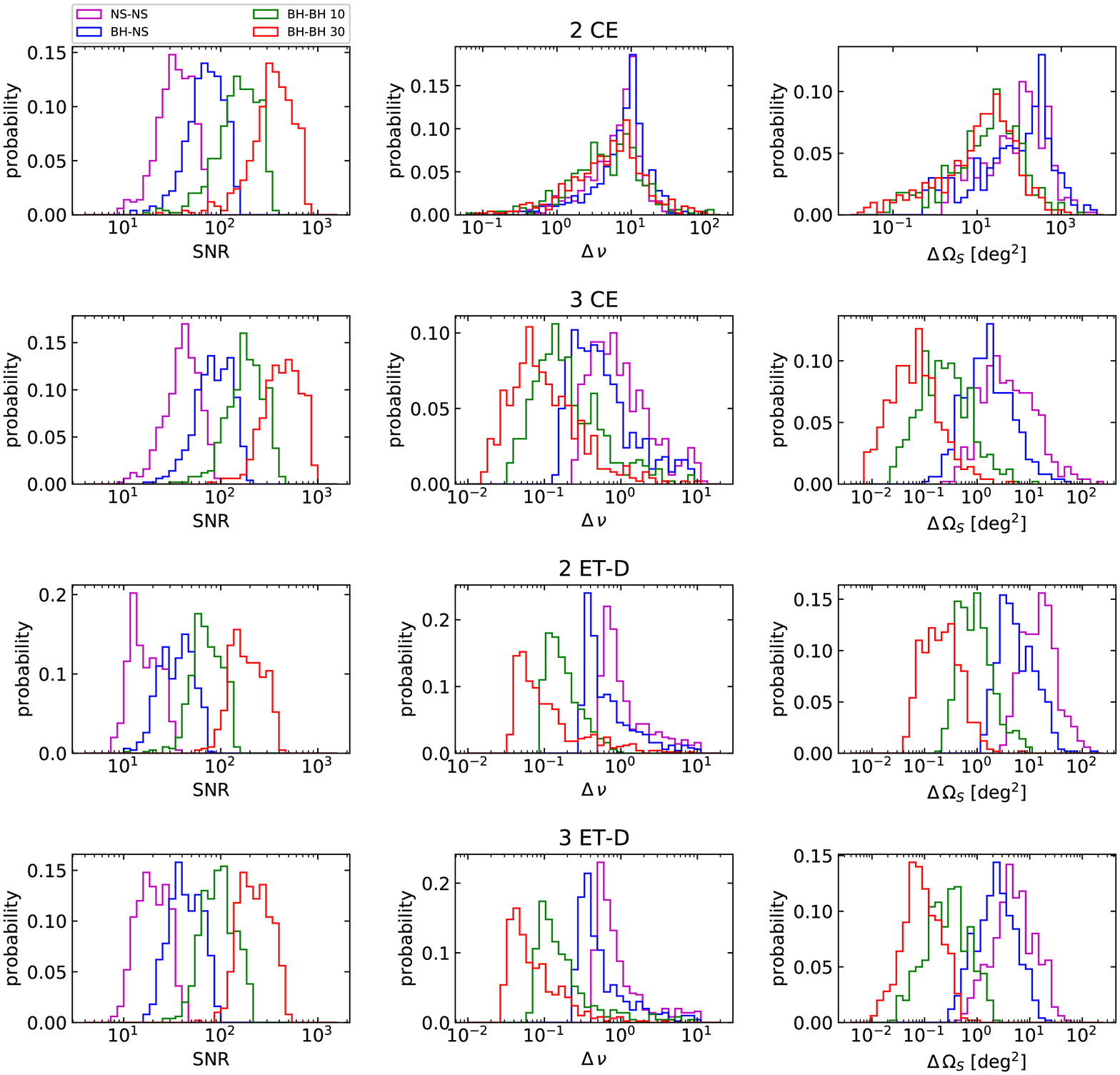}
\caption{Source dependence of SNR and parameter estimation errors for a source at $z=0.5$, showing mass dependence: $30M_{\odot}$-$30M_{\odot}$ (red), $10M_{\odot}$-$10M_{\odot}$ (green), $10M_{\odot}$-$1.4M_{\odot}$ (blue), $1.4M_{\odot}$-$1.4M_{\odot}$ (purple).} 
\label{fig:PE-dist-m-z05}
\end{center}
\end{figure*}

In Figs.~\ref{fig:PE-dist-m-z01} and \ref{fig:PE-dist-m-z05} and Tables~\ref{tab1} and \ref{tab2}, we show the source and detector dependences of the parameter estimation errors. The common feature among detectors and GW sources is that the parameter estimation errors are proportional to the inverse of SNR. For all detectors except for 2 CE, the sky localization errors $\Delta \Omega_{\rm S}$ at $z=0.1$ and $z=0.5$ follow the SNR scaling law, $\propto {\rm SNR}^{-2}$. The $\nu$ errors for sources at $z=0.5$ in Fig.~\ref{fig:PE-dist-m-z05} also obey the scaling law, $\propto {\rm SNR}^{-1}$. But at $z=0.1$ in Fig.~\ref{fig:PE-dist-m-z01}, $30M_{\odot}$-$30M_{\odot}$ BBHs give larger error in $\nu$. This is caused by the parameter degeneracy problem that we mentioned in the previous subsection. The shorter GW signal from a heavier compact binary makes the determination of the chirp mass in GW phase more difficult and prevents $\nu$ error from improving at lower redshifts. Another exception is the 2-CE case, in which it is relatively difficult to localize the source direction by triangulation. On the other hand, ET has triangle shape and is composed of effectively two orthogonal detectors at the same cite. Therefore, the case of 2 ET-D detectors has four orthogonal detectors in effect and enables to point the sky direction.

In summary, heavier sources give a tighter constraint on $\nu$. Among the sources we studied, $30M_{\odot}$-$30M_{\odot}$ BBH is the best source to measure $\nu$, irrespective of a source redshift. Compared the sensitivities to $\nu$ in Table~\ref{tab1} with the previous results from the second-generation detectors such as aLIGO, aVIRGO, and KAGRA ($\Delta \nu \approx 1.2$) \cite{Nishizawa:2017nef}, the detector networks of 3 Voyager, 2 ET-D, and 3 CE (or 3 ET-D) can reach about 4, 30, 60 times better sensitivities ($\Delta \nu \approx 0.3$, $0.04$, $0.02$) to $\nu$, respectively, with a single GW event among top 1\% events. While 2 CE cannot improve the error because of poor sky localization. As shown in Table~\ref{tab2}, the third-generation detectors can measure graviton mass with $\Delta \mu \approx 1.0\times 10^{-23}\,{\rm eV}$, which is only 4-5 times better than the measurement with the second-generation detectors.

\begin{table*}[t]
\begin{center}
\begin{tabular}{cccccc}
\hline \hline
detectors \& & & $z=0.1$ & & $z=0.5$ & \\
source redshift & source & $\;\;\Delta \nu$ (top 1\%) \;\; & $\;\;\Delta \nu$ (median) \;\; & $\;\;\Delta \nu$ (top 1\%) \;\; & $\;\;\Delta \nu$ (median) \;\; \\
\hline \hline  
3 Voyager & $30\,M_{\odot}$ BBH & 0.259 & 0.807 & --- & --- \\
& $10\,M_{\odot}$ BBH & 0.396 & 1.63 & --- & --- \\
& BH-NS & 1.24 & 3.97 & --- & --- \\
& BNS & 1.75 & 6.82 & --- & --- \\
\hline 
2 CE & $30\,M_{\odot}$ BBH & 0.238 & 6.95 & 0.175 & 5.55 \\
& $10\,M_{\odot}$ BBH & 0.156 & 5.98 & 0.201 & 4.75 \\
& BH-NS & 0.429 & 13.3 & 0.389 & 8.91 \\
& BNS & 0.488 & 8.13 & 0.702 & 7.39 \\
\hline
3 CE & $30\,M_{\odot}$ BBH & 0.215 & 0.231 & 0.0184 & 0.0844 \\
& $10\,M_{\odot}$ BBH & 0.0732 & 0.171 & 0.0381 & 0.162 \\
& BH-NS & 0.161 & 0.423 & 0.160 & 0.478 \\
& BNS & 0.299 & 0.768 & 0.254 & 0.826 \\
\hline  
2 ET-D & $30\,M_{\odot}$ BBH & 0.0967 & 0.137 & 0.0369 & 0.0794 \\
& $10\,M_{\odot}$ BBH & 0.0794 & 0.156 & 0.0902 & 0.159 \\
& BH-NS & 0.278 & 0.465 & 0.313 & 0.500 \\
& BNS & 0.613 & 0.942 & 0.536 & 0.862 \\
\hline
3 ET-D & $30\,M_{\odot}$ BBH & 0.0797 & 0.117 & 0.0300 & 0.0589 \\
& $10\,M_{\odot}$ BBH & 0.0625 & 0.113 & 0.0693 & 0.136 \\
& BH-NS & 0.218 & 0.343 & 0.248 & 0.427 \\
& BNS & 0.452 & 0.732 & 0.438 & 0.723 \\
\hline  
\end{tabular}
\end{center}
\caption{Top 1\% and median errors of $\nu$ for sources at redshifts $z=0.1$ and $z=0.5$.}
\label{tab1}
\end{table*}

\begin{table*}[t]
\begin{center}
\begin{tabular}{cccccc}
\hline \hline
detectors \& & & $z=0.1$ & & $z=0.5$ & \\
source redshift & source &  $\;\;\Delta \mu$ (top 1\%) \;\; & $\;\;\Delta \mu$ (median) \;\; & $\;\;\Delta \mu$ (top 1\%) \;\; & $\;\;\Delta \mu$ (median) \;\; \\
\hline \hline  
3 Voyager & $30\,M_{\odot}$ BBH & 2.34 & 3.26 & --- & --- \\
& $10\,M_{\odot}$ BBH & 7.52 & 10.2 & --- & --- \\
& BH-NS & 8.96 & 12.0 & --- & --- \\
& BNS & 16.6 & 17.5 & --- & --- \\
\hline 
2 CE & $30\,M_{\odot}$ BBH & 0.963 & 1.38 & 0.897 & 1.29 \\
& $10\,M_{\odot}$ BBH & 1.76 & 2.56 & 2.35 & 3.31 \\
& BH-NS & 4.07 & 5.59 & 3.55 & 4.19 \\
& BNS & 4.80 & 7.51 & 5.95 & 7.01 \\
\hline
3 CE & $30\,M_{\odot}$ BBH & 0.904 & 1.23 & 0.817 & 1.14 \\
& $10\,M_{\odot}$ BBH & 1.68 & 2.30 & 2.17 & 3.07 \\
& BH-NS & 3.96 & 5.17 & 3.20 & 3.79 \\
& BNS & 5.28 & 7.01 & 5.38 & 6.59 \\
\hline  
2 ET-D & $30\,M_{\odot}$ BBH & 0.967 & 1.38 & 1.07 & 1.55 \\
& $10\,M_{\odot}$ BBH & 1.13 & 1.63 & 2.00 & 2.55 \\
& BH-NS & 3.83 & 5.28 & 3.69 & 4.14 \\
& BNS & 4.97 & 7.05 & 5.21 & 6.47 \\
\hline
3 ET-D & $30\,M_{\odot}$ BBH & 0.843 & 1.25 & 0.984 & 1.41 \\
& $10\,M_{\odot}$ BBH & 1.06 & 1.45 & 1.24 & 1.82 \\
& BH-NS & 3.32 & 4.95 & 3.45 & 4.03 \\
& BNS & 4.51 & 6.48 & 4.58 & 6.04 \\
\hline 
\end{tabular}
\end{center}
\caption{Top 1\% and median errors of $\mu$ for sources at redshifts $z=0.1$ and $z=0.5$. The figures are in the unit of $10^{-23}\,{\rm eV}$.}
\label{tab2}
\end{table*}

\subsection{Redshift identification and multiple sources}

In the subsections above, we considered sources with known redshifts. In reality, obtaining a source redshift is more difficult at higher redshifts, though we observe more sources there. It indicates that there is a trade-off relation between the number of sources and redshift identification. We will discuss this issue here from more practical side. 

There are two ways to obtain a source redshift for an individual GW event: observing an electromagnetic counterpart \cite{GW170817:multimessenger} or identifying a unique host galaxy \cite{Cutler:2009qv}. The success fraction of redshift identification is still highly uncertain and depends on the emission mechanisms of electromagnetic waves or the properties of host galaxies in which GW sources reside. Here we introduce the success fraction (efficiency) of redshift identification as a redshift-dependent parameter $\epsilon(z)$.
Then the number of sources available for the measurement of GW amplitude damping at redshift $z$ is written as
\begin{equation}
\frac{dN}{dz} = \epsilon(z) \frac{4\pi \chi^2(z) \dot{n}(z) T_{\rm obs}}{(1+z)H(z)}  \;,
\end{equation} 
where $T_{\rm obs}$ is the observation time and
$\dot{n}(z)$ is the merger rate per unit comoving volume and unit
proper time at redshift $z$. The factor $\epsilon(z)$ still has large uncertainty, but it is convenient for theoretical studies to parametrize
\begin{equation}
\epsilon(z) = \epsilon_0 \Theta (z_{\rm max}-z) \;, 
\end{equation}
where $z_{\rm max}$ is the maximum redshift beyond which the redshift identification fails, and $\Theta (\cdot)$ is the step function.

\begin{figure*}[t]
\begin{center} 
\includegraphics[width=6.5cm]{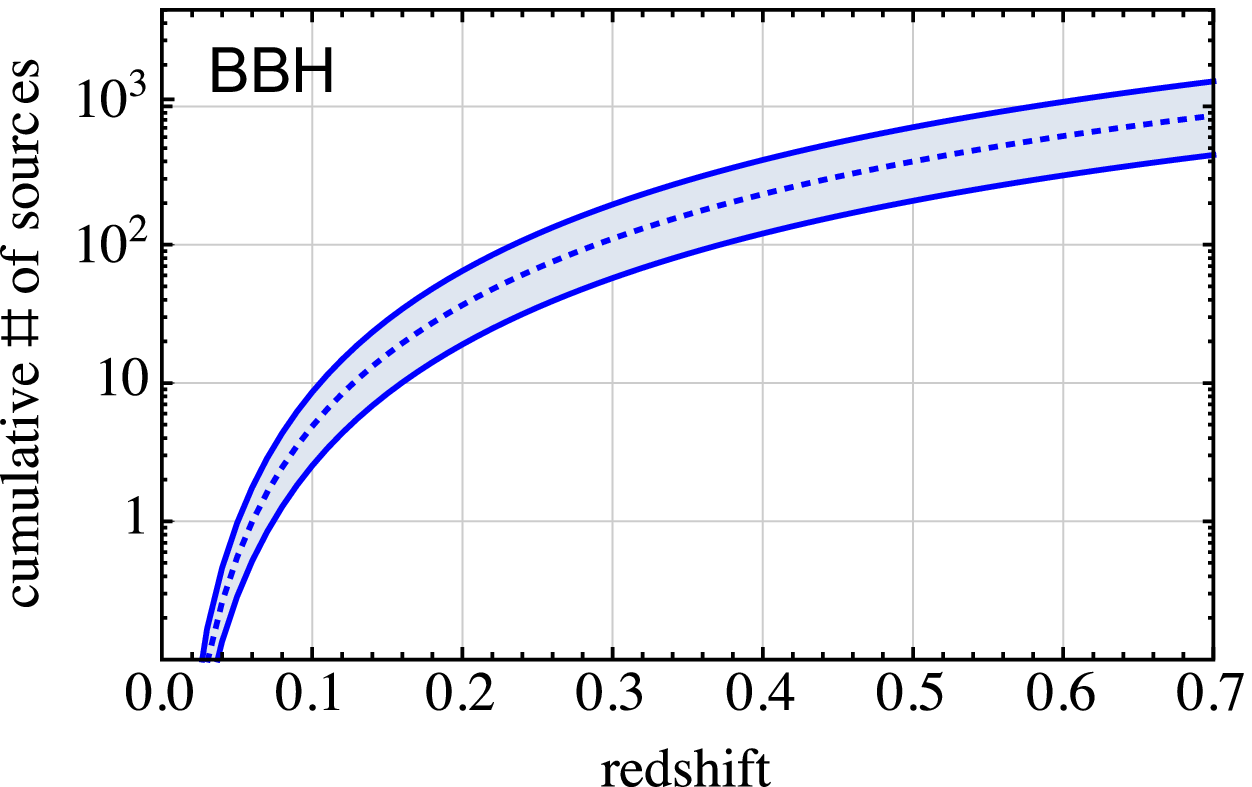}
\hspace{3mm}
\includegraphics[width=6.5cm]{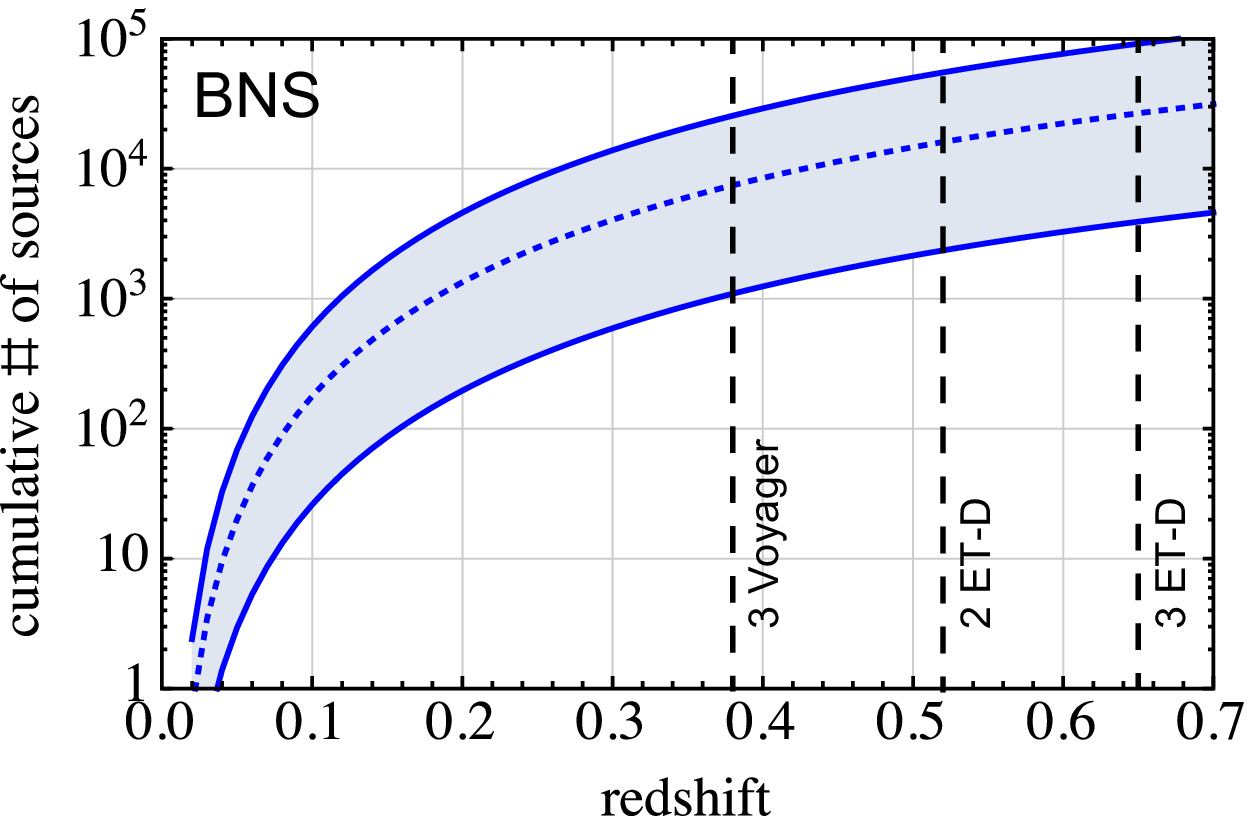}
\caption{Cumulative number of sources in $T_{\rm obs}=1\,{\rm yr}$ as a function of redshift, considering the observational uncertainty in merger rates. The dotted lines in the bands are those with the intermediate values of merger rates. The vertical lines at $z=0.38$, $0.52$, and $0.65$ in the right panel (BNS) are the horizon redshifts defined by ${\rm SNR}=12$ with the detector networks shown when averaged over all angular parameters. For the left panel (BBH), there is no horizon redshifts below $z=0.7$.} 
\label{fig:number-of-sources}
\end{center}
\end{figure*}

The merger rates have been constrained by GW observations in the ranges, $\dot{n}(0)=18.1^{+13.9}_{-8.7}\,{\rm Gpc}^{-3}\,{\rm yr}^{-1}$ for BBH with the uniform-in-log mass distribution (GstLAL) and $\dot{n}(0)=662^{+1609}_{-565}\,{\rm Gpc}^{-3}\,{\rm yr}^{-1}$ for BNS with the uniform mass set (GstLAL) \cite{LIGOScientific:2018mvr}. We take into account the ranges of the merger rates but assume that the rates are constant in redshift. Based on this assumption, the cumulative number of sources as a function of redshift during the observation of $T_{\rm obs}=1\,{\rm yr}$ is shown in Fig.~\ref{fig:number-of-sources}. Note that this number of sources is nothing to do with a detection process but the intrinsic number of mergers. From Fig.~\ref{fig:number-of-sources}, sources at higher redshifts are likely to be used for constraining modified gravity parameters merely because the measurement errors of gravity modification parameters are nearly independent of redshifts, as seen in Sec.~\ref{sec:err-redshift}. However, the success fraction of the redshift identification would be significantly lower at higher redshifts. Consequently, there must be a typical source redshift for testing GW propagation.

For BBH, a source redshift can be obtained from the identification of a host galaxy. According to the previous study \cite{Nishizawa:2016ood}, BBH with ${\rm SNR}>200$ have the sky localization volume small enough to identify a unique host galaxy. Here we define the maximum redshift for BBH as the redshift at which the angular-averaged SNR (averaged over the parameters, $\theta_{\rm S}$, $\phi_{\rm S}$, $\iota$, and $\psi$) is 200. Once the condition of ${\rm SNR}>200$ is satisfied, it would be possible to obtain the source redshift in most cases by an electromagnetic follow-up spectroscopic observation of galaxies. Therefore, for BBH we take $\epsilon_0=0.5$ and $z_{\rm max}$ as in Table~\ref{tab3}. For BNS, it would be easier to find an electromagnetic transient counterpart to determine a source redshift. However, the detectable distance of the electromagnetic counterpart, short gamma-ray bursts or kilonovae, is highly uncertain because of large uncertainty in modeling. Therefore, we keep $z_{\rm max}$ free up to the horizon distance of BNS for each detector network and consider $\epsilon_0$ from the optimistic value 0.3 to the pessimistic 0.03, including possible astrophysical uncertainties, e.~g.~\cite{Gupte:2018pht,Howell:2018nhu,Mogushi:2018ufy}. Our choices of the model parameters for the redshift identification is summarized in Table~\ref{tab3}.

\begin{table}[h]
\begin{center}
\begin{tabular}{cccc}
\hline \hline
detectors  & sources & $\epsilon_0 $ & $z_{\rm max}$ \\
\hline \hline  
3 Voyager & $30\,M_{\odot}$ BBH & 0.5 & 0.10 \\
& $10\,M_{\odot}$ BBH & 0.5 & 0.046 \\
& BNS & 0.03 - 0.3 & 0.38 \\
3 CE & $30\,M_{\odot}$ BBH & 0.5 & 0.80 \\
& $10\,M_{\odot}$ BBH & 0.5 & 0.39 \\
& BNS & 0.03 - 0.3 & 2.8 \\
3 ET-D & $30\,M_{\odot}$ BBH & 0.5 & 0.45 \\
& $10\,M_{\odot}$ BBH & 0.5 & 0.20 \\
& BNS & 0.03 - 0.3 & 0.65 \\
\hline 
\end{tabular}
\end{center}
\caption{Parameters for the redshift identification model $\epsilon(z)$.}
\label{tab3}
\end{table}

\begin{figure*}[t]
\begin{center} 
\includegraphics[width=5.35cm]{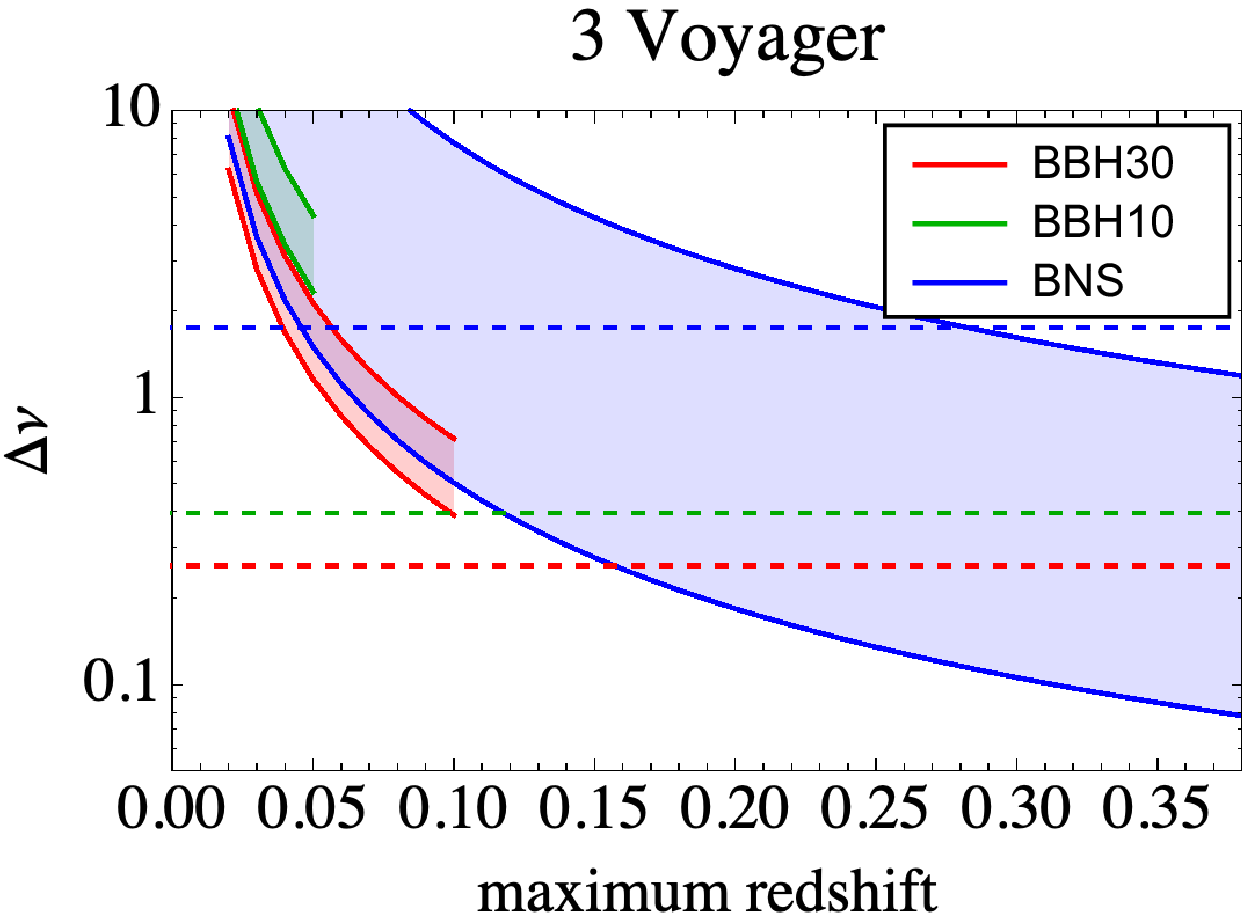}
\hspace{0.7mm}
\includegraphics[width=5.5cm]{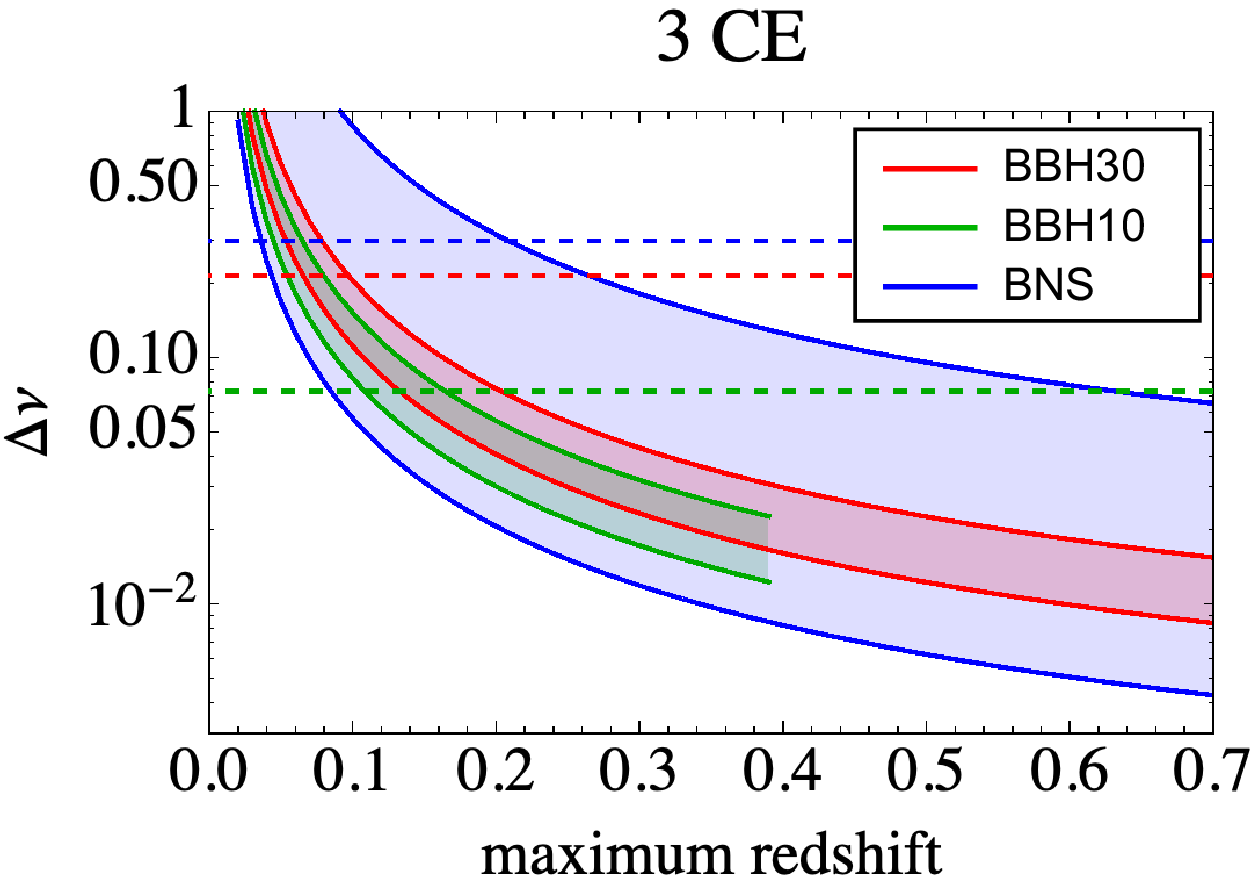}
\includegraphics[width=5.4cm]{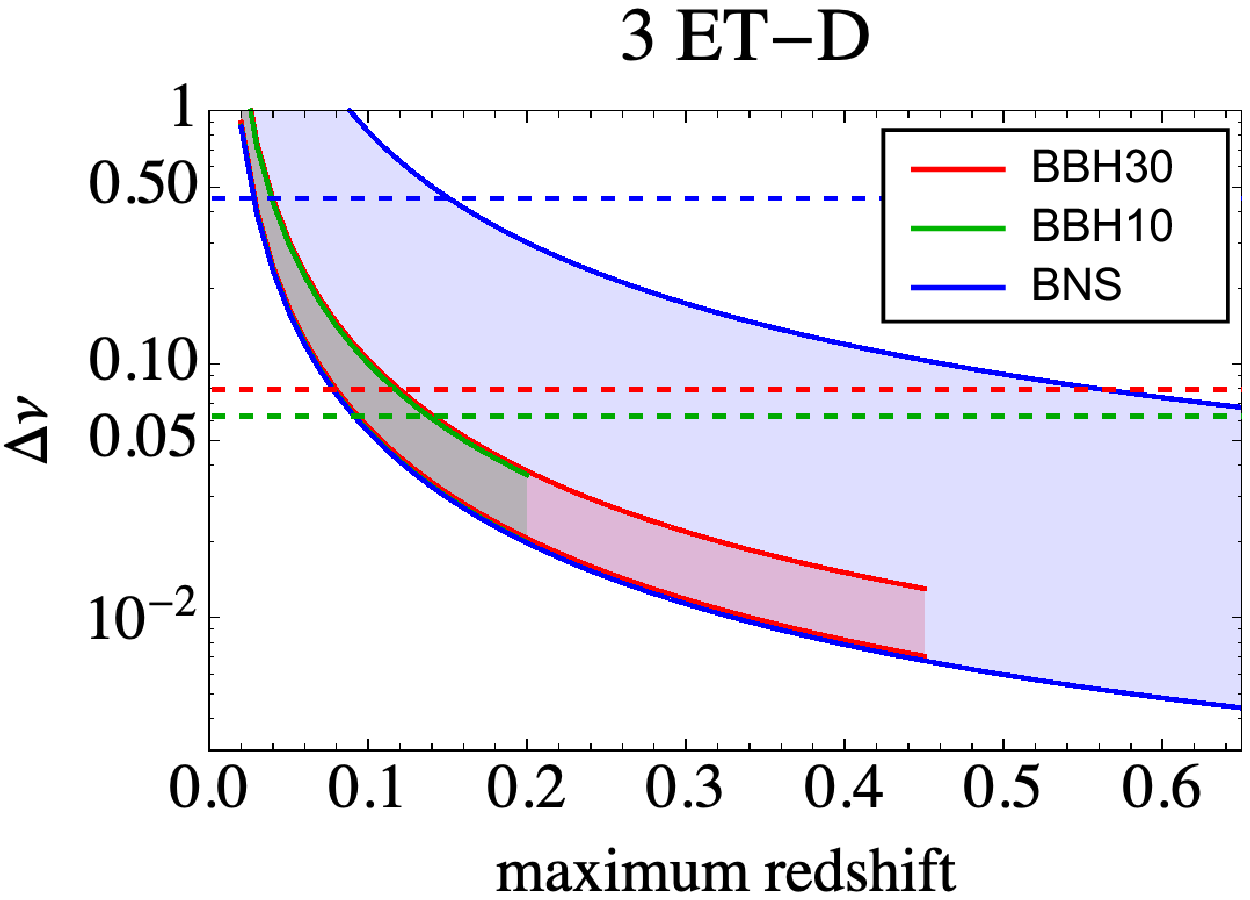}
\caption{Errors in $\nu$ with multiple sources as a function of the maximum redshift for redshift identification. From the left, the detector networks are 3 Voyager, 3 CE, and 3 ET-D. Each color bands represent $30M_{\odot}$-$30M_{\odot}$ BBH (red), $10M_{\odot}$-$10M_{\odot}$ BBH (green), and BNS (blue). For BBH, we choose $\epsilon_0=0.5$. For BNS, the range corresponds to from the lowest merger rate with $\epsilon_0=0.03$ to the highest merger rate with $\epsilon_0=0.3$. The horizontal lines in colors corresponding to each source are the top 1\% errors at $z=0.1$ from Table~\ref{tab1}.}
\label{fig:nu-err-multiple-sources}
\end{center}
\end{figure*}

Figure~\ref{fig:nu-err-multiple-sources} shows the estimation errors of $\nu$ with multiple sources for the detector networks of 3 Voyagers, 3 CEs, and 3 ET-Ds, respectively. Although $z_{\rm max}$ is highly uncertain, the errors with multiple sources can be smaller than the top 1\% errors of a single source by following-up the sources out to relatively low $z_{\rm max}$ except for the BBH cases with 3 Voyagers, in which the number of sources is relatively smaller. The best errors achieved, assuming the source redshift identification is well done, are 0.1 - 0.7 for 3 Voyagers, 0.005 - 0.020 for 3 CEs, and 0.005 - 0.015 for 3 ET-Ds, respectively. These errors are consistent with the previous study with a single ET but with redshift-identified sources out to much higher redshifts beyond $z_{\rm max}\sim1$ \cite{Belgacem:2018lbp}. It should be emphasized that the achieved sensitivity to $\nu$ significantly depends on the number of sources with redshift information, that is, $z_{\rm max}$ because the parameter estimation error is almost independent of a redshift and we have much more events at higher redshifts. 

We note that the sensitivities above should also be compared with those at $z=0.5$ with a single redshift-identified source: $\Delta \nu \approx 0.02$ with 3 CEs and $\Delta \nu \approx  0.03$ with 3 ET-Ds. However, it would quite difficult to obtain redshift information for a source at $z=0.5$. For this reason, we introduced $z_{\rm max}$ and adapted the median error at $z=0.1$ for the improvement with multiple sources. In conclusion, in either case, GW observation can reach the measurement error of $\Delta \nu \approx 0.02$.

\section{Application to Horndeski theory}
\label{sec:param_Horn}

We discuss the impact of the observational constraint on $\nu$ forecasted in the previous section on modified gravity theories. To be concrete, we consider Horndeski theory, in which $\mu=0$ and other modification parameters $\delta_g$ and $\nu$ have already been constrained from GW170817/GRB170817A \cite{GW170817:GRB} in our previous work \cite{Arai:2017hxj}. The constraint on $\delta_g$ is extremely tight and rules out some functions in the full Lagrangian of the Horndeski theory \cite{Baker:2017hug,Creminelli:2017sry,Ezquiaga:2017ekz,Sakstein:2017xjx,Arai:2017hxj}. However, the constraint on $\nu$ is $-75.3 \leq \nu \leq 78.4$ for constant $\nu$ and is still too weak to limit modified gravity theories meaningfully. Here we focus on the Horndeski theory with $\delta_g=0$ and clarify which parameter range of $\nu$ can be tested with future GW observations.

\subsection{Numerical formulation}
\label{ssec:setup}

We study the time evolutions of $\nu=\alpha_M$, $G_{\rm matter}$, and $G_{\rm light}$ in the Horndeski theory given in Eq.~(\ref{action}) with the numerical method we developed previously in \cite{Arai:2017hxj}. In the method, we compute the time evolutions of physical quantities consistently with the cosmological background. We apply the same method here, but what makes a difference from the previous work is the range of redshift and the computation of $G_{\rm matter}$ and $G_{\rm light}$. We extend the redshift out to $z=1$, where the future GW detectors are able to probe, and discuss prospective constraints on the gravitational couplings.

\begin{figure}[ht]
  \begin{center} 
    \includegraphics[width=7cm]{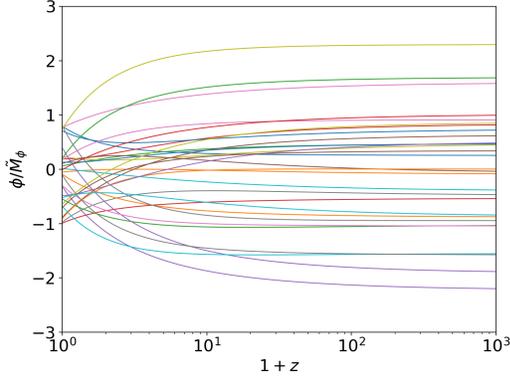}
   \caption{Time variation of $\phi$ with different random coefficients, $c^{(n)} (n=0,1,2,3)$.}
   \label{fig:phi_app}
  \end{center}
\end{figure}

We define a parameterization of the Horndeski theory in the flat FLRW Universe
\begin{align}
ds^2 = -dt^2 + a^2(t)\delta_{ij}dx^idx^j\,.\label{FRW}
\end{align}
Using the look-back conformal time $\tau_{\rm LB}$
\begin{align}
& \tau_{\rm LB}(a) = \int^{1}_a{\frac{da'}{a'^2H(a')}}\,, \label{tau_LB}
\end{align}
as a time variable, the time dependence of $\phi$ is expanded as the Taylor series to the $N$th order
\begin{align}
\phi(\tau_{\rm LB}) = M_\phi \sum^N_{n=0}{\frac{\phi^{(n)}}{n !} \tau^n_{\rm LB}} \,,\label{phi}
\end{align}
where $\phi^{(n)} \equiv d^n \phi/ d \tau^n_{\rm LB}$ and $M_{\phi}$ is the normalization of $\phi$ at $\tau_{\rm LB}=0$, being unfixed. Hereafter we assume $N=3$ to make $\phi$ slowly varying out to higher redshifts.
When we take the cosmic expansion in the $\Lambda$CDM model $H_{\rm \Lambda CDM}(a)$ as in Eq.~(\ref{Hz}), 
the look-back time in Eq.~(\ref{tau_LB}) is expanded around $a=0$ as
\begin{align}
\tau_{\rm LB}(a) = \tau_{\rm LB}(0) &- \frac{1}{H_0 a_t \sqrt{1-\Omega_{{\rm m}}}} \nonumber \\
& \times \left \{ 2{\left ( \frac{a}{a_t} \right )}^{1/2} + {\cal O}{\left (\left ( \frac{a}{a_t} \right )^{7/2} \right )}\right \} \,.\label{tau_app}
\end{align}
Here we introduce the scale factor at which the energy density of matter equals to that of the cosmological constant, $a_t \equiv (\Omega_{{\rm m}}/(1-\Omega_{{\rm m}}))^{1/3}$. Notice that $\tau_{LB}(0)$ and $a_t$ are determined once we fix $\Omega_{\rm m}$ and $H_0$. Hereafter we assume $\Omega_{\rm m}=0.3080$ to be consistent with the Planck observation of cosmic microwave background (CMB) \cite{Planck2015cosmology}. Then we obtain $H_0 \tau_{LB}(0) \sim 3.27$ and $a_t \sim 0.76$, i.~e., $z_t = a^{-1}_t - 1 \sim 0.31$. Equation~(\ref{tau_app}) gives $\tau_{\rm LB}$ with respect to $a$ for $a \ll a_t$, representing the time in the matter-dominated Universe. The approximation for $\tau_{\rm LB}$ breaks down at $a > a_t$ ($z<z_t$) and may lose an accuracy in computations at low redshifts. For instance, the exact value is $H_0\tau_{\rm LB}(z=0.1) = 0.10$, while the approximated value from Eq.~(\ref{tau_app}) is $H_0\tau_{\rm LB}(z=0.1) = 0.19$. However, the discrepancy in the approximation of $\tau_{LB}$ is absorbed by the coefficients $\phi^{(n)}$ and the normalization of $\phi$, causing no inconsistency. We choose the normalization of $\phi$ in the following way. Substituting Eq.~(\ref{tau_app}) into Eq.~(\ref{phi}), we approximate $\phi$ as
\begin{align}
\phi(a) \simeq \tilde{M}_\phi \left \{ c^{(0)}_\phi + \sum^N_{n=1}{ c^{(n)}_\phi (1-a^{n/2})} \right \}\,.\label{phi_app}
\end{align}
Here we normalize the coefficients of $\phi$ with its asymptotic value at $a=0$, that is, $\phi(\tau_{\rm LB}(0))$ in Eq.~(\ref{phi}) so that
\begin{align}
&\tilde{M}_\phi = M_\phi \frac{\sum^N_{n=0}{\frac{\phi^{(n)}}{n !} \tau^n_{\rm LB}(0)}}{ c^{(0)}_\phi + \sum^N_{n=1}{ c^{(n)}_\phi }}\,. \label{tildeMphi}
\end{align}
Notice that $N=3$ is the same as in Eq.~(\ref{phi}) to guarantee the smoothness of the functional shape. 
The time evolution of $\phi(a)$ is controlled by the coefficients $c^{(n)}_\phi (n = 0,1,2,3)$ instead of $\phi^{(n)}$. We assume without loss of generality that the coefficients $c^{(n)}_\phi$ (n=0,1,2,3) span in the range $[-1,1]$. This is because the energy scale of $\phi$, namely, $\tilde{M}_{\phi}$ determines the normalization of $\phi$. The approximation of $\phi$ in Eq.~(\ref{phi_app}) traces the models such that $\phi$ changes in time at intermediate redshifts, $z \lesssim 10$ as shown in Fig.~\ref{fig:phi_app}. At low redshifts, $\phi$ diversely fluctuates, depending on the random coefficients. On the contrary, at higher redshifts $z \gtrsim 10$, $\phi$ converges to its initial value and $\dot{\phi}$ derived from Eq.~(\ref{phi_app}) by differentiating with respect to $t$ in both sides of the equation universally scales, regardless of the random parameters, as $\dot{\phi}/H \propto -\sqrt{a} \propto -(1+z)^{-1/2}$. Thus, the time evolution of $\phi$ becomes relatively slower as a redshift increases.
In our previous work \cite{Arai:2017hxj}, the applicable range of a redshift was limited to $z \lesssim 1$ and is now extended to higher redshifts due to the different parametrization of time.

Next we move on to the parameterization of arbitrary functions $G_i\, (i = 2,3,4)$. To trace various types of $G_i$, we parameterize $G_i$ as  
\begin{align}
&G^{\rm app}_i (\phi, X) \equiv {\cal G}_i (M_{\rm pl}, \tilde{M}_\phi, H_0) \sum^{N_{\rm app}}_{m,n = 0}{\frac{g_{i,mn}}{m!n!}\hat{\phi}^m \hat{X}^n} \,. \label{G_i}
\end{align}
Here $N_{\rm app}$ controls the truncation order for the expansion of the $G^{\rm app}_{i}$ with respect to $\hat{\phi}$ and $\hat{X}$. $\hat{\phi}$ and $\hat{X}$ are the dimensionless quantities given as $\hat{\phi} \equiv \phi/\tilde{M}_\phi$ and $\hat{X} \equiv \dot{\phi}^2/2H^2\tilde{M}_{\phi}^2$. The dot denotes the derivative with respect to physical time $t$. We normalize $X$ with the Hubble parameter $H=H(\phi,X; t)$, keeping $\hat{X}$ a small expansion parameter at any redshifts we consider.\footnote{The inclusion of $H=H(\phi,X;t)$ in $G^{\rm app}_i$ in Eq.~(\ref{G_i}) could violate a property of the Horndeski theory such that the $G$ function is an implicit function of $t$, i.~e., $\partial_t G_i = 0$. Since the approximation of the $G$ functions with Eq.~(\ref{G_i}) is applied after the derivation of the equations of motion and the physical quantities, there is no inconsistency in the computation of the physical quantities.} Hereafter, we take $H$ in the normalization of $X$ as $H=H_{\rm \Lambda CDM}$. This is justified from the model filtering condition in the next subsection to require that $H$ should be close to that in the $\Lambda$CDM cosmology. We assume that $N_{\rm app}=3$ to guarantee for $G^{\rm app}_i$ to change slowly in time, compared with the cosmic expansion.
Since we consider the action (\ref{action}) of the Horndeski theory after GW170817, $g_{4,mn}=0$ for $n \geq 1$.
The coefficients ${\cal G}_i$ are the time-independent normalization factors such as
\begin{align}
{\cal G}_2 = M_c^4\,,\ {\cal G}_3 =  \frac{M_c^4}{\tilde{M}_\phi H^2_0}\,,\ {\cal G}_4 =  \frac{M_c^4}{H^2_0}\,, \label{calg_i}
\end{align}
where $M_c \equiv \sqrt{M_{\rm pl} H_0}$ is the critical energy associated with the cosmic acceleration. With the normalizations in Eq.~(\ref{calg_i}), all the terms in each $G^{\rm app}_i$ are potentially relevant to the dynamics of the cosmic expansion at late time. In other words, we can take the dimensionless coefficients in Eq.~(\ref{G_i}) as random numbers in the range of $[-1,1]$. Therefore, given all the coefficients in the expansions, we can determine the time evolution of $\phi$ by Eq.~(\ref{phi_app}) and then $G^{\rm app}_i$ by Eq.~(\ref{G_i}) as a function of time. Hereafter $G_i$ (i=2,3,4) are replaced with $G^{(\rm app)}_i$ in all the following equations.

\subsection{Consistency conditions for model extraction}
\label{ssec:filtering}
 
In the process of producing models above, we do not solve the equations of motion. To check the validity of models, we filter them with the two following criteria: consistency and stability.

\begin{enumerate}
\item Consistency with $\Lambda$CDM cosmology: \\
We collect the models whose cosmological time evolution, $H_{\rm Horn}$ and $\dot{H}_{\rm Horn}$, is close to that of $\Lambda$CDM cosmology. The Hubble parameter and its time derivative, $H_{\rm Horn}$ and $\dot{H}_{\rm Horn}$, are given by the Friedmann equations in Eqs.~(\ref{F1}) and (\ref{F2}) in Appendix~\ref{app:computations}. To obtain them, we substitute $H_{\Lambda \rm CDM}$ and $\phi(t)$ for the right-hand side of Eqs.~(\ref{E}) and (\ref{P}). Then we impose consistency criteria as
\begin{align}
 {\rm FH}: \biggl | 1- H_{\rm Horn}/H_{\rm \Lambda CDM} \biggr | < 20\%\,, \label{H20}\\
{\rm FdH}: \biggl | 1- \dot{H}_{\rm Horn}/\dot{H}_{\rm \Lambda CDM} \biggr | < 20\%\,. \label{dH20} 
\end{align}
The abbreviations ``FH" and ``EdH" represent ``Filter of the Hubble parameter" and ``Filter of the derivative of the Hubble parameter", respectively. Equations~(\ref{H20}) and ~(\ref{dH20}) work so that only the models whose cosmic expansions are similar to $H_{\Lambda \rm CDM}$ and $\dot{H}_{\Lambda \rm CDM}$ are allowed to pass through. We choose the allowed ranges of the deviation from the $\Lambda \rm CDM$ model as 20$\%$, based on current various observations of the Hubble parameter shown in the Table~I of \cite{Farooq:2016zwm}. The condition for $\dot{H}_{\rm Horn}$ controls the deviation of $\dot{H}_{\rm Horn}$ from $\dot{H}_{\rm \Lambda CDM}$ within the same error as the Hubble parameter and filters the rapid changes of the Hubble parameter. Besides, the conditions guarantee that a given $\phi(t)$ is a solution of the equations of motion within the observational error of the Hubble parameter. We check the consistency conditions at specific redshifts: z=0, 0.1, 0.5, 1.0, 1.5, and 2.0, where the constraints on the Hubble parameter exist \cite{Farooq:2016zwm}.

\item Stability of the theory: \\
To avoid ghost and gradient instabilities for the perturbations of scalar and tensor modes, the conditions
\begin{align}
{\rm stab}: Q_S >0,\quad  c^2_S > 0, \quad Q_ T> 0 \,,\label{stability}
\end{align} 
must be satisfied (We already set to $c_{\rm T}=1$). All the quantities are given in Appendix {\ref{app:computations}}. For the computation, we substitute $H = H_{\Lambda \rm CDM}$, $\dot{H} = \dot{H}_{\Lambda \rm CDM}$ into the quantities. Matter density $\tilde{\rho}_{\rm m}$ and pressure $\tilde{p}_{\rm m}$ are identified with matter density such as $\tilde{\rho}_{\rm m} = 3M^2_{\rm pl} H^2_0\Omega_{{\rm m}}a^{-3}/M^2_*$ and $\tilde{p}_{\rm m} = 0$, respectively. Again we impose the stability conditions at specific redshifts: z=0, 0.1, 0.5, 1.0, 1.5, and 2.0.

\end{enumerate}

\begin{figure*}[t]
  \vspace{-2mm}
  \begin{center} 
     \includegraphics[width=13.8cm]{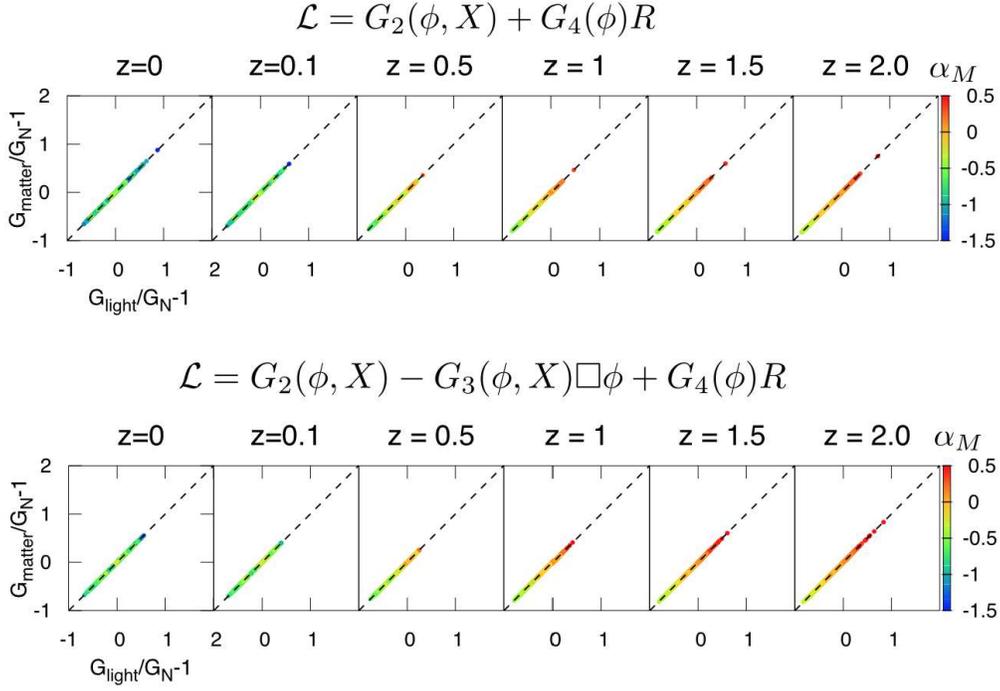}
   \caption{Time evolution of the correlation between $G_{\rm light}$ and $G_{\rm matter}$ at super-Compton limit. The color bar shows the value of $\alpha_M$. The diagonal dashed lines show $G_{\rm matter} = G_{\rm light}$, i.e.~$\Delta \gamma=0$. Top: models without $G_3$ term. Bottom: models with $G_3$ term. The range of $\alpha_M$ covers over $95\%$ of all the filtered models.} 
    \label{fig:super-aT0}
  \end{center}
  \vspace{-6mm}
\end{figure*}

\begin{figure*}[t]
  \begin{center} 
    \includegraphics[width=13.8cm]{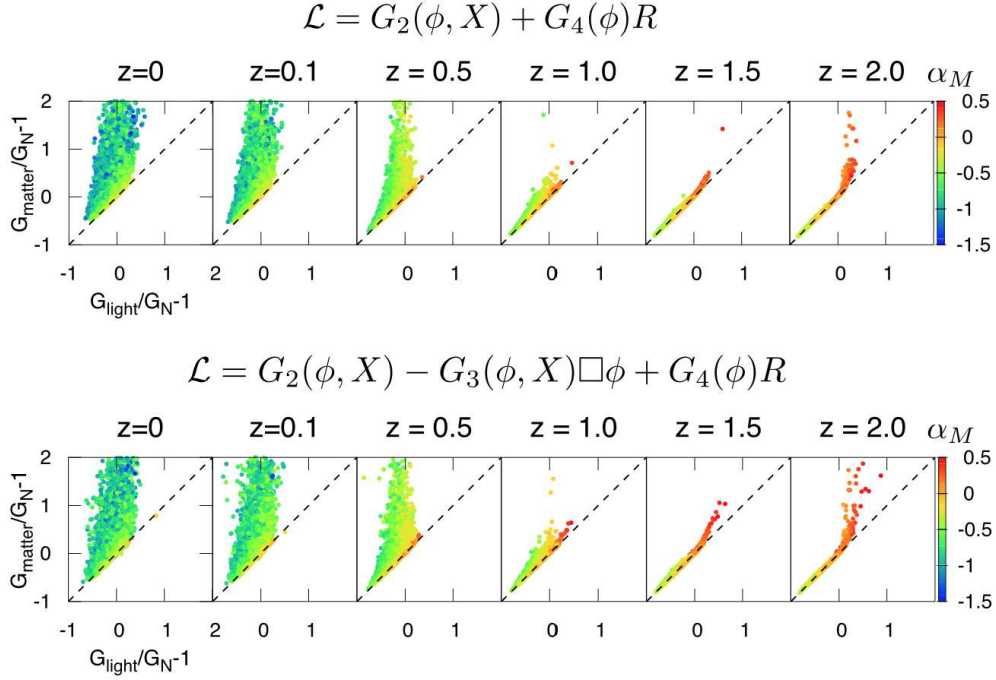}
   \caption{Time evolution of the correlation between $G_{\rm light}$ and $G_{\rm matter}$ at sub-Compton limit. The color shows the value of $\alpha_M$. The diagonal dashed lines show $G_{\rm matter} = G_{\rm light}$, i.e.~$\Delta \gamma=0$. Top: models without $G_3$ term. Bottom: models with $G_3$ term. The range of $\alpha_M$ covers over $95\%$ of all the filtered models.} 
    \label{fig:sub-aT0}
  \end{center}
\end{figure*}

\subsection{Model distributions}
\label{ssec:model_dis}
From Eqs.~(\ref{supG_aT0})-(\ref{subGlight_aT0}), we obtain $G_{\rm matter} = G_{\rm light}$ at the super-Compton scales, while $G_{\rm light}$ and $G_{\rm matter}$ are not equivalent at the sub-Compton scales. To see these behaviors, it is useful to see the correlation between $G_{\rm matter}$ and $G_{\rm light}$. In addition to that, we are interested in how $\alpha_M$ is distributed and related to $G_{\rm matter}$ and $G_{\rm light}$. 
In Figs.~\ref{fig:super-aT0} and \ref{fig:sub-aT0}, we distribute the models filtered by the conditions in Sec.~\ref{ssec:filtering} and show $\alpha_M$ in color on the $G_{\rm light}$-$G_{\rm matter}$ plane at different redshifts for two representative models of the Horndeski theory.

At first glance, there is little difference between the Horndeski Lagrangian with/ without the $G_3$ function. This explicitly shows that the $G_3$ term does not play any significant role to distribute models in the parameter space.

At the super-Compton scales in Fig.~\ref{fig:super-aT0}, all the models are aligned along the diagonal line, while at the sub-Compton scales in Fig.~\ref{fig:sub-aT0}, the off-diagonal scatter is apparent.  This trend at sub-Compton scales is expected since the fluctuations of a scalar field become significant, as discussed in Sec.~\ref{sec:Gs_Horn}. The offset trend is traced back to the third term in Eq.~(\ref{subGlight_aT0}). 

For convenience to discuss the offset trend, we introduce the gravitational slip parameter $\gamma$ \cite{Hu:2007pj,Jain:2007yk,Ade:2015rim}\footnote{{In the literature \cite{Amendola:2007rr, Bertschinger:2008zb,Daniel:2008et}, the gravitational slip parameter has different definitions.}} as
\begin{align}
\Phi = \gamma \Psi\,, \label{gslip}
\end{align}
where $\Phi$ and $\Psi$ are the linear perturbations in Eq.~(\ref{FRW_metric}). In general relativity, $\gamma = 1$, while in general theories of modified gravity, $\gamma \neq 1$. Therefore, $\gamma \neq 1$ explicitly captures the modification of gravity.  We further introduce the deviation parameter $\Delta \gamma \equiv \gamma -1$. Let us focus on the sub-Compton scales. By using Eqs.~(\ref{Poisson_Gmatter}), (\ref{Poisson_Glight}), and (\ref{gslip}), $\Delta \gamma$ relates to the gravitational couplings as
\begin{align}
\frac{G_{\rm light}}{G_{\rm matter}} = 1+\frac{\Delta \gamma}{2}\,, \label{offset}
\end{align}
and from Eqs.~(\ref{subGmatter_aT0}) and (\ref{subGlight_aT0}),
\begin{align}
\Delta \gamma = \sqrt{\frac{2}{c^2_S D}}\frac{\alpha_M \beta_\xi}{1+\beta^2_\xi}\,.\label{eta_aT0}
\end{align}
When $\Delta \gamma = 0$, the offset disappears and there are two different branches $\alpha_M = 0$ or $\beta_\xi = 0$. From Eq.~(\ref{beta_xi_aT0}), the latter is the case of $\alpha_B = -2\alpha_M$, known as No Slip Gravity \cite{Linder:2018jil}.

Taking a closer look at low redshifts below $z=1$ in Fig.~\ref{fig:sub-aT0}, the offset trend we observe implies $\Delta \gamma < 0$, consequently $\alpha_M \beta_\xi<0$ from Eq.~(\ref{eta_aT0}). To understand the condition $\alpha_M \beta_\xi < 0$, we recall the relation among $\alpha_M$, $\alpha_B$, and $G_3$, which comes from Eqs.~(\ref{aM_app}) and (\ref{aB_app}), 
\begin{align}
\alpha_B =  -\alpha_M + \frac{\dot{\phi}X G_{3X}}{HG_4}\,. \label{aB_explicit}
\end{align}
The case when $G_3 = 0$ (the top panels in Fig.~\ref{fig:sub-aT0}), we obtain $\alpha_B = -\alpha_M$.
In this case, by substituting $\beta_\xi$ in Eq.~(\ref{beta_xi_aT0}) for Eq.~(\ref{eta_aT0}) and using $\alpha_B= -\alpha_M$, $\Delta \gamma$ becomes
\begin{align}
\Delta \gamma =  -\frac{\alpha^2_M}{c^2_S D(1+\beta^2_\xi)}\,.\label{eta_aT0_G2G4phi}
\end{align}
Since we impose the stability conditions, $c^2_S>0$ and $D>0$, $\Delta \gamma < 0$ is always satisfied for a non-zero $\alpha_M$. Consequently, the offset scatters above the diagonal line, as seen in the top panels of Fig.~\ref{fig:sub-aT0}. \\

\begin{figure}[h]
  \begin{center}
  \includegraphics[width=7cm]{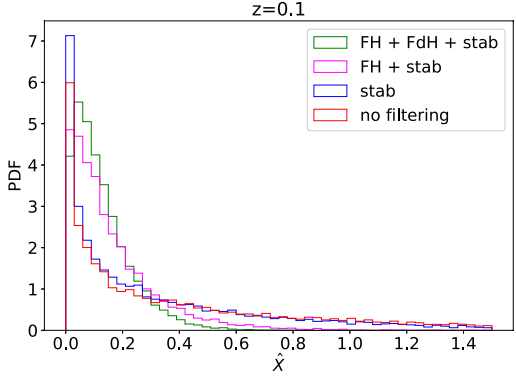}
  \end{center}
    \caption{The probability distribution of $\hat{X}$, showing the roles of the filters at $z=0.1$. In the legend, ``FH", ``FdH", and ``stab" denote the filters in Eqs.~(\ref{H20}), ~(\ref{dH20}), and ~(\ref{stability}), respectively. ``no filtering" denotes the distribution without any filter.} 
    \label{fig:Xhist}
\end{figure}

The opposite case $\Delta \gamma > 0$ with a nonzero $G_3$ is also possible in principle. However, the bottom panels of Fig.~\ref{fig:sub-aT0} in the presence of $G_3$ show no trend of $\Delta \gamma > 0$. To have a positive $\Delta \gamma$, the second term on the right-hand side in Eq.~(\ref{aB_explicit}) should be negative and dominate the first term. In other words, $\phi$ should decrease in time as rapid as the cosmic expansion. However, this is not the case, indicating that the models such that $\phi$ changes rapidly is less supported by our filtering conditions we imposed.
The absence of the contribution from the $G_3$ term in Eq.~(\ref{aB_explicit}) is because of the small value of $X$. In fact, the derivative of $G_3$ with respect to $X$ and the multiplication of $X$ in the term $\dot{\phi}XG_{3X}/HG_4$ bring the suppression factor proportional to $\hat{X}^{3/2}$. The smallness of $\hat{X}$ is due to the filtering conditions on the cosmic expansion history in Eqs.~(\ref{H20}) and (\ref{dH20}).  As explicitly shown in Fig.~\ref{fig:Xhist}, the filters in Eqs.~(\ref{H20}) and (\ref{dH20}) preferencially choose the models with smaller magnitude of $\hat{X}$. This is because the time variation of the energy density on the right-hand side in Eq.~(\ref{F1}) is slow to keep the agreement with the $\Lambda \rm CDM$ model. For these reasons, the models with $\Delta \gamma > 0$ do not appear.

The interesting feature is the signature of $\alpha_M$. We clearly see that the trend $\alpha_M \lesssim 0$ at low redshifts, that is, $M^2_*$ decreases in time. In addition, the magnitude of $\alpha_M$ is of the order of 0.1. At the super-Compton scales, the models with negative $\alpha_M$ have smaller $G_{\rm matter}$ and $G_{\rm light}$, namely, $M^2_*$ larger than $M^2_{\rm pl}$ from Eq.~(\ref{supG_aT0}). On the other hand, at the sub-Compton scales, the values of $G_{\rm matter}$ and $G_{\rm light}$ distribute more widely from smaller to larger and the offset scatter significantly correlates with the negative values of $\alpha_M$, unlike the super-Compton case. This difference arises since the magnitude of $\beta_\xi$ is larger as that of $\alpha_M$ is larger. In other words, $G_{\rm matter}$ and $G_{\rm light}$ at the sub-Compton scales are significantly diversified by the larger magnitude of $\alpha_M$, explicitly breaking the equivalence principle of gravity.\\

\subsection{Negative sign of $\alpha_M$}
\begin{figure}[h]
  \begin{center}
  \includegraphics[width=7cm]{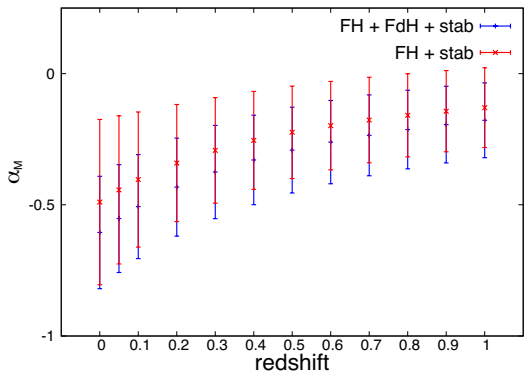}
    \caption{The effect of the consistency filters. The points at the middle of the lines represent the mean values. The ranges of the lines correspond to the standard deviation. Note that the stability conditions are already imposed on the both cases. The legends ``FH" and ``EdH", and ``stab" denote the filters in Eqs.~(\ref{H20}), ~(\ref{dH20}), and ~(\ref{stability}), respectively.}
    \label{fig:effectoffilters}
     \end{center}
\end{figure}

Remarkably, the origin of the negative value of $\alpha_M$ is the conditions that the cosmic expansion history should be similar to that of the ${\rm \Lambda CDM}$ model.  As shown in Fig.~\ref{fig:effectoffilters}, the consistency conditions for $H_{\rm Horn}$ and $\dot{H}_{\rm Horn}$ are essential to bias $\alpha_M$ toward the negative side as the redshift becomes smaller. Looking at the Friedmann equation in Eq.~(\ref{F1}) divided by $3H^2$ for the both sides of the equation, we obtain
\begin{align}
1 = \frac{V_{\rm eff}}{3M^2_*H^2} + \frac{\rho_m}{3M^2_*H^2} + {\cal O}(\hat{X})\,, \label{H_app}
\end{align}
where we omit the kinetic terms and define $V_{\rm eff}$ as
\begin{align}
V_{\rm eff} = -3M^2_*H^2 \alpha_M + V(\phi)\,.\label{Veff}
\end{align}
Here $V(\phi)$ denotes the terms in $G_2$ depending only on $\phi$. When the Universe is accelerating and the kinetic energy $\hat{X}$ is small, i.~e., the second and last terms in Eq.~(\ref{H_app}) are negligible, $V_{\rm eff} \sim 3M^2_*H^2$. The both terms in Eq.~(\ref{Veff}) equivalently contribute to $V_{\rm eff}$ because there is no prior knowledge about which term is more significant than the other. Therefore, it is probabilistically reasonable to assume $-3M^2_*H^2 \alpha_M \sim V(\phi) \sim 0.5V_{\rm eff} > 0$. As a result, $\alpha_M$ stays negative. 

The other evidence for $\alpha_M <0$ is the signature of $\dot{H}$. From Fig.~\ref{fig:effectoffilters}, we find that the time variation of $\alpha_M$ with redshifts is small, i.e., $|\dot{\alpha}_M/H\alpha_M| \ll 1$. Since $\dot{\alpha}_M$ is negligibly small, we can drop the term with $\dot{\alpha}_M$ from Eq.~(\ref{F3}) and obtain
\begin{align}
(\alpha_M+2)\frac{\dot{H}}{H^2} = \alpha_M -\frac{\rho_m}{H^2M^2_*} + \left( \rm{kinetic\; terms\; for\;} \phi \right)\,, \label{dH_app}
\end{align}
where we omit $\alpha^2_M$ and take $p_{\rm m}=0$. When the second and last terms in Eq.~(\ref{dH_app}) are negligibly small compared to $\alpha_M$, namely, corresponding to the epoch when the Universe is accelerating with the slow-rolling scalar field, the signature of the $\dot{H}$ is the same as $\alpha_M/(\alpha_M+2)$. Since the range of $\alpha_M$ is $|\alpha_M| < 1$ in Fig.~\ref{fig:effectoffilters}, the consistency condition in Eq.~(\ref{dH20}) selects $\dot{H}<0$ and consequently $\alpha_M<0$. 

\begin{figure}[h]
  \begin{center}
  \includegraphics[width=7cm]{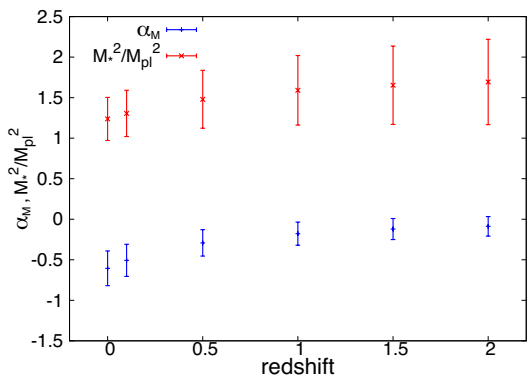}
    \caption{The ranges of $\alpha_M$ and $M^2_*/M^2_{\rm pl}$ at $z=0, 0.1,0.5,1.0,1.5$, and $2.0$. The mean value and the standard deviation are shown.}
    \label{fig:aM-Mg2}
    \end{center}
\end{figure}

We conclude for the reasons above that $\alpha_M < 0$ is statistically favored as a general trend of the viable models in the Horndeski theory. 
By fitting the mean values of $\alpha_M$ as a function of redshifts in Fig.~\ref{fig:aM-Mg2} with a commonly-used fitting formula $\alpha_M=\alpha_{M0}a^s$, we obtain the time evolution of $\alpha_M$ as 
\begin{align}
\alpha_M = -0.5980 \cdot a^{1.753}\,, \label{aM_fit}
\end{align}
namely, $\alpha_{M0} = -0.5980$ and $s=1.753$.

We comment the following two points on the negativeness of $\alpha_M$. Firstly, we can show that the signature of $\alpha_M$ does not affect the condition $c^2_S >0$ at the leading order. By substituting Eq.~(\ref{F3}) into Eq.~(\ref{cs2}), we obtain
\begin{align}
&c^2_S = \frac{2X(G_{2X}-2G_{3\phi})/H^2M^2_* + 3\alpha^2_M/2 + \Delta}{\alpha_K + 3\alpha^2_B/2}\,, \label{cs2-a}
\end{align}
where $\Delta$ in the numerator denotes
\begin{align}
\Delta &= (\alpha_M+\alpha_B)\left \{\frac{\dot{H}}{H^2}-\frac{\ddot{\phi}}{H\dot{\phi}}+ 4-\frac{\alpha_M+\alpha_B}{2} \right\} \nonumber \\
&\quad  + \frac{\dot{\alpha}_M+\dot{\alpha}_B}{H} \;. \label{Delta}
\end{align}
Under the approximation $\hat{X}\ll1$ obtained from the conditions in Eqs.~(\ref{H20}) and ~(\ref{dH20}) (more directly see Fig.~\ref{fig:Xhist}), we obtain the following equations from Eqs.~(\ref{aB_explicit}) and (\ref{aK_app}),
\begin{align}
&\alpha_B = -\alpha_M + {\cal O}(\hat{X}^{3/2})\,, \label{aB_app_cs2} \\
&\alpha_K = \frac{2X(G_{2X}-2G_{3\phi})}{H^2M^2_*} +  {\cal O}(\hat{X}^{3/2}), \label{aK_app_cs2} 
\end{align}
By using Eqs.~(\ref{aB_app_cs2}) and (\ref{aK_app_cs2}), $c^2_S$ is given by
\begin{align}
c^2_S = 1 + {\cal O}(\hat{X}^{1/2})\,, \label{cs2_app}
\end{align}
where we use $\alpha_K + 3\alpha^2_B/2 = {\cal O}(\hat{X})$ and $\Delta = {\cal O}(\hat{X}^{3/2})$. The formula in Eq.~(\ref{cs2_app}) explicitly states that the condition $c^2_S > 0$ is nothing to do with the value of $\alpha_M$. Secondly, one might consider that the negative $\alpha_M$ or the decrease of $M^2_*$ seem to be counter-intuitive as a behavior of the cosmic acceleration because a larger gravitational coupling could decelerate the Universe more by stronger gravitational attraction. However, we find that $M_*$ larger than $M_{\rm pl}$ is realized in the filtered solutions. As a result, $G_{\rm matter}$ and $G_{\rm light}$ remain smaller than the Newton constant at the super-Compton scale. In Fig.~\ref{fig:aM-Mg2}, at low redshifts, $M^2_*$ mostly stays larger than $M^2_{\rm pl}$, while $\alpha_M$ is negative. Therefore, $\alpha_M \lesssim 0$ and weaker gravitational couplings are compatible. 

\subsection{Impliction for future GW observation}

\begin{figure}[h]
  \begin{center}
  \includegraphics[width=7.5cm]{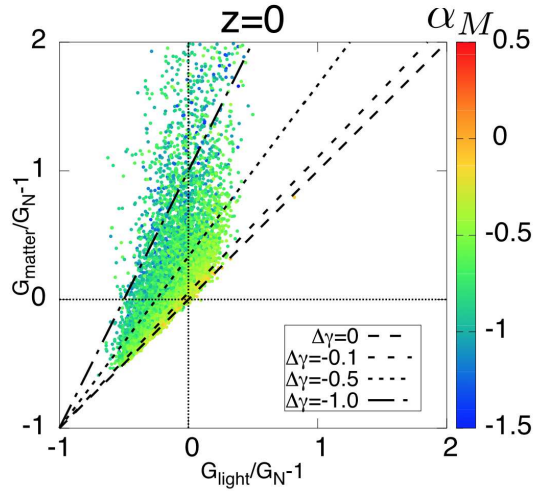}
    \caption{The contour plot of $\Delta \gamma$ on the $G_{\rm matter} - G_{\rm light}$ plane at $z=0$.}
    \label{fig:Gm-Gl_summary}
    \end{center}
\end{figure}

The gravitational slip parameter $\Delta \gamma$ is positively correlated with $\alpha_M$, as we see in Fig.~\ref{fig:Gm-Gl_summary}. More quantitatively, both of $\Delta \gamma$ and $\alpha_M$ are of the order of $-0.1$ at $z=0$. Observationally, the negativeness of $\alpha_M$ is interesting because GW amplitude is enhanced than in general relativity, as seen from Eqs.~(\ref{eq3})-(\ref{eq:damping-factor}). Since the possible range of $\alpha_M$ parameter will be significantly constrained at the level of $\sim 0.01$ by the third-generation GW detectors, most models of the Horndeski theory with $\Delta \gamma \lesssim -0.01$ (most models we obtained numerically) will be tested. If we can measure on $\nu$ at the levels of $0.5$, $0.1$, and $0.02$ and no deviation from GR is found, rejection fractions out of all models plotted in Fig.~\ref{fig:Gm-Gl_summary} (corresponding lower limits on $\Delta \gamma$) are 65.37\% ($\gtrsim-1.5$), 99.88\% ($\gtrsim-0.031$), and 99.99\% ($\gtrsim-10^{-3}$), respectively. We emphasize that positive $\alpha_M$ and $\Delta \gamma$ are hardly realized in our numerical model sampling, as we explained in the previous subsection. Therefore, without a positive detection, almost all models we generated will be ruled out, resulting in the test of the equivalence principle at cosmological distance with an unprecedented precision.

\section{Discussions}
\label{sec:discussions}

Here we discuss the current constraints on $\alpha_M$ and compare sensitivities of different observations to $\alpha_M$.

\subsection{Local measurements}

Although in the model of the Horndeski theory in Eq.~(\ref{action}), the Vainshtein mechanism produces the Newtonian law of gravity at small scales, whereas the time variation of the gravitational couplings are allowed at cosmological scales~\cite{Kimura:2011dc}. However, the direct measurements of the gravitational couplings with local astronomical objects can give the constraint on the present value of $\alpha_M$, denoted by $\alpha_{M0}$, by connecting a local solution of a scalar field to a cosmological solution. For instance, the observations of the binary pulsars \cite{Zhu:2018etc} and the lunar laser ranging experiments \cite{Williams:2004qba} currently give the constraints\footnote{The upper limit can be stronger by one order of magnitude by assuming the advanced models of the lunar core rotation for lunar laser ranging \cite{Hofmann:2018myc} and of solar mass loss for Mercury's ephemeris \cite{Genova:2018NatComm}.} $|\dot{G}/G| \approx 0.02 H_0$. As pointed out by \cite{Kimura:2011dc}, these observations directly measure $\alpha_M$, namely $|\alpha_{M0}| < 0.02$, which gives the tightest constraint on $\alpha_M$ so far. As we discussed in this paper, the GW observation can constrain $\alpha_M$ at the order of 0.01, which is comparable with the local measurements such as the binary pulsar and the lunar laser ranging. More importantly, the observation of GW propagation does not rely on gravity at local scales but can measure modification of gravity at cosmological scales directly. Potentially, the GW observation allows us to measure not only the time dependence of $\alpha_M$ but also the scale dependence. For these reasons, the GW observation combined with the local measurements is significant to check the consistency of a gravity theory over the wide ranges of space and time. 

\subsection{Cosmological measurements}

It is known that cosmological observations also put bounds on the gravitational couplings. For instance, the gravitational constant $G$ is constrained at the time of the Big Bang nucleosynthesis, $|1-G/G_N|<20\%$ \cite{Accetta:1990au,Uzan:2010pm}. Moreover, $G$ has been constrained by the detailed analyses of the CMB anisotropy \cite{Zahn:2002rr,Umezu:2005ee,Galli:2009pr}. However, these constraints are implicitly based on that the equivalence principle of gravity holds through the past of the Universe, which is in general not the case among the modified gravity theories. Recently, the constraint on $\alpha_{M0}$ has been obtained from the CMB observation by Planck \cite{Ade:2015rim}, by jointly analyzing the galaxy survey data \cite{Noller:2018wyv} and the recent cosmic shear measurement data by KiDS and GAMA observations \cite{SpurioMancini:2019rxy}. In these studies, the violation of the equivalence principle is taken into account by implementing $G_{\rm matter}$ and $G_{\rm light}$ for cosmological perturbations. The current stringent bound on $\alpha_{M0}$ is $|\alpha_{M0}|<0.04$ \cite{Ade:2015rim}. However, in order to put the bounds on $G_{\rm matter}$, $G_{\rm light}$, and $\alpha_M$ by cosmological observations, it is crucial to assume simple forms of the time evolutions for them, except for specific models such as the Jordan-Brans-Dicke theory \cite{Nagata:2002tm,Ooba:2016PRD}. In this sense, it is difficult to compare the constraining power of these cosmological observations with GW observations. In addition, the simple parametrization may be problematic in that it cannot cover the whole parameter space of the Horndeski theory.

The recent paper \cite{Denissenya:2018mqs} points out that the stable region of the Horndeski theory significantly depends on the parametrization taken, showing explicitly with a common parameterization $\alpha_M = \alpha_{M0}a^s$. Additionally, the paper \cite{Kreisch:2017uet} argues that stability conditions affect the possible range of $\alpha_M$, depending on its parameterization. As shown in Sec.~\ref{ssec:model_dis}, the fitting formula in Eq.~(\ref{aM_fit}) satisfies the stability condition in Eq.~(\ref{stability}). Nevertheless, the fitting coefficients in the formula violate the stability criteria in \cite{Denissenya:2018mqs} (see F3 condition in Sec.~3). This implies that the parameterization for $\alpha_M$ crucially drops the physical information of the Horndeski theory.


Irrespective of the parametrization issue, combining GW observations with the cosmological ones such as Euclid \cite{Amendola:2016saw}, LSST\cite{Abate:2012za}, and SKA \cite{Bacon:2018dui,Bull:2018lat} is important because they are complimentary and can break degeneracies in the large parameter space of $\alpha_M$, $\alpha_B$, and $\alpha_K$. The GW observation itself can also measure these parameters through $G_{\rm matter}$ and $G_{\rm light}$ by observing the lensing signal of GWs induced by the large-scale structures of the Universe \cite{Camera:2013xfa}. We keep the detailed study with multiple tracers for the future work.

\section{Conclusions}
\label{sec:conclusions}

In this paper, we have studied how modification of gravity, particularly, in Horndeski theory with $c_{\rm T}=1$, affects the properties of GW propagation. In the former part, we have estimated the measurement errors of the modification parameters with Voyager and the third generation detectors such as CE and ET, showing that
\begin{itemize}
\item the measurement errors of the gravity modification parameters, $\nu$ and $\mu$, hardly depend on a redshift due to the accumulation effect during propagation, 
\item a heavier source in general gives a smaller error, 
\item the future GW observation can reach the measurement error of $\Delta \nu \approx 0.02$ or less, significantly depending on the maximum redshift at which a source redshift can be identified with electromagnetic observations and on the intrinsic merger rates of binary sources.
\end{itemize}
In the latter part, we have studied the model distribution of the Horndeski theory with a numerical approach. We performed a Monte Carlo-based numerical simulation and computed $G_{\rm matter}$, $G_{\rm light}$, and $\alpha_M$. We found that 
\begin{itemize}
\item $G_{\rm matter} \approx G_{\rm light}$ in the super-Compton case, while $G_{\rm matter} \geq G_{\rm light}$ in the sub-Compton case,
\item model-filtering conditions consistent with $\Lambda$CDM cosmology preferentially select the negative sign of $\alpha_M$ at lower redshifts $z < 1$, indicating that the observed amplitude of a GW is relatively enhanced. 
\end{itemize}
Thus, the future GW observations can constrain $\nu$ in the general formalism of GW propagation and equivalently $\alpha_M$ in the Horndeski theory at the precision of ${\cal O}(0.01)$, which is comparable with the local measurements such as the binary pulsars and the lunar laser ranging. The strength of the GW observations is that it does not rely on gravity at local scales but can measure modification of gravity at cosmological scales directly, allowing us to measure not only the time dependence of $\alpha_M$ but also the scale dependence of $\alpha_M$. In the future, the GW observations combined with the local and cosmological measurements play a significant role to check the consistency of a gravity theory at cosmological distance.


\begin{acknowledgments}
 We thank K.~Ichiki, A.~Taylor, and M.~Yamaguchi for fruitful discussions. A.N. is supported by JSPS KAKENHI Grant Nos.~JP17H06358 and JP18H04581. S.A is supported by  Research Fellow of the Japan Society for the Promotion of Science. No.~17J04978.
\end{acknowledgments}

\appendix

\section{Computation of model parameters}
\label{app:computations}

Here we use the $\alpha$ parametrization for the Horndeski theory by Bellini and Sawicki \cite{Bellini2014JCAP} and introduce the explicit expressions of physical quantities necessary for the computations in the main text. Based on the Lagrangian of the Horndeski theory after GW170817 ($c_{\rm T} = 1$), the time-evolving fundamental parameters are reduced to
\vspace{-0.2cm}
\begin{align}
&HM^2_*\alpha_{\rm M} = \frac{d}{dt}M^2_* = 2\dot{\phi} G_{4\phi}\,, \label{aM_app}\\
&H^2M^2_*\alpha_{\rm K} = 2X(G_{2X}+2XG_{2XX}-2G_{3\phi}-2XG_{3\phi X}) \cr
&\qquad \qquad \;\; +12\dot{\phi}XH(G_{3X}+XG_{3XX}) \,, \label{aK_app} \\
&HM^2_*\alpha_{\rm B} = 2\dot{\phi}(XG_{3X}-G_{4\phi}) \,, \label{aB_app}
\end{align}
with $M^2_* = 2 G_4$.

The Friedmann equations in the Horndeski theory are given by
\begin{align}
&3H^2 = \tilde{\rho}_{\rm m} + \tilde{\cal E}\,, \label{F1}\\
&2\dot{H} + 3H^2 = -\tilde{p}_{\rm m} - \tilde{\cal P}\,, \label{F2}
\end{align}
where the matter energy density and pressure are $\tilde{\rho}_{\rm m} \equiv \rho_{\rm m}/M^2_*$ and $\tilde{p}_{\rm m} \equiv p_{\rm m}/M^2_*$. The quantities $\tilde{\cal E}$ and $\tilde{\cal P}$ are given by
\begin{align}
M^2_*\tilde{\cal E} &= -G_2+2X(G_{2X}-G_{3\phi}) \nonumber \\
\qquad &\;\;+6\dot{\phi} H(XG_{3X}-G_{4\phi}) \,, \label{E}\\
M^2_*\tilde{\cal P} &= G_2-2X(G_{3\phi}-2G_{4\phi\phi}) \nonumber \\
\qquad &\;\;+4\dot{\phi}H G_{4\phi} -M^2_*\alpha_B H\frac{\ddot{\phi}}{\dot{\phi}}\,.\label{P}
\end{align}

It is useful to present an additional equation for $\dot{H}$ from Eqs.~(\ref{F1}) - (\ref{P}) as
\begin{align}
&(2+\alpha_M)M^2_*\dot{H} = -HM^2_*\dot{\alpha}_M +H^2M^2_*\alpha_M (1-\alpha_M) \cr
& \qquad \qquad \qquad \qquad -2X(G_{2X}-2G_{3\phi})-6\dot{\phi}H X G_{3X} \cr
& \qquad \qquad \qquad \qquad +(\alpha_M + \alpha_B)\frac{M^2_*H\ddot{\phi}}{\dot{\phi}}  -\rho_{\rm m}-p_{\rm m} \,, \label{F3}
\end{align}
where we replaced $G_{4\phi\phi}$ with $\dot{\alpha}_M$ by using the relation
\begin{align}
\dot{\alpha}_M = H \left \{\frac{4XG_{4\phi\phi}}{H^2M^2_*} + \left (\frac{\ddot{\phi}}{H\dot{\phi}}-\frac{\dot{H}}{H^2} \right )\alpha_M - \alpha^2_M \right \}\,. 
\end{align}

The action of a scalar field $\zeta$ and tensor modes $h_{ij}$ at the quadratic order is given by  
\vspace{-0.2cm}
\begin{align}
&S_2 = \int{dtd^3xa^3}\Biggl [Q_S \left(\dot{\zeta}^2-\frac{c^2_S}{a^2}(\partial_i \zeta)^2\right) \cr 
& \qquad \qquad \quad \quad+ Q_T \left(\dot{h}_{ij}^2-\frac{c^2_T}{a^2}(\partial_k h_{ij})^2\right)  \Biggr]\,, \label{S2}
\end{align}
where 
\begin{align}
&Q_S = \frac{2M^2_*D}{(2-\alpha_B)^2}, \label{Qs}\\
&c^2_S = -\frac{1}{H^2D}\Biggl \{(2-\alpha_B)\biggl [ \dot{H}-\frac{1}{2}H^2\alpha_B -H^2 \alpha_M \biggr] \cr
&\qquad \qquad \qquad -H\dot{\alpha}_B+\tilde{\rho}_{\rm m} + \tilde{p}_{\rm m} \Biggr \}\,,\label{cs2}\\
&D = \alpha_K + \frac{3}{2}\alpha^2_B\,,\nonumber \\
&Q_T = \frac{M^2_*}{8}\,.\label{QT}
\end{align}
To avoid the ghost and gradient instabilities, we should impose the conditions: $Q_S > 0$, $c^2_S > 0$, and $Q_T > 0$.  

A combination of the above functions defines
\begin{equation}
\beta_\xi = -\sqrt{\frac{2}{c^2_S D}} \left ( \frac{\alpha_{\rm B}}{2} + \alpha_{\rm M} \right )\,.\label{beta_xi_aT0} 
\end{equation}

\section{Detector noise power spectra}
\label{sec:AppB}
We give the fitting formulas to the original power spectra of detector noise: 

\begin{itemize}
\item{Voyager}
\begin{align}
S_h(f) &= \exp \bigl[ 114.158 - 239.608\, (\log f) \nonumber \\
 & \quad \quad  + 106.701\, (\log f)^2 - 25.1711\, (\log f)^3 \nonumber \\  
 & \quad \quad  + 3.28936\, (\log f)^4 - 2.24500 \times 10^{-1}\, (\log f)^5 \nonumber \\
 & \quad \quad  + 6.24738 \times 10^{-3}\, (\log f)^6 \bigr] \;.
\end{align}
\item{CE}
\begin{align}
S_h(f) &= \exp \bigl[ 13.2133 - 147.068\,(\log f)  \nonumber \\
 &\quad \quad  + 68.7631\, (\log f)^2 - 16.6009\, (\log f)^3 \nonumber \\
 &\quad \quad  + 2.17634\, (\log f)^4 - 1.46744 \times 10^{-1}\, (\log f)^5 \nonumber \\
 & \quad \quad + 3.99167 \times 10^{-3}\, (\log f)^6 \bigr] \;. 
\end{align}
\item{ET-D}
\begin{align}
S_h(f) &= \exp \bigl[ -77.040758821 + 49.059400375\,(\log f) \nonumber \\
&\quad \quad  - 574.22111339\,(\log f)^2 + 1458.0537777\,(\log f)^3 \nonumber \\
&\quad \quad  - 1945.8076716\,(\log f)^4 + 1624.3455366\,(\log f)^5 \nonumber \\
&\quad \quad  - 919.88895662\,(\log f)^6 + 370.29239747\,(\log f)^7 \nonumber \\
&\quad \quad  - 108.96849033\,(\log f)^8 + 23.811578869\,(\log f)^9 \nonumber \\
&\quad \quad  - 3.8858541906\,(\log f)^{10} \nonumber \\
&\quad \quad  + 4.7183060556\times 10^{-1}\,(\log f)^{11} \nonumber \\
&\quad \quad  - 4.2012506692\times 10^{-2}\,(\log f)^{12} \nonumber \\
&\quad \quad  + 2.6632249588\times 10^{-4}\,(\log f)^{13} \nonumber \\
&\quad \quad  - 1.1374031387\times 10^{-4}\,(\log f)^{14} \nonumber \\
&\quad \quad  + 2.9321596012\times 10^{-6}\,(\log f)^{15} \nonumber \\
&\quad \quad  - 3.4453217899\times 10^{-8}\,(\log f)^{16} \bigr] \;. 
\end{align}
\end{itemize}

\section{Time-dependent detector response functions}
\label{sec:AppC}

\begin{figure*}[t]
\begin{center} 
\includegraphics[width=13.8cm]{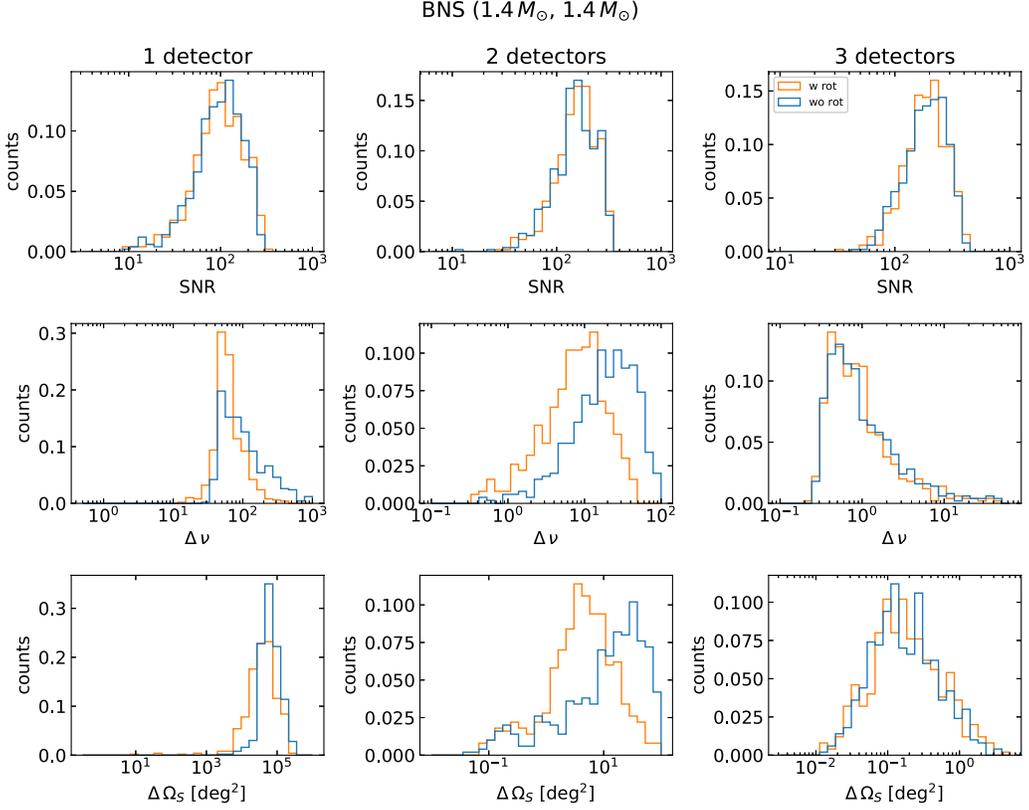}
\caption{SNR and parameter estimation errors for BNS at $z=0.1$ without (blue) and with (orange) time-dependent response functions of CE due to the Earth's rotation.} 
\label{fig:PE-dist-r-z01-CE}
\end{center}
\end{figure*}

\begin{figure*}[t]
\begin{center} 
\includegraphics[width=13.8cm]{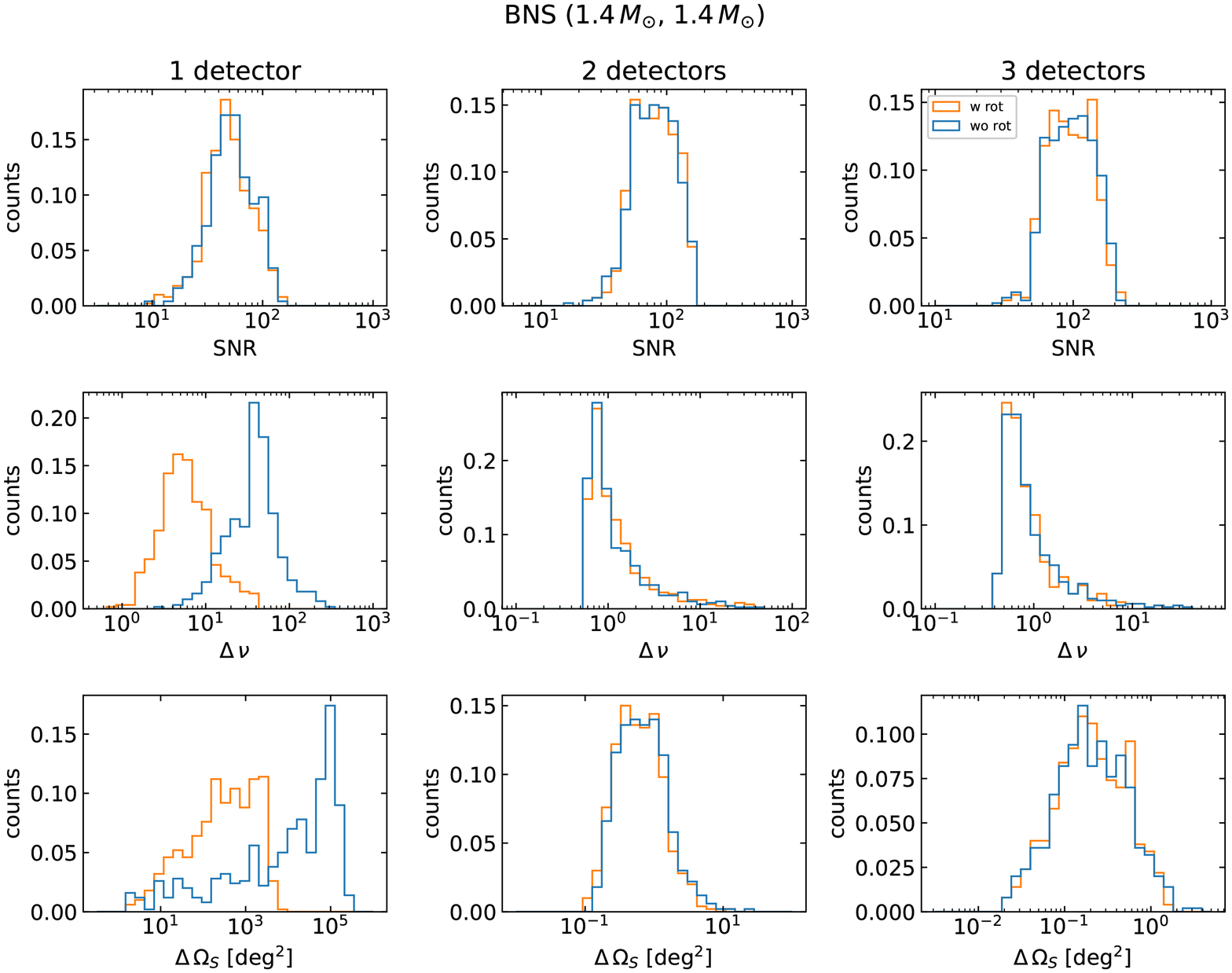}
\caption{SNR and parameter estimation errors for BNS at $z=0.1$ without (blue) and with (orange) time-dependent response functions of ET-D due to the Earth's rotation.} 
\label{fig:PE-dist-r-z01-ETD}
\end{center}
\end{figure*}

Here we show for which detector network the time-evolving response functions affect parameter estimation. Figures~\ref{fig:PE-dist-r-z01-CE} and \ref{fig:PE-dist-r-z01-ETD} show SNR and the parameter estimation errors for BNS at $z=0.1$ detected by the detector networks composed of CE or ET-D without and with time-dependent response functions due to the Earth's rotation. For a single detector of CE and ET-D, the Earth's rotation should be considered because $\nu$ error and $\Omega_{\rm S}$ error are improved due to the time evolution. However, with three detectors, a source direction is well determined by triangulation and the parameter estimation errors are not improved by the time-dependent response functions. In the two-detector case, the results depend on CE or ET-D. ET has a triangle shape and is composed of effectively two orthogonal detectors at the same cite. Thus, the two-ET-D case has four orthogonal detectors in effect and enables to point the sky direction, while the two-CE case, whose arms are physically orthogonal, has only two detectors and fails to triangulate the source direction. Therefore, we have to take into account the effect of the Earth rotation for one-CE, one-ET-D, and two-CE cases.

\bibliography{/Volumes/SSD2/bibliography}

\begin{thebibliography}{87}%
\makeatletter
\providecommand \@ifxundefined [1]{%
 \@ifx{#1\undefined}
}%
\providecommand \@ifnum [1]{%
 \ifnum #1\expandafter \@firstoftwo
 \else \expandafter \@secondoftwo
 \fi
}%
\providecommand \@ifx [1]{%
 \ifx #1\expandafter \@firstoftwo
 \else \expandafter \@secondoftwo
 \fi
}%
\providecommand \natexlab [1]{#1}%
\providecommand \enquote  [1]{``#1''}%
\providecommand \bibnamefont  [1]{#1}%
\providecommand \bibfnamefont [1]{#1}%
\providecommand \citenamefont [1]{#1}%
\providecommand \href@noop [0]{\@secondoftwo}%
\providecommand \href [0]{\begingroup \@sanitize@url \@href}%
\providecommand \@href[1]{\@@startlink{#1}\@@href}%
\providecommand \@@href[1]{\endgroup#1\@@endlink}%
\providecommand \@sanitize@url [0]{\catcode `\\12\catcode `\$12\catcode
  `\&12\catcode `\#12\catcode `\^12\catcode `\_12\catcode `\%12\relax}%
\providecommand \@@startlink[1]{}%
\providecommand \@@endlink[0]{}%
\providecommand \url  [0]{\begingroup\@sanitize@url \@url }%
\providecommand \@url [1]{\endgroup\@href {#1}{\urlprefix }}%
\providecommand \urlprefix  [0]{URL }%
\providecommand \Eprint [0]{\href }%
\providecommand \doibase [0]{http://dx.doi.org/}%
\providecommand \selectlanguage [0]{\@gobble}%
\providecommand \bibinfo  [0]{\@secondoftwo}%
\providecommand \bibfield  [0]{\@secondoftwo}%
\providecommand \translation [1]{[#1]}%
\providecommand \BibitemOpen [0]{}%
\providecommand \bibitemStop [0]{}%
\providecommand \bibitemNoStop [0]{.\EOS\space}%
\providecommand \EOS [0]{\spacefactor3000\relax}%
\providecommand \BibitemShut  [1]{\csname bibitem#1\endcsname}%
\let\auto@bib@innerbib\@empty
\bibitem [{\citenamefont {Tsujikawa}(2010)}]{Tsujikawa:2010zza}%
  \BibitemOpen
  \bibfield  {author} {\bibinfo {author} {\bibfnamefont {S.}~\bibnamefont
  {Tsujikawa}},\ }\href {\doibase 10.1007/978-3-642-10598-2_3} {\bibfield
  {journal} {\bibinfo  {journal} {Lect. Notes Phys.}\ }\textbf {\bibinfo
  {volume} {800}},\ \bibinfo {pages} {99} (\bibinfo {year} {2010})},\ \Eprint
  {http://arxiv.org/abs/1101.0191} {arXiv:1101.0191 [gr-qc]} \BibitemShut
  {NoStop}%
\bibitem [{\citenamefont {Nojiri}\ and\ \citenamefont
  {Odintsov}(2011)}]{Nojiri:2010wj}%
  \BibitemOpen
  \bibfield  {author} {\bibinfo {author} {\bibfnamefont {S.}~\bibnamefont
  {Nojiri}}\ and\ \bibinfo {author} {\bibfnamefont {S.~D.}\ \bibnamefont
  {Odintsov}},\ }\href {\doibase 10.1016/j.physrep.2011.04.001} {\bibfield
  {journal} {\bibinfo  {journal} {Phys.Rept.}\ }\textbf {\bibinfo {volume}
  {505}},\ \bibinfo {pages} {59} (\bibinfo {year} {2011})},\ \Eprint
  {http://arxiv.org/abs/1011.0544} {arXiv:1011.0544 [gr-qc]} \BibitemShut
  {NoStop}%
\bibitem [{\citenamefont {Clifton}\ \emph {et~al.}(2012)\citenamefont
  {Clifton}, \citenamefont {Ferreira}, \citenamefont {Padilla},\ and\
  \citenamefont {Skordis}}]{Clifton:2011jh}%
  \BibitemOpen
  \bibfield  {author} {\bibinfo {author} {\bibfnamefont {T.}~\bibnamefont
  {Clifton}}, \bibinfo {author} {\bibfnamefont {P.~G.}\ \bibnamefont
  {Ferreira}}, \bibinfo {author} {\bibfnamefont {A.}~\bibnamefont {Padilla}}, \
  and\ \bibinfo {author} {\bibfnamefont {C.}~\bibnamefont {Skordis}},\ }\href
  {\doibase 10.1016/j.physrep.2012.01.001} {\bibfield  {journal} {\bibinfo
  {journal} {Phys. Rept.}\ }\textbf {\bibinfo {volume} {513}},\ \bibinfo
  {pages} {1} (\bibinfo {year} {2012})},\ \Eprint
  {http://arxiv.org/abs/1106.2476} {arXiv:1106.2476 [astro-ph.CO]} \BibitemShut
  {NoStop}%
\bibitem [{\citenamefont {Joyce}\ \emph {et~al.}(2016)\citenamefont {Joyce},
  \citenamefont {Lombriser},\ and\ \citenamefont {Schmidt}}]{Joyce:2016vqv}%
  \BibitemOpen
  \bibfield  {author} {\bibinfo {author} {\bibfnamefont {A.}~\bibnamefont
  {Joyce}}, \bibinfo {author} {\bibfnamefont {L.}~\bibnamefont {Lombriser}}, \
  and\ \bibinfo {author} {\bibfnamefont {F.}~\bibnamefont {Schmidt}},\ }\href
  {\doibase 10.1146/annurev-nucl-102115-044553} {\bibfield  {journal} {\bibinfo
   {journal} {Ann. Rev. Nucl. Part. Sci.}\ }\textbf {\bibinfo {volume} {66}},\
  \bibinfo {pages} {95} (\bibinfo {year} {2016})},\ \Eprint
  {http://arxiv.org/abs/1601.06133} {arXiv:1601.06133 [astro-ph.CO]}
  \BibitemShut {NoStop}%
\bibitem [{\citenamefont {Nojiri}\ \emph {et~al.}(2017)\citenamefont {Nojiri},
  \citenamefont {Odintsov},\ and\ \citenamefont {Oikonomou}}]{Nojiri:2017ncd}%
  \BibitemOpen
  \bibfield  {author} {\bibinfo {author} {\bibfnamefont {S.}~\bibnamefont
  {Nojiri}}, \bibinfo {author} {\bibfnamefont {S.~D.}\ \bibnamefont
  {Odintsov}}, \ and\ \bibinfo {author} {\bibfnamefont {V.~K.}\ \bibnamefont
  {Oikonomou}},\ }\href {\doibase 10.1016/j.physrep.2017.06.001} {\bibfield
  {journal} {\bibinfo  {journal} {Phys. Rept.}\ }\textbf {\bibinfo {volume}
  {692}},\ \bibinfo {pages} {1} (\bibinfo {year} {2017})},\ \Eprint
  {http://arxiv.org/abs/1705.11098} {arXiv:1705.11098 [gr-qc]} \BibitemShut
  {NoStop}%
\bibitem [{\citenamefont {Ade}\ \emph {et~al.}(2016{\natexlab{a}})\citenamefont
  {Ade} \emph {et~al.}}]{Ade:2015rim}%
  \BibitemOpen
  \bibfield  {author} {\bibinfo {author} {\bibfnamefont {P.~A.~R.}\
  \bibnamefont {Ade}} \emph {et~al.} (\bibinfo {collaboration} {Planck}),\
  }\href {\doibase 10.1051/0004-6361/201525814} {\bibfield  {journal} {\bibinfo
   {journal} {Astron. Astrophys.}\ }\textbf {\bibinfo {volume} {594}},\
  \bibinfo {pages} {A14} (\bibinfo {year} {2016}{\natexlab{a}})},\ \Eprint
  {http://arxiv.org/abs/1502.01590} {arXiv:1502.01590 [astro-ph.CO]}
  \BibitemShut {NoStop}%
\bibitem [{\citenamefont {Bellini}\ \emph {et~al.}(2016)\citenamefont
  {Bellini}, \citenamefont {Cuesta}, \citenamefont {Jimenez},\ and\
  \citenamefont {Verde}}]{Bellini:2015xja}%
  \BibitemOpen
  \bibfield  {author} {\bibinfo {author} {\bibfnamefont {E.}~\bibnamefont
  {Bellini}}, \bibinfo {author} {\bibfnamefont {A.~J.}\ \bibnamefont {Cuesta}},
  \bibinfo {author} {\bibfnamefont {R.}~\bibnamefont {Jimenez}}, \ and\
  \bibinfo {author} {\bibfnamefont {L.}~\bibnamefont {Verde}},\ }\href
  {\doibase 10.1088/1475-7516/2016/06/E01, 10.1088/1475-7516/2016/02/053}
  {\bibfield  {journal} {\bibinfo  {journal} {JCAP}\ }\textbf {\bibinfo
  {volume} {1602}},\ \bibinfo {pages} {053} (\bibinfo {year} {2016})},\
  \bibinfo {note} {[Erratum: JCAP1606,no.06,E01(2016)]},\ \Eprint
  {http://arxiv.org/abs/1509.07816} {arXiv:1509.07816 [astro-ph.CO]}
  \BibitemShut {NoStop}%
\bibitem [{\citenamefont {Renk}\ \emph {et~al.}(2016)\citenamefont {Renk},
  \citenamefont {Zumalacarregui},\ and\ \citenamefont
  {Montanari}}]{Renk:2016olm}%
  \BibitemOpen
  \bibfield  {author} {\bibinfo {author} {\bibfnamefont {J.}~\bibnamefont
  {Renk}}, \bibinfo {author} {\bibfnamefont {M.}~\bibnamefont
  {Zumalacarregui}}, \ and\ \bibinfo {author} {\bibfnamefont {F.}~\bibnamefont
  {Montanari}},\ }\href {\doibase 10.1088/1475-7516/2016/07/040} {\bibfield
  {journal} {\bibinfo  {journal} {JCAP}\ }\textbf {\bibinfo {volume} {1607}},\
  \bibinfo {pages} {040} (\bibinfo {year} {2016})},\ \Eprint
  {http://arxiv.org/abs/1604.03487} {arXiv:1604.03487 [astro-ph.CO]}
  \BibitemShut {NoStop}%
\bibitem [{\citenamefont {Kreisch}\ and\ \citenamefont
  {Komatsu}(2018)}]{Kreisch:2017uet}%
  \BibitemOpen
  \bibfield  {author} {\bibinfo {author} {\bibfnamefont {C.~D.}\ \bibnamefont
  {Kreisch}}\ and\ \bibinfo {author} {\bibfnamefont {E.}~\bibnamefont
  {Komatsu}},\ }\href {\doibase 10.1088/1475-7516/2018/12/030} {\bibfield
  {journal} {\bibinfo  {journal} {JCAP}\ }\textbf {\bibinfo {volume} {1812}},\
  \bibinfo {pages} {030} (\bibinfo {year} {2018})},\ \Eprint
  {http://arxiv.org/abs/1712.02710} {arXiv:1712.02710 [astro-ph.CO]}
  \BibitemShut {NoStop}%
\bibitem [{\citenamefont {Nersisyan}\ \emph {et~al.}(2018)\citenamefont
  {Nersisyan}, \citenamefont {Lima},\ and\ \citenamefont
  {Amendola}}]{Nersisyan:2018auj}%
  \BibitemOpen
  \bibfield  {author} {\bibinfo {author} {\bibfnamefont {H.}~\bibnamefont
  {Nersisyan}}, \bibinfo {author} {\bibfnamefont {N.~A.}\ \bibnamefont {Lima}},
  \ and\ \bibinfo {author} {\bibfnamefont {L.}~\bibnamefont {Amendola}},\
  }\href@noop {} {\  (\bibinfo {year} {2018})},\ \Eprint
  {http://arxiv.org/abs/1801.06683} {arXiv:1801.06683 [astro-ph.CO]}
  \BibitemShut {NoStop}%
\bibitem [{\citenamefont {Peirone}\ \emph {et~al.}(2018)\citenamefont
  {Peirone}, \citenamefont {Koyama}, \citenamefont {Pogosian}, \citenamefont
  {Raveri},\ and\ \citenamefont {Silvestri}}]{Peirone:2017ywi}%
  \BibitemOpen
  \bibfield  {author} {\bibinfo {author} {\bibfnamefont {S.}~\bibnamefont
  {Peirone}}, \bibinfo {author} {\bibfnamefont {K.}~\bibnamefont {Koyama}},
  \bibinfo {author} {\bibfnamefont {L.}~\bibnamefont {Pogosian}}, \bibinfo
  {author} {\bibfnamefont {M.}~\bibnamefont {Raveri}}, \ and\ \bibinfo {author}
  {\bibfnamefont {A.}~\bibnamefont {Silvestri}},\ }\href {\doibase
  10.1103/PhysRevD.97.043519} {\bibfield  {journal} {\bibinfo  {journal} {Phys.
  Rev.}\ }\textbf {\bibinfo {volume} {D97}},\ \bibinfo {pages} {043519}
  (\bibinfo {year} {2018})},\ \Eprint {http://arxiv.org/abs/1712.00444}
  {arXiv:1712.00444 [astro-ph.CO]} \BibitemShut {NoStop}%
\bibitem [{\citenamefont {Espejo}\ \emph {et~al.}(2019)\citenamefont {Espejo},
  \citenamefont {Peirone}, \citenamefont {Raveri}, \citenamefont {Koyama},
  \citenamefont {Pogosian},\ and\ \citenamefont {Silvestri}}]{Espejo:2018hxa}%
  \BibitemOpen
  \bibfield  {author} {\bibinfo {author} {\bibfnamefont {J.}~\bibnamefont
  {Espejo}}, \bibinfo {author} {\bibfnamefont {S.}~\bibnamefont {Peirone}},
  \bibinfo {author} {\bibfnamefont {M.}~\bibnamefont {Raveri}}, \bibinfo
  {author} {\bibfnamefont {K.}~\bibnamefont {Koyama}}, \bibinfo {author}
  {\bibfnamefont {L.}~\bibnamefont {Pogosian}}, \ and\ \bibinfo {author}
  {\bibfnamefont {A.}~\bibnamefont {Silvestri}},\ }\href {\doibase
  10.1103/PhysRevD.99.023512} {\bibfield  {journal} {\bibinfo  {journal} {Phys.
  Rev.}\ }\textbf {\bibinfo {volume} {D99}},\ \bibinfo {pages} {023512}
  (\bibinfo {year} {2019})},\ \Eprint {http://arxiv.org/abs/1809.01121}
  {arXiv:1809.01121 [astro-ph.CO]} \BibitemShut {NoStop}%
\bibitem [{\citenamefont {Frusciante}\ \emph {et~al.}(2018)\citenamefont
  {Frusciante}, \citenamefont {Peirone}, \citenamefont {Casas},\ and\
  \citenamefont {Lima}}]{Frusciante:2018jzw}%
  \BibitemOpen
  \bibfield  {author} {\bibinfo {author} {\bibfnamefont {N.}~\bibnamefont
  {Frusciante}}, \bibinfo {author} {\bibfnamefont {S.}~\bibnamefont {Peirone}},
  \bibinfo {author} {\bibfnamefont {S.}~\bibnamefont {Casas}}, \ and\ \bibinfo
  {author} {\bibfnamefont {N.~A.}\ \bibnamefont {Lima}},\ }\href@noop {} {\
  (\bibinfo {year} {2018})},\ \Eprint {http://arxiv.org/abs/1810.10521}
  {arXiv:1810.10521 [astro-ph.CO]} \BibitemShut {NoStop}%
\bibitem [{\citenamefont {Abbott}\ \emph
  {et~al.}(2018{\natexlab{a}})\citenamefont {Abbott} \emph
  {et~al.}}]{Abbott:2018xao}%
  \BibitemOpen
  \bibfield  {author} {\bibinfo {author} {\bibfnamefont {T.~M.~C.}\
  \bibnamefont {Abbott}} \emph {et~al.} (\bibinfo {collaboration} {DES}),\
  }\href@noop {} {\  (\bibinfo {year} {2018}{\natexlab{a}})},\ \Eprint
  {http://arxiv.org/abs/1810.02499} {arXiv:1810.02499 [astro-ph.CO]}
  \BibitemShut {NoStop}%
\bibitem [{\citenamefont {Noller}\ and\ \citenamefont
  {Nicola}(2018)}]{Noller:2018wyv}%
  \BibitemOpen
  \bibfield  {author} {\bibinfo {author} {\bibfnamefont {J.}~\bibnamefont
  {Noller}}\ and\ \bibinfo {author} {\bibfnamefont {A.}~\bibnamefont
  {Nicola}},\ }\href@noop {} {\  (\bibinfo {year} {2018})},\ \Eprint
  {http://arxiv.org/abs/1811.12928} {arXiv:1811.12928 [astro-ph.CO]}
  \BibitemShut {NoStop}%
\bibitem [{\citenamefont {Camera}\ and\ \citenamefont
  {Nishizawa}(2013)}]{Camera:2013xfa}%
  \BibitemOpen
  \bibfield  {author} {\bibinfo {author} {\bibfnamefont {S.}~\bibnamefont
  {Camera}}\ and\ \bibinfo {author} {\bibfnamefont {A.}~\bibnamefont
  {Nishizawa}},\ }\href {\doibase 10.1103/PhysRevLett.110.151103} {\bibfield
  {journal} {\bibinfo  {journal} {Phys.Rev.Lett.}\ }\textbf {\bibinfo {volume}
  {110}},\ \bibinfo {pages} {151103} (\bibinfo {year} {2013})},\ \Eprint
  {http://arxiv.org/abs/1303.5446} {arXiv:1303.5446 [astro-ph.CO]} \BibitemShut
  {NoStop}%
\bibitem [{\citenamefont {Spurio~Mancini}\ \emph {et~al.}(2019)\citenamefont
  {Spurio~Mancini}, \citenamefont {Kohlinger}, \citenamefont {Joachimi},
  \citenamefont {Pettorino}, \citenamefont {Schafer}, \citenamefont {Reischke},
  \citenamefont {Brieden}, \citenamefont {Archidiacono},\ and\ \citenamefont
  {Lesgourgues}}]{SpurioMancini:2019rxy}%
  \BibitemOpen
  \bibfield  {author} {\bibinfo {author} {\bibfnamefont {A.}~\bibnamefont
  {Spurio~Mancini}}, \bibinfo {author} {\bibfnamefont {F.}~\bibnamefont
  {Kohlinger}}, \bibinfo {author} {\bibfnamefont {B.}~\bibnamefont {Joachimi}},
  \bibinfo {author} {\bibfnamefont {V.}~\bibnamefont {Pettorino}}, \bibinfo
  {author} {\bibfnamefont {B.~M.}\ \bibnamefont {Schafer}}, \bibinfo {author}
  {\bibfnamefont {R.}~\bibnamefont {Reischke}}, \bibinfo {author}
  {\bibfnamefont {S.}~\bibnamefont {Brieden}}, \bibinfo {author} {\bibfnamefont
  {M.}~\bibnamefont {Archidiacono}}, \ and\ \bibinfo {author} {\bibfnamefont
  {J.}~\bibnamefont {Lesgourgues}},\ }\href@noop {} {\  (\bibinfo {year}
  {2019})},\ \Eprint {http://arxiv.org/abs/1901.03686} {arXiv:1901.03686
  [astro-ph.CO]} \BibitemShut {NoStop}%
\bibitem [{\citenamefont {Ishak}(2019)}]{Ishak:2018his}%
  \BibitemOpen
  \bibfield  {author} {\bibinfo {author} {\bibfnamefont {M.}~\bibnamefont
  {Ishak}},\ }\href {\doibase 10.1007/s41114-018-0017-4} {\bibfield  {journal}
  {\bibinfo  {journal} {Living Rev. Rel.}\ }\textbf {\bibinfo {volume} {22}},\
  \bibinfo {pages} {1} (\bibinfo {year} {2019})},\ \Eprint
  {http://arxiv.org/abs/1806.10122} {arXiv:1806.10122 [astro-ph.CO]}
  \BibitemShut {NoStop}%
\bibitem [{\citenamefont {Saltas}\ \emph {et~al.}(2014)\citenamefont {Saltas},
  \citenamefont {Sawicki}, \citenamefont {Amendola},\ and\ \citenamefont
  {Kunz}}]{Saltas:2014dha}%
  \BibitemOpen
  \bibfield  {author} {\bibinfo {author} {\bibfnamefont {I.~D.}\ \bibnamefont
  {Saltas}}, \bibinfo {author} {\bibfnamefont {I.}~\bibnamefont {Sawicki}},
  \bibinfo {author} {\bibfnamefont {L.}~\bibnamefont {Amendola}}, \ and\
  \bibinfo {author} {\bibfnamefont {M.}~\bibnamefont {Kunz}},\ }\href {\doibase
  10.1103/PhysRevLett.113.191101} {\bibfield  {journal} {\bibinfo  {journal}
  {Phys. Rev. Lett.}\ }\textbf {\bibinfo {volume} {113}},\ \bibinfo {pages}
  {191101} (\bibinfo {year} {2014})},\ \Eprint {http://arxiv.org/abs/1406.7139}
  {arXiv:1406.7139 [astro-ph.CO]} \BibitemShut {NoStop}%
\bibitem [{\citenamefont {Nishizawa}(2018)}]{Nishizawa:2017nef}%
  \BibitemOpen
  \bibfield  {author} {\bibinfo {author} {\bibfnamefont {A.}~\bibnamefont
  {Nishizawa}},\ }\href {\doibase 10.1103/PhysRevD.97.104037} {\bibfield
  {journal} {\bibinfo  {journal} {Phys. Rev.}\ }\textbf {\bibinfo {volume}
  {D97}},\ \bibinfo {pages} {104037} (\bibinfo {year} {2018})},\ \Eprint
  {http://arxiv.org/abs/1710.04825} {arXiv:1710.04825 [gr-qc]} \BibitemShut
  {NoStop}%
\bibitem [{\citenamefont {Abbott}\ \emph
  {et~al.}(2017{\natexlab{a}})\citenamefont {Abbott} \emph
  {et~al.}}]{GW170817:detection}%
  \BibitemOpen
  \bibfield  {author} {\bibinfo {author} {\bibfnamefont {B.~P.}\ \bibnamefont
  {Abbott}} \emph {et~al.} (\bibinfo {collaboration} {Virgo, LIGO
  Scientific}),\ }\href {\doibase 10.1103/PhysRevLett.119.161101} {\bibfield
  {journal} {\bibinfo  {journal} {Phys. Rev. Lett.}\ }\textbf {\bibinfo
  {volume} {119}},\ \bibinfo {pages} {161101} (\bibinfo {year}
  {2017}{\natexlab{a}})},\ \Eprint {http://arxiv.org/abs/1710.05832}
  {arXiv:1710.05832 [gr-qc]} \BibitemShut {NoStop}%
\bibitem [{\citenamefont {Abbott}\ \emph
  {et~al.}(2017{\natexlab{b}})\citenamefont {Abbott} \emph
  {et~al.}}]{GW170817:GRB}%
  \BibitemOpen
  \bibfield  {author} {\bibinfo {author} {\bibfnamefont {B.~P.}\ \bibnamefont
  {Abbott}} \emph {et~al.} (\bibinfo {collaboration} {Virgo, Fermi-GBM,
  INTEGRAL, LIGO Scientific}),\ }\href {\doibase 10.3847/2041-8213/aa920c}
  {\bibfield  {journal} {\bibinfo  {journal} {Astrophys. J.}\ }\textbf
  {\bibinfo {volume} {848}},\ \bibinfo {pages} {L13} (\bibinfo {year}
  {2017}{\natexlab{b}})},\ \Eprint {http://arxiv.org/abs/1710.05834}
  {arXiv:1710.05834 [astro-ph.HE]} \BibitemShut {NoStop}%
\bibitem [{\citenamefont {Baker}\ \emph {et~al.}(2017)\citenamefont {Baker},
  \citenamefont {Bellini}, \citenamefont {Ferreira}, \citenamefont {Lagos},
  \citenamefont {Noller},\ and\ \citenamefont {Sawicki}}]{Baker:2017hug}%
  \BibitemOpen
  \bibfield  {author} {\bibinfo {author} {\bibfnamefont {T.}~\bibnamefont
  {Baker}}, \bibinfo {author} {\bibfnamefont {E.}~\bibnamefont {Bellini}},
  \bibinfo {author} {\bibfnamefont {P.~G.}\ \bibnamefont {Ferreira}}, \bibinfo
  {author} {\bibfnamefont {M.}~\bibnamefont {Lagos}}, \bibinfo {author}
  {\bibfnamefont {J.}~\bibnamefont {Noller}}, \ and\ \bibinfo {author}
  {\bibfnamefont {I.}~\bibnamefont {Sawicki}},\ }\href {\doibase
  10.1103/PhysRevLett.119.251301} {\bibfield  {journal} {\bibinfo  {journal}
  {Phys. Rev. Lett.}\ }\textbf {\bibinfo {volume} {119}},\ \bibinfo {pages}
  {251301} (\bibinfo {year} {2017})},\ \Eprint
  {http://arxiv.org/abs/1710.06394} {arXiv:1710.06394 [astro-ph.CO]}
  \BibitemShut {NoStop}%
\bibitem [{\citenamefont {Creminelli}\ and\ \citenamefont
  {Vernizzi}(2017)}]{Creminelli:2017sry}%
  \BibitemOpen
  \bibfield  {author} {\bibinfo {author} {\bibfnamefont {P.}~\bibnamefont
  {Creminelli}}\ and\ \bibinfo {author} {\bibfnamefont {F.}~\bibnamefont
  {Vernizzi}},\ }\href {\doibase 10.1103/PhysRevLett.119.251302} {\bibfield
  {journal} {\bibinfo  {journal} {Phys. Rev. Lett.}\ }\textbf {\bibinfo
  {volume} {119}},\ \bibinfo {pages} {251302} (\bibinfo {year} {2017})},\
  \Eprint {http://arxiv.org/abs/1710.05877} {arXiv:1710.05877 [astro-ph.CO]}
  \BibitemShut {NoStop}%
\bibitem [{\citenamefont {Sakstein}\ and\ \citenamefont
  {Jain}(2017)}]{Sakstein:2017xjx}%
  \BibitemOpen
  \bibfield  {author} {\bibinfo {author} {\bibfnamefont {J.}~\bibnamefont
  {Sakstein}}\ and\ \bibinfo {author} {\bibfnamefont {B.}~\bibnamefont
  {Jain}},\ }\href {\doibase 10.1103/PhysRevLett.119.251303} {\bibfield
  {journal} {\bibinfo  {journal} {Phys. Rev. Lett.}\ }\textbf {\bibinfo
  {volume} {119}},\ \bibinfo {pages} {251303} (\bibinfo {year} {2017})},\
  \Eprint {http://arxiv.org/abs/1710.05893} {arXiv:1710.05893 [astro-ph.CO]}
  \BibitemShut {NoStop}%
\bibitem [{\citenamefont {Ezquiaga}\ and\ \citenamefont
  {Zumalacarregui}(2017)}]{Ezquiaga:2017ekz}%
  \BibitemOpen
  \bibfield  {author} {\bibinfo {author} {\bibfnamefont {J.~M.}\ \bibnamefont
  {Ezquiaga}}\ and\ \bibinfo {author} {\bibfnamefont {M.}~\bibnamefont
  {Zumalacarregui}},\ }\href {\doibase 10.1103/PhysRevLett.119.251304}
  {\bibfield  {journal} {\bibinfo  {journal} {Phys. Rev. Lett.}\ }\textbf
  {\bibinfo {volume} {119}},\ \bibinfo {pages} {251304} (\bibinfo {year}
  {2017})},\ \Eprint {http://arxiv.org/abs/1710.05901} {arXiv:1710.05901
  [astro-ph.CO]} \BibitemShut {NoStop}%
\bibitem [{\citenamefont {Arai}\ and\ \citenamefont
  {Nishizawa}(2018)}]{Arai:2017hxj}%
  \BibitemOpen
  \bibfield  {author} {\bibinfo {author} {\bibfnamefont {S.}~\bibnamefont
  {Arai}}\ and\ \bibinfo {author} {\bibfnamefont {A.}~\bibnamefont
  {Nishizawa}},\ }\href {\doibase 10.1103/PhysRevD.97.104038} {\bibfield
  {journal} {\bibinfo  {journal} {Phys. Rev.}\ }\textbf {\bibinfo {volume}
  {D97}},\ \bibinfo {pages} {104038} (\bibinfo {year} {2018})},\ \Eprint
  {http://arxiv.org/abs/1711.03776} {arXiv:1711.03776 [gr-qc]} \BibitemShut
  {NoStop}%
\bibitem [{\citenamefont {Emir~Gumrukcuoglu}\ \emph {et~al.}(2018)\citenamefont
  {Emir~Gumrukcuoglu}, \citenamefont {Saravani},\ and\ \citenamefont
  {Sotiriou}}]{Gumrukcuoglu:2017ijh}%
  \BibitemOpen
  \bibfield  {author} {\bibinfo {author} {\bibfnamefont {A.}~\bibnamefont
  {Emir~Gumrukcuoglu}}, \bibinfo {author} {\bibfnamefont {M.}~\bibnamefont
  {Saravani}}, \ and\ \bibinfo {author} {\bibfnamefont {T.~P.}\ \bibnamefont
  {Sotiriou}},\ }\href {\doibase 10.1103/PhysRevD.97.024032} {\bibfield
  {journal} {\bibinfo  {journal} {Phys. Rev.}\ }\textbf {\bibinfo {volume}
  {D97}},\ \bibinfo {pages} {024032} (\bibinfo {year} {2018})},\ \Eprint
  {http://arxiv.org/abs/1711.08845} {arXiv:1711.08845 [gr-qc]} \BibitemShut
  {NoStop}%
\bibitem [{\citenamefont {Oost}\ \emph {et~al.}(2018)\citenamefont {Oost},
  \citenamefont {Mukohyama},\ and\ \citenamefont {Wang}}]{Oost:2018tcv}%
  \BibitemOpen
  \bibfield  {author} {\bibinfo {author} {\bibfnamefont {J.}~\bibnamefont
  {Oost}}, \bibinfo {author} {\bibfnamefont {S.}~\bibnamefont {Mukohyama}}, \
  and\ \bibinfo {author} {\bibfnamefont {A.}~\bibnamefont {Wang}},\ }\href
  {\doibase 10.1103/PhysRevD.97.124023} {\bibfield  {journal} {\bibinfo
  {journal} {Phys. Rev.}\ }\textbf {\bibinfo {volume} {D97}},\ \bibinfo {pages}
  {124023} (\bibinfo {year} {2018})},\ \Eprint
  {http://arxiv.org/abs/1802.04303} {arXiv:1802.04303 [gr-qc]} \BibitemShut
  {NoStop}%
\bibitem [{\citenamefont {Boran}\ \emph {et~al.}(2018)\citenamefont {Boran},
  \citenamefont {Desai}, \citenamefont {Kahya},\ and\ \citenamefont
  {Woodard}}]{Boran:2017rdn}%
  \BibitemOpen
  \bibfield  {author} {\bibinfo {author} {\bibfnamefont {S.}~\bibnamefont
  {Boran}}, \bibinfo {author} {\bibfnamefont {S.}~\bibnamefont {Desai}},
  \bibinfo {author} {\bibfnamefont {E.~O.}\ \bibnamefont {Kahya}}, \ and\
  \bibinfo {author} {\bibfnamefont {R.~P.}\ \bibnamefont {Woodard}},\ }\href
  {\doibase 10.1103/PhysRevD.97.041501} {\bibfield  {journal} {\bibinfo
  {journal} {Phys. Rev.}\ }\textbf {\bibinfo {volume} {D97}},\ \bibinfo {pages}
  {041501} (\bibinfo {year} {2018})},\ \Eprint
  {http://arxiv.org/abs/1710.06168} {arXiv:1710.06168 [astro-ph.HE]}
  \BibitemShut {NoStop}%
\bibitem [{\citenamefont {Nojiri}\ and\ \citenamefont
  {Odintsov}(2018)}]{Nojiri:2017hai}%
  \BibitemOpen
  \bibfield  {author} {\bibinfo {author} {\bibfnamefont {S.}~\bibnamefont
  {Nojiri}}\ and\ \bibinfo {author} {\bibfnamefont {S.~D.}\ \bibnamefont
  {Odintsov}},\ }\href {\doibase 10.1016/j.physletb.2018.01.078} {\bibfield
  {journal} {\bibinfo  {journal} {Phys. Lett.}\ }\textbf {\bibinfo {volume}
  {B779}},\ \bibinfo {pages} {425} (\bibinfo {year} {2018})},\ \Eprint
  {http://arxiv.org/abs/1711.00492} {arXiv:1711.00492 [astro-ph.CO]}
  \BibitemShut {NoStop}%
\bibitem [{\citenamefont {Nishizawa}\ and\ \citenamefont
  {Kobayashi}(2018)}]{Nishizawa:2018srh}%
  \BibitemOpen
  \bibfield  {author} {\bibinfo {author} {\bibfnamefont {A.}~\bibnamefont
  {Nishizawa}}\ and\ \bibinfo {author} {\bibfnamefont {T.}~\bibnamefont
  {Kobayashi}},\ }\href {\doibase 10.1103/PhysRevD.98.124018} {\bibfield
  {journal} {\bibinfo  {journal} {Phys. Rev.}\ }\textbf {\bibinfo {volume}
  {D98}},\ \bibinfo {pages} {124018} (\bibinfo {year} {2018})},\ \Eprint
  {http://arxiv.org/abs/1809.00815} {arXiv:1809.00815 [gr-qc]} \BibitemShut
  {NoStop}%
\bibitem [{\citenamefont {Casalino}\ \emph {et~al.}(2018)\citenamefont
  {Casalino}, \citenamefont {Rinaldi}, \citenamefont {Sebastiani},\ and\
  \citenamefont {Vagnozzi}}]{Casalino:2018tcd}%
  \BibitemOpen
  \bibfield  {author} {\bibinfo {author} {\bibfnamefont {A.}~\bibnamefont
  {Casalino}}, \bibinfo {author} {\bibfnamefont {M.}~\bibnamefont {Rinaldi}},
  \bibinfo {author} {\bibfnamefont {L.}~\bibnamefont {Sebastiani}}, \ and\
  \bibinfo {author} {\bibfnamefont {S.}~\bibnamefont {Vagnozzi}},\ }\href
  {\doibase 10.1016/j.dark.2018.10.001} {\bibfield  {journal} {\bibinfo
  {journal} {Phys. Dark Univ.}\ }\textbf {\bibinfo {volume} {22}},\ \bibinfo
  {pages} {108} (\bibinfo {year} {2018})},\ \Eprint
  {http://arxiv.org/abs/1803.02620} {arXiv:1803.02620 [gr-qc]} \BibitemShut
  {NoStop}%
\bibitem [{\citenamefont {Song}\ \emph {et~al.}(2011)\citenamefont {Song},
  \citenamefont {Zhao}, \citenamefont {Bacon}, \citenamefont {Koyama},
  \citenamefont {Nichol},\ and\ \citenamefont {Pogosian}}]{Song:2010fg}%
  \BibitemOpen
  \bibfield  {author} {\bibinfo {author} {\bibfnamefont {Y.-S.}\ \bibnamefont
  {Song}}, \bibinfo {author} {\bibfnamefont {G.-B.}\ \bibnamefont {Zhao}},
  \bibinfo {author} {\bibfnamefont {D.}~\bibnamefont {Bacon}}, \bibinfo
  {author} {\bibfnamefont {K.}~\bibnamefont {Koyama}}, \bibinfo {author}
  {\bibfnamefont {R.~C.}\ \bibnamefont {Nichol}}, \ and\ \bibinfo {author}
  {\bibfnamefont {L.}~\bibnamefont {Pogosian}},\ }\href {\doibase
  10.1103/PhysRevD.84.083523} {\bibfield  {journal} {\bibinfo  {journal} {Phys.
  Rev.}\ }\textbf {\bibinfo {volume} {D84}},\ \bibinfo {pages} {083523}
  (\bibinfo {year} {2011})},\ \Eprint {http://arxiv.org/abs/1011.2106}
  {arXiv:1011.2106 [astro-ph.CO]} \BibitemShut {NoStop}%
\bibitem [{\citenamefont {Johnson}\ \emph {et~al.}(2016)\citenamefont
  {Johnson}, \citenamefont {Blake}, \citenamefont {Dossett}, \citenamefont
  {Koda}, \citenamefont {Parkinson},\ and\ \citenamefont
  {Joudaki}}]{Johnson:2015aaa}%
  \BibitemOpen
  \bibfield  {author} {\bibinfo {author} {\bibfnamefont {A.}~\bibnamefont
  {Johnson}}, \bibinfo {author} {\bibfnamefont {C.}~\bibnamefont {Blake}},
  \bibinfo {author} {\bibfnamefont {J.}~\bibnamefont {Dossett}}, \bibinfo
  {author} {\bibfnamefont {J.}~\bibnamefont {Koda}}, \bibinfo {author}
  {\bibfnamefont {D.}~\bibnamefont {Parkinson}}, \ and\ \bibinfo {author}
  {\bibfnamefont {S.}~\bibnamefont {Joudaki}},\ }\href {\doibase
  10.1093/mnras/stw447} {\bibfield  {journal} {\bibinfo  {journal} {Mon. Not.
  Roy. Astron. Soc.}\ }\textbf {\bibinfo {volume} {458}},\ \bibinfo {pages}
  {2725} (\bibinfo {year} {2016})},\ \Eprint {http://arxiv.org/abs/1504.06885}
  {arXiv:1504.06885 [astro-ph.CO]} \BibitemShut {NoStop}%
\bibitem [{\citenamefont {Horndeski}(1974)}]{Horndeski:1974wa}%
  \BibitemOpen
  \bibfield  {author} {\bibinfo {author} {\bibfnamefont {G.~W.}\ \bibnamefont
  {Horndeski}},\ }\href {\doibase 10.1007/BF01807638} {\bibfield  {journal}
  {\bibinfo  {journal} {Int. J. Theor. Phys.}\ }\textbf {\bibinfo {volume}
  {10}},\ \bibinfo {pages} {363} (\bibinfo {year} {1974})}\BibitemShut
  {NoStop}%
\bibitem [{\citenamefont {Deffayet}\ \emph {et~al.}(2011)\citenamefont
  {Deffayet}, \citenamefont {Gao}, \citenamefont {Steer},\ and\ \citenamefont
  {Zahariade}}]{Deffayet:2011gz}%
  \BibitemOpen
  \bibfield  {author} {\bibinfo {author} {\bibfnamefont {C.}~\bibnamefont
  {Deffayet}}, \bibinfo {author} {\bibfnamefont {X.}~\bibnamefont {Gao}},
  \bibinfo {author} {\bibfnamefont {D.~A.}\ \bibnamefont {Steer}}, \ and\
  \bibinfo {author} {\bibfnamefont {G.}~\bibnamefont {Zahariade}},\ }\href
  {\doibase 10.1103/PhysRevD.84.064039} {\bibfield  {journal} {\bibinfo
  {journal} {Phys. Rev.}\ }\textbf {\bibinfo {volume} {D84}},\ \bibinfo {pages}
  {064039} (\bibinfo {year} {2011})},\ \Eprint {http://arxiv.org/abs/1103.3260}
  {arXiv:1103.3260 [hep-th]} \BibitemShut {NoStop}%
\bibitem [{\citenamefont {Kobayashi}\ \emph {et~al.}(2011)\citenamefont
  {Kobayashi}, \citenamefont {Yamaguchi},\ and\ \citenamefont
  {Yokoyama}}]{Kobayashi:2011nu}%
  \BibitemOpen
  \bibfield  {author} {\bibinfo {author} {\bibfnamefont {T.}~\bibnamefont
  {Kobayashi}}, \bibinfo {author} {\bibfnamefont {M.}~\bibnamefont
  {Yamaguchi}}, \ and\ \bibinfo {author} {\bibfnamefont {J.}~\bibnamefont
  {Yokoyama}},\ }\href {\doibase 10.1143/PTP.126.511} {\bibfield  {journal}
  {\bibinfo  {journal} {Prog. Theor. Phys.}\ }\textbf {\bibinfo {volume}
  {126}},\ \bibinfo {pages} {511} (\bibinfo {year} {2011})},\ \Eprint
  {http://arxiv.org/abs/1105.5723} {arXiv:1105.5723 [hep-th]} \BibitemShut
  {NoStop}%
\bibitem [{\citenamefont {De~Felice}\ \emph {et~al.}(2011)\citenamefont
  {De~Felice}, \citenamefont {Kobayashi},\ and\ \citenamefont
  {Tsujikawa}}]{DeFelice:2011hq}%
  \BibitemOpen
  \bibfield  {author} {\bibinfo {author} {\bibfnamefont {A.}~\bibnamefont
  {De~Felice}}, \bibinfo {author} {\bibfnamefont {T.}~\bibnamefont
  {Kobayashi}}, \ and\ \bibinfo {author} {\bibfnamefont {S.}~\bibnamefont
  {Tsujikawa}},\ }\href {\doibase 10.1016/j.physletb.2011.11.028} {\bibfield
  {journal} {\bibinfo  {journal} {Phys. Lett.}\ }\textbf {\bibinfo {volume}
  {B706}},\ \bibinfo {pages} {123} (\bibinfo {year} {2011})},\ \Eprint
  {http://arxiv.org/abs/1108.4242} {arXiv:1108.4242 [gr-qc]} \BibitemShut
  {NoStop}%
\bibitem [{\citenamefont {Tsujikawa}(2015)}]{Tsujikawa:2014mba}%
  \BibitemOpen
  \bibfield  {author} {\bibinfo {author} {\bibfnamefont {S.}~\bibnamefont
  {Tsujikawa}},\ }\href {\doibase 10.1007/978-3-319-10070-8_4} {\bibfield
  {journal} {\bibinfo  {journal} {Lect. Notes Phys.}\ }\textbf {\bibinfo
  {volume} {892}},\ \bibinfo {pages} {97} (\bibinfo {year} {2015})},\ \Eprint
  {http://arxiv.org/abs/1404.2684} {arXiv:1404.2684 [gr-qc]} \BibitemShut
  {NoStop}%
\bibitem [{\citenamefont {Gleyzes}\ \emph {et~al.}(2016)\citenamefont
  {Gleyzes}, \citenamefont {Langlois}, \citenamefont {Mancarella},\ and\
  \citenamefont {Vernizzi}}]{Gleyzes:2015rua}%
  \BibitemOpen
  \bibfield  {author} {\bibinfo {author} {\bibfnamefont {J.}~\bibnamefont
  {Gleyzes}}, \bibinfo {author} {\bibfnamefont {D.}~\bibnamefont {Langlois}},
  \bibinfo {author} {\bibfnamefont {M.}~\bibnamefont {Mancarella}}, \ and\
  \bibinfo {author} {\bibfnamefont {F.}~\bibnamefont {Vernizzi}},\ }\href
  {\doibase 10.1088/1475-7516/2016/02/056} {\bibfield  {journal} {\bibinfo
  {journal} {JCAP}\ }\textbf {\bibinfo {volume} {1602}},\ \bibinfo {pages}
  {056} (\bibinfo {year} {2016})},\ \Eprint {http://arxiv.org/abs/1509.02191}
  {arXiv:1509.02191 [astro-ph.CO]} \BibitemShut {NoStop}%
\bibitem [{\citenamefont {Belgacem}\ \emph
  {et~al.}(2018{\natexlab{a}})\citenamefont {Belgacem}, \citenamefont {Dirian},
  \citenamefont {Foffa},\ and\ \citenamefont {Maggiore}}]{Belgacem:2017ihm}%
  \BibitemOpen
  \bibfield  {author} {\bibinfo {author} {\bibfnamefont {E.}~\bibnamefont
  {Belgacem}}, \bibinfo {author} {\bibfnamefont {Y.}~\bibnamefont {Dirian}},
  \bibinfo {author} {\bibfnamefont {S.}~\bibnamefont {Foffa}}, \ and\ \bibinfo
  {author} {\bibfnamefont {M.}~\bibnamefont {Maggiore}},\ }\href {\doibase
  10.1103/PhysRevD.97.104066} {\bibfield  {journal} {\bibinfo  {journal} {Phys.
  Rev.}\ }\textbf {\bibinfo {volume} {D97}},\ \bibinfo {pages} {104066}
  (\bibinfo {year} {2018}{\natexlab{a}})},\ \Eprint
  {http://arxiv.org/abs/1712.08108} {arXiv:1712.08108 [astro-ph.CO]}
  \BibitemShut {NoStop}%
\bibitem [{\citenamefont {Amendola}\ \emph
  {et~al.}(2018{\natexlab{a}})\citenamefont {Amendola}, \citenamefont
  {Sawicki}, \citenamefont {Kunz},\ and\ \citenamefont
  {Saltas}}]{Amendola:2017ovw}%
  \BibitemOpen
  \bibfield  {author} {\bibinfo {author} {\bibfnamefont {L.}~\bibnamefont
  {Amendola}}, \bibinfo {author} {\bibfnamefont {I.}~\bibnamefont {Sawicki}},
  \bibinfo {author} {\bibfnamefont {M.}~\bibnamefont {Kunz}}, \ and\ \bibinfo
  {author} {\bibfnamefont {I.~D.}\ \bibnamefont {Saltas}},\ }\href {\doibase
  10.1088/1475-7516/2018/08/030} {\bibfield  {journal} {\bibinfo  {journal}
  {JCAP}\ }\textbf {\bibinfo {volume} {1808}},\ \bibinfo {pages} {030}
  (\bibinfo {year} {2018}{\natexlab{a}})},\ \Eprint
  {http://arxiv.org/abs/1712.08623} {arXiv:1712.08623 [astro-ph.CO]}
  \BibitemShut {NoStop}%
\bibitem [{\citenamefont {Nunes}\ \emph {et~al.}(2018)\citenamefont {Nunes},
  \citenamefont {Alves},\ and\ \citenamefont {de~Araujo}}]{Nunes:2018zot}%
  \BibitemOpen
  \bibfield  {author} {\bibinfo {author} {\bibfnamefont {R.~C.}\ \bibnamefont
  {Nunes}}, \bibinfo {author} {\bibfnamefont {M.~E.~S.}\ \bibnamefont {Alves}},
  \ and\ \bibinfo {author} {\bibfnamefont {J.~C.~N.}\ \bibnamefont
  {de~Araujo}},\ }\href@noop {} {\  (\bibinfo {year} {2018})},\ \Eprint
  {http://arxiv.org/abs/1811.12760} {arXiv:1811.12760 [gr-qc]} \BibitemShut
  {NoStop}%
\bibitem [{\citenamefont {Khoury}\ and\ \citenamefont
  {Weltman}(2004{\natexlab{a}})}]{Khoury:2003aq}%
  \BibitemOpen
  \bibfield  {author} {\bibinfo {author} {\bibfnamefont {J.}~\bibnamefont
  {Khoury}}\ and\ \bibinfo {author} {\bibfnamefont {A.}~\bibnamefont
  {Weltman}},\ }\href {\doibase 10.1103/PhysRevLett.93.171104} {\bibfield
  {journal} {\bibinfo  {journal} {Phys.Rev.Lett.}\ }\textbf {\bibinfo {volume}
  {93}},\ \bibinfo {pages} {171104} (\bibinfo {year} {2004}{\natexlab{a}})},\
  \Eprint {http://arxiv.org/abs/astro-ph/0309300} {arXiv:astro-ph/0309300
  [astro-ph]} \BibitemShut {NoStop}%
\bibitem [{\citenamefont {Khoury}\ and\ \citenamefont
  {Weltman}(2004{\natexlab{b}})}]{Khoury:2003rn}%
  \BibitemOpen
  \bibfield  {author} {\bibinfo {author} {\bibfnamefont {J.}~\bibnamefont
  {Khoury}}\ and\ \bibinfo {author} {\bibfnamefont {A.}~\bibnamefont
  {Weltman}},\ }\href {\doibase 10.1103/PhysRevD.69.044026} {\bibfield
  {journal} {\bibinfo  {journal} {Phys.Rev.}\ }\textbf {\bibinfo {volume}
  {D69}},\ \bibinfo {pages} {044026} (\bibinfo {year} {2004}{\natexlab{b}})},\
  \Eprint {http://arxiv.org/abs/astro-ph/0309411} {arXiv:astro-ph/0309411
  [astro-ph]} \BibitemShut {NoStop}%
\bibitem [{\citenamefont {Vainshtein}(1972)}]{Vainshtein:1972sx}%
  \BibitemOpen
  \bibfield  {author} {\bibinfo {author} {\bibfnamefont {A.}~\bibnamefont
  {Vainshtein}},\ }\href {\doibase 10.1016/0370-2693(72)90147-5} {\bibfield
  {journal} {\bibinfo  {journal} {Phys.Lett.}\ }\textbf {\bibinfo {volume}
  {B39}},\ \bibinfo {pages} {393} (\bibinfo {year} {1972})}\BibitemShut
  {NoStop}%
\bibitem [{\citenamefont {Pogosian}\ and\ \citenamefont
  {Silvestri}(2016)}]{Pogosian:2016pwr}%
  \BibitemOpen
  \bibfield  {author} {\bibinfo {author} {\bibfnamefont {L.}~\bibnamefont
  {Pogosian}}\ and\ \bibinfo {author} {\bibfnamefont {A.}~\bibnamefont
  {Silvestri}},\ }\href {\doibase 10.1103/PhysRevD.94.104014} {\bibfield
  {journal} {\bibinfo  {journal} {Phys. Rev.}\ }\textbf {\bibinfo {volume}
  {D94}},\ \bibinfo {pages} {104014} (\bibinfo {year} {2016})},\ \Eprint
  {http://arxiv.org/abs/1606.05339} {arXiv:1606.05339 [astro-ph.CO]}
  \BibitemShut {NoStop}%
\bibitem [{\citenamefont {Abbott}\ \emph
  {et~al.}(2017{\natexlab{c}})\citenamefont {Abbott} \emph
  {et~al.}}]{GW170817:multimessenger}%
  \BibitemOpen
  \bibfield  {author} {\bibinfo {author} {\bibfnamefont {B.~P.}\ \bibnamefont
  {Abbott}} \emph {et~al.} (\bibinfo {collaboration} {GROND, SALT Group,
  OzGrav, DFN, INTEGRAL, Virgo, Insight-Hxmt, MAXI Team, Fermi-LAT, J-GEM,
  RATIR, IceCube, CAASTRO, LWA, ePESSTO, GRAWITA, RIMAS, SKA South
  Africa/MeerKAT, H.E.S.S., 1M2H Team, IKI-GW Follow-up, Fermi GBM, Pi of Sky,
  DWF (Deeper Wider Faster Program), Dark Energy Survey, MASTER, AstroSat
  Cadmium Zinc Telluride Imager Team, Swift, Pierre Auger, ASKAP, VINROUGE,
  JAGWAR, Chandra Team at McGill University, TTU-NRAO, GROWTH, AGILE Team, MWA,
  ATCA, AST3, TOROS, Pan-STARRS, NuSTAR, ATLAS Telescopes, BOOTES, CaltechNRAO,
  LIGO Scientific, High Time Resolution Universe Survey, Nordic Optical
  Telescope, Las Cumbres Observatory Group, TZAC Consortium, LOFAR, IPN, DLT40,
  Texas Tech University, HAWC, ANTARES, KU, Dark Energy Camera GW-EM, CALET,
  Euro VLBI Team, ALMA}),\ }\href {\doibase 10.3847/2041-8213/aa91c9}
  {\bibfield  {journal} {\bibinfo  {journal} {Astrophys. J.}\ }\textbf
  {\bibinfo {volume} {848}},\ \bibinfo {pages} {L12} (\bibinfo {year}
  {2017}{\natexlab{c}})},\ \Eprint {http://arxiv.org/abs/1710.05833}
  {arXiv:1710.05833 [astro-ph.HE]} \BibitemShut {NoStop}%
\bibitem [{\citenamefont {Belgacem}\ \emph
  {et~al.}(2018{\natexlab{b}})\citenamefont {Belgacem}, \citenamefont {Dirian},
  \citenamefont {Foffa},\ and\ \citenamefont {Maggiore}}]{Belgacem:2018lbp}%
  \BibitemOpen
  \bibfield  {author} {\bibinfo {author} {\bibfnamefont {E.}~\bibnamefont
  {Belgacem}}, \bibinfo {author} {\bibfnamefont {Y.}~\bibnamefont {Dirian}},
  \bibinfo {author} {\bibfnamefont {S.}~\bibnamefont {Foffa}}, \ and\ \bibinfo
  {author} {\bibfnamefont {M.}~\bibnamefont {Maggiore}},\ }\href {\doibase
  10.1103/PhysRevD.98.023510} {\bibfield  {journal} {\bibinfo  {journal} {Phys.
  Rev.}\ }\textbf {\bibinfo {volume} {D98}},\ \bibinfo {pages} {023510}
  (\bibinfo {year} {2018}{\natexlab{b}})},\ \Eprint
  {http://arxiv.org/abs/1805.08731} {arXiv:1805.08731 [gr-qc]} \BibitemShut
  {NoStop}%
\bibitem [{\citenamefont {{Khan}}\ \emph {et~al.}(2016)\citenamefont {{Khan}},
  \citenamefont {{Husa}}, \citenamefont {{Hannam}}, \citenamefont {{Ohme}},
  \citenamefont {{P{\"u}rrer}}, \citenamefont {{Forteza}},\ and\ \citenamefont
  {{Boh{\'e}}}}]{Khan:2016PRD}%
  \BibitemOpen
  \bibfield  {author} {\bibinfo {author} {\bibfnamefont {S.}~\bibnamefont
  {{Khan}}}, \bibinfo {author} {\bibfnamefont {S.}~\bibnamefont {{Husa}}},
  \bibinfo {author} {\bibfnamefont {M.}~\bibnamefont {{Hannam}}}, \bibinfo
  {author} {\bibfnamefont {F.}~\bibnamefont {{Ohme}}}, \bibinfo {author}
  {\bibfnamefont {M.}~\bibnamefont {{P{\"u}rrer}}}, \bibinfo {author}
  {\bibfnamefont {X.~J.}\ \bibnamefont {{Forteza}}}, \ and\ \bibinfo {author}
  {\bibfnamefont {A.}~\bibnamefont {{Boh{\'e}}}},\ }\href {\doibase
  10.1103/PhysRevD.93.044007} {\bibfield  {journal} {\bibinfo  {journal}
  {\prd}\ }\textbf {\bibinfo {volume} {93}},\ \bibinfo {eid} {044007} (\bibinfo
  {year} {2016})},\ \Eprint {http://arxiv.org/abs/1508.07253} {arXiv:1508.07253
  [gr-qc]} \BibitemShut {NoStop}%
\bibitem [{\citenamefont {Ade}\ \emph {et~al.}(2016{\natexlab{b}})\citenamefont
  {Ade} \emph {et~al.}}]{Planck2015cosmology}%
  \BibitemOpen
  \bibfield  {author} {\bibinfo {author} {\bibfnamefont {P.~A.~R.}\
  \bibnamefont {Ade}} \emph {et~al.} (\bibinfo {collaboration} {Planck
  collaboration}),\ }\href {\doibase 10.1051/0004-6361/201525830} {\bibfield
  {journal} {\bibinfo  {journal} {Astron. Astrophys.}\ }\textbf {\bibinfo
  {volume} {594}},\ \bibinfo {pages} {A13} (\bibinfo {year}
  {2016}{\natexlab{b}})},\ \Eprint {http://arxiv.org/abs/1502.01589}
  {arXiv:1502.01589 [astro-ph.CO]} \BibitemShut {NoStop}%
\bibitem [{\citenamefont {Finn}(1992)}]{Finn:1992wt}%
  \BibitemOpen
  \bibfield  {author} {\bibinfo {author} {\bibfnamefont {L.~S.}\ \bibnamefont
  {Finn}},\ }\href {\doibase 10.1103/PhysRevD.46.5236} {\bibfield  {journal}
  {\bibinfo  {journal} {Phys.Rev.}\ }\textbf {\bibinfo {volume} {D46}},\
  \bibinfo {pages} {5236} (\bibinfo {year} {1992})},\ \Eprint
  {http://arxiv.org/abs/gr-qc/9209010} {arXiv:gr-qc/9209010 [gr-qc]}
  \BibitemShut {NoStop}%
\bibitem [{\citenamefont {Cutler}\ and\ \citenamefont
  {Flanagan}(1994)}]{Cutler:1994ys}%
  \BibitemOpen
  \bibfield  {author} {\bibinfo {author} {\bibfnamefont {C.}~\bibnamefont
  {Cutler}}\ and\ \bibinfo {author} {\bibfnamefont {E.~E.}\ \bibnamefont
  {Flanagan}},\ }\href {\doibase 10.1103/PhysRevD.49.2658} {\bibfield
  {journal} {\bibinfo  {journal} {Phys.Rev.}\ }\textbf {\bibinfo {volume}
  {D49}},\ \bibinfo {pages} {2658} (\bibinfo {year} {1994})},\ \Eprint
  {http://arxiv.org/abs/gr-qc/9402014} {arXiv:gr-qc/9402014 [gr-qc]}
  \BibitemShut {NoStop}%
\bibitem [{\citenamefont {Zhao}\ and\ \citenamefont
  {Wen}(2018)}]{Zhao:2017cbb}%
  \BibitemOpen
  \bibfield  {author} {\bibinfo {author} {\bibfnamefont {W.}~\bibnamefont
  {Zhao}}\ and\ \bibinfo {author} {\bibfnamefont {L.}~\bibnamefont {Wen}},\
  }\href {\doibase 10.1103/PhysRevD.97.064031} {\bibfield  {journal} {\bibinfo
  {journal} {Phys. Rev.}\ }\textbf {\bibinfo {volume} {D97}},\ \bibinfo {pages}
  {064031} (\bibinfo {year} {2018})},\ \Eprint
  {http://arxiv.org/abs/1710.05325} {arXiv:1710.05325 [astro-ph.CO]}
  \BibitemShut {NoStop}%
\bibitem [{\citenamefont {Takeda}\ \emph {et~al.}(2018)\citenamefont {Takeda},
  \citenamefont {Nishizawa}, \citenamefont {Michimura}, \citenamefont {Nagano},
  \citenamefont {Komori}, \citenamefont {Ando},\ and\ \citenamefont
  {Hayama}}]{Takeda:2018uai}%
  \BibitemOpen
  \bibfield  {author} {\bibinfo {author} {\bibfnamefont {H.}~\bibnamefont
  {Takeda}}, \bibinfo {author} {\bibfnamefont {A.}~\bibnamefont {Nishizawa}},
  \bibinfo {author} {\bibfnamefont {Y.}~\bibnamefont {Michimura}}, \bibinfo
  {author} {\bibfnamefont {K.}~\bibnamefont {Nagano}}, \bibinfo {author}
  {\bibfnamefont {K.}~\bibnamefont {Komori}}, \bibinfo {author} {\bibfnamefont
  {M.}~\bibnamefont {Ando}}, \ and\ \bibinfo {author} {\bibfnamefont
  {K.}~\bibnamefont {Hayama}},\ }\href {\doibase 10.1103/PhysRevD.98.022008}
  {\bibfield  {journal} {\bibinfo  {journal} {Phys. Rev.}\ }\textbf {\bibinfo
  {volume} {D98}},\ \bibinfo {pages} {022008} (\bibinfo {year} {2018})},\
  \Eprint {http://arxiv.org/abs/1806.02182} {arXiv:1806.02182 [gr-qc]}
  \BibitemShut {NoStop}%
\bibitem [{\citenamefont {Cutler}\ and\ \citenamefont
  {Holz}(2009)}]{Cutler:2009qv}%
  \BibitemOpen
  \bibfield  {author} {\bibinfo {author} {\bibfnamefont {C.}~\bibnamefont
  {Cutler}}\ and\ \bibinfo {author} {\bibfnamefont {D.~E.}\ \bibnamefont
  {Holz}},\ }\href {\doibase 10.1103/PhysRevD.80.104009} {\bibfield  {journal}
  {\bibinfo  {journal} {Phys. Rev.}\ }\textbf {\bibinfo {volume} {D80}},\
  \bibinfo {pages} {104009} (\bibinfo {year} {2009})},\ \Eprint
  {http://arxiv.org/abs/0906.3752} {arXiv:0906.3752 [astro-ph.CO]} \BibitemShut
  {NoStop}%
\bibitem [{\citenamefont {Abbott}\ \emph
  {et~al.}(2018{\natexlab{b}})\citenamefont {Abbott} \emph
  {et~al.}}]{LIGOScientific:2018mvr}%
  \BibitemOpen
  \bibfield  {author} {\bibinfo {author} {\bibfnamefont {B.~P.}\ \bibnamefont
  {Abbott}} \emph {et~al.} (\bibinfo {collaboration} {LIGO Scientific,
  Virgo}),\ }\href@noop {} {\  (\bibinfo {year} {2018}{\natexlab{b}})},\
  \Eprint {http://arxiv.org/abs/1811.12907} {arXiv:1811.12907 [astro-ph.HE]}
  \BibitemShut {NoStop}%
\bibitem [{\citenamefont {Nishizawa}(2017)}]{Nishizawa:2016ood}%
  \BibitemOpen
  \bibfield  {author} {\bibinfo {author} {\bibfnamefont {A.}~\bibnamefont
  {Nishizawa}},\ }\href {\doibase 10.1103/PhysRevD.96.101303} {\bibfield
  {journal} {\bibinfo  {journal} {Phys. Rev.}\ }\textbf {\bibinfo {volume}
  {D96}},\ \bibinfo {pages} {101303} (\bibinfo {year} {2017})},\ \Eprint
  {http://arxiv.org/abs/1612.06060} {arXiv:1612.06060 [astro-ph.CO]}
  \BibitemShut {NoStop}%
\bibitem [{\citenamefont {Gupte}\ and\ \citenamefont
  {Bartos}(2018)}]{Gupte:2018pht}%
  \BibitemOpen
  \bibfield  {author} {\bibinfo {author} {\bibfnamefont {N.}~\bibnamefont
  {Gupte}}\ and\ \bibinfo {author} {\bibfnamefont {I.}~\bibnamefont {Bartos}},\
  }\href@noop {} {\  (\bibinfo {year} {2018})},\ \Eprint
  {http://arxiv.org/abs/1808.06238} {arXiv:1808.06238 [astro-ph.HE]}
  \BibitemShut {NoStop}%
\bibitem [{\citenamefont {Howell}\ \emph {et~al.}(2018)\citenamefont {Howell},
  \citenamefont {Ackley}, \citenamefont {Rowlinson},\ and\ \citenamefont
  {Coward}}]{Howell:2018nhu}%
  \BibitemOpen
  \bibfield  {author} {\bibinfo {author} {\bibfnamefont {E.~J.}\ \bibnamefont
  {Howell}}, \bibinfo {author} {\bibfnamefont {K.}~\bibnamefont {Ackley}},
  \bibinfo {author} {\bibfnamefont {A.}~\bibnamefont {Rowlinson}}, \ and\
  \bibinfo {author} {\bibfnamefont {D.}~\bibnamefont {Coward}},\ }\href@noop {}
  {\  (\bibinfo {year} {2018})},\ \Eprint {http://arxiv.org/abs/1811.09168}
  {arXiv:1811.09168 [astro-ph.HE]} \BibitemShut {NoStop}%
\bibitem [{\citenamefont {Mogushi}\ \emph {et~al.}(2018)\citenamefont
  {Mogushi}, \citenamefont {Cavaglia},\ and\ \citenamefont
  {Siellez}}]{Mogushi:2018ufy}%
  \BibitemOpen
  \bibfield  {author} {\bibinfo {author} {\bibfnamefont {K.}~\bibnamefont
  {Mogushi}}, \bibinfo {author} {\bibfnamefont {M.}~\bibnamefont {Cavaglia}}, \
  and\ \bibinfo {author} {\bibfnamefont {K.}~\bibnamefont {Siellez}},\
  }\href@noop {} {\  (\bibinfo {year} {2018})},\ \Eprint
  {http://arxiv.org/abs/1811.08542} {arXiv:1811.08542 [astro-ph.HE]}
  \BibitemShut {NoStop}%
\bibitem [{\citenamefont {Farooq}\ \emph {et~al.}(2017)\citenamefont {Farooq},
  \citenamefont {Madiyar}, \citenamefont {Crandall},\ and\ \citenamefont
  {Ratra}}]{Farooq:2016zwm}%
  \BibitemOpen
  \bibfield  {author} {\bibinfo {author} {\bibfnamefont {O.}~\bibnamefont
  {Farooq}}, \bibinfo {author} {\bibfnamefont {F.~R.}\ \bibnamefont {Madiyar}},
  \bibinfo {author} {\bibfnamefont {S.}~\bibnamefont {Crandall}}, \ and\
  \bibinfo {author} {\bibfnamefont {B.}~\bibnamefont {Ratra}},\ }\href
  {\doibase 10.3847/1538-4357/835/1/26} {\bibfield  {journal} {\bibinfo
  {journal} {Astrophys. J.}\ }\textbf {\bibinfo {volume} {835}},\ \bibinfo
  {pages} {26} (\bibinfo {year} {2017})},\ \Eprint
  {http://arxiv.org/abs/1607.03537} {arXiv:1607.03537 [astro-ph.CO]}
  \BibitemShut {NoStop}%
\bibitem [{\citenamefont {Hu}\ and\ \citenamefont {Sawicki}(2007)}]{Hu:2007pj}%
  \BibitemOpen
  \bibfield  {author} {\bibinfo {author} {\bibfnamefont {W.}~\bibnamefont
  {Hu}}\ and\ \bibinfo {author} {\bibfnamefont {I.}~\bibnamefont {Sawicki}},\
  }\href {\doibase 10.1103/PhysRevD.76.104043} {\bibfield  {journal} {\bibinfo
  {journal} {Phys. Rev.}\ }\textbf {\bibinfo {volume} {D76}},\ \bibinfo {pages}
  {104043} (\bibinfo {year} {2007})},\ \Eprint {http://arxiv.org/abs/0708.1190}
  {arXiv:0708.1190 [astro-ph]} \BibitemShut {NoStop}%
\bibitem [{\citenamefont {Jain}\ and\ \citenamefont
  {Zhang}(2008)}]{Jain:2007yk}%
  \BibitemOpen
  \bibfield  {author} {\bibinfo {author} {\bibfnamefont {B.}~\bibnamefont
  {Jain}}\ and\ \bibinfo {author} {\bibfnamefont {P.}~\bibnamefont {Zhang}},\
  }\href {\doibase 10.1103/PhysRevD.78.063503} {\bibfield  {journal} {\bibinfo
  {journal} {Phys. Rev.}\ }\textbf {\bibinfo {volume} {D78}},\ \bibinfo {pages}
  {063503} (\bibinfo {year} {2008})},\ \Eprint {http://arxiv.org/abs/0709.2375}
  {arXiv:0709.2375 [astro-ph]} \BibitemShut {NoStop}%
\bibitem [{\citenamefont {Amendola}\ \emph {et~al.}(2008)\citenamefont
  {Amendola}, \citenamefont {Kunz},\ and\ \citenamefont
  {Sapone}}]{Amendola:2007rr}%
  \BibitemOpen
  \bibfield  {author} {\bibinfo {author} {\bibfnamefont {L.}~\bibnamefont
  {Amendola}}, \bibinfo {author} {\bibfnamefont {M.}~\bibnamefont {Kunz}}, \
  and\ \bibinfo {author} {\bibfnamefont {D.}~\bibnamefont {Sapone}},\ }\href
  {\doibase 10.1088/1475-7516/2008/04/013} {\bibfield  {journal} {\bibinfo
  {journal} {JCAP}\ }\textbf {\bibinfo {volume} {0804}},\ \bibinfo {pages}
  {013} (\bibinfo {year} {2008})},\ \Eprint {http://arxiv.org/abs/0704.2421}
  {arXiv:0704.2421 [astro-ph]} \BibitemShut {NoStop}%
\bibitem [{\citenamefont {Bertschinger}\ and\ \citenamefont
  {Zukin}(2008)}]{Bertschinger:2008zb}%
  \BibitemOpen
  \bibfield  {author} {\bibinfo {author} {\bibfnamefont {E.}~\bibnamefont
  {Bertschinger}}\ and\ \bibinfo {author} {\bibfnamefont {P.}~\bibnamefont
  {Zukin}},\ }\href {\doibase 10.1103/PhysRevD.78.024015} {\bibfield  {journal}
  {\bibinfo  {journal} {Phys. Rev.}\ }\textbf {\bibinfo {volume} {D78}},\
  \bibinfo {pages} {024015} (\bibinfo {year} {2008})},\ \Eprint
  {http://arxiv.org/abs/0801.2431} {arXiv:0801.2431 [astro-ph]} \BibitemShut
  {NoStop}%
\bibitem [{\citenamefont {Daniel}\ \emph {et~al.}(2008)\citenamefont {Daniel},
  \citenamefont {Caldwell}, \citenamefont {Cooray},\ and\ \citenamefont
  {Melchiorri}}]{Daniel:2008et}%
  \BibitemOpen
  \bibfield  {author} {\bibinfo {author} {\bibfnamefont {S.~F.}\ \bibnamefont
  {Daniel}}, \bibinfo {author} {\bibfnamefont {R.~R.}\ \bibnamefont
  {Caldwell}}, \bibinfo {author} {\bibfnamefont {A.}~\bibnamefont {Cooray}}, \
  and\ \bibinfo {author} {\bibfnamefont {A.}~\bibnamefont {Melchiorri}},\
  }\href {\doibase 10.1103/PhysRevD.77.103513} {\bibfield  {journal} {\bibinfo
  {journal} {Phys. Rev.}\ }\textbf {\bibinfo {volume} {D77}},\ \bibinfo {pages}
  {103513} (\bibinfo {year} {2008})},\ \Eprint {http://arxiv.org/abs/0802.1068}
  {arXiv:0802.1068 [astro-ph]} \BibitemShut {NoStop}%
\bibitem [{\citenamefont {Linder}(2018)}]{Linder:2018jil}%
  \BibitemOpen
  \bibfield  {author} {\bibinfo {author} {\bibfnamefont {E.~V.}\ \bibnamefont
  {Linder}},\ }\href {\doibase 10.1088/1475-7516/2018/03/005} {\bibfield
  {journal} {\bibinfo  {journal} {JCAP}\ }\textbf {\bibinfo {volume} {1803}},\
  \bibinfo {pages} {005} (\bibinfo {year} {2018})},\ \Eprint
  {http://arxiv.org/abs/1801.01503} {arXiv:1801.01503 [astro-ph.CO]}
  \BibitemShut {NoStop}%
\bibitem [{\citenamefont {Kimura}\ \emph {et~al.}(2012)\citenamefont {Kimura},
  \citenamefont {Kobayashi},\ and\ \citenamefont {Yamamoto}}]{Kimura:2011dc}%
  \BibitemOpen
  \bibfield  {author} {\bibinfo {author} {\bibfnamefont {R.}~\bibnamefont
  {Kimura}}, \bibinfo {author} {\bibfnamefont {T.}~\bibnamefont {Kobayashi}}, \
  and\ \bibinfo {author} {\bibfnamefont {K.}~\bibnamefont {Yamamoto}},\ }\href
  {\doibase 10.1103/PhysRevD.85.024023} {\bibfield  {journal} {\bibinfo
  {journal} {Phys. Rev.}\ }\textbf {\bibinfo {volume} {D85}},\ \bibinfo {pages}
  {024023} (\bibinfo {year} {2012})},\ \Eprint {http://arxiv.org/abs/1111.6749}
  {arXiv:1111.6749 [astro-ph.CO]} \BibitemShut {NoStop}%
\bibitem [{\citenamefont {Zhu}\ \emph {et~al.}(2019)\citenamefont {Zhu} \emph
  {et~al.}}]{Zhu:2018etc}%
  \BibitemOpen
  \bibfield  {author} {\bibinfo {author} {\bibfnamefont {W.~W.}\ \bibnamefont
  {Zhu}} \emph {et~al.},\ }\href {\doibase 10.1093/mnras/sty2905} {\bibfield
  {journal} {\bibinfo  {journal} {Mon. Not. Roy. Astron. Soc.}\ }\textbf
  {\bibinfo {volume} {482}},\ \bibinfo {pages} {3249} (\bibinfo {year}
  {2019})},\ \Eprint {http://arxiv.org/abs/1802.09206} {arXiv:1802.09206
  [astro-ph.HE]} \BibitemShut {NoStop}%
\bibitem [{\citenamefont {Williams}\ \emph {et~al.}(2004)\citenamefont
  {Williams}, \citenamefont {Turyshev},\ and\ \citenamefont
  {Boggs}}]{Williams:2004qba}%
  \BibitemOpen
  \bibfield  {author} {\bibinfo {author} {\bibfnamefont {J.~G.}\ \bibnamefont
  {Williams}}, \bibinfo {author} {\bibfnamefont {S.~G.}\ \bibnamefont
  {Turyshev}}, \ and\ \bibinfo {author} {\bibfnamefont {D.~H.}\ \bibnamefont
  {Boggs}},\ }\href {\doibase 10.1103/PhysRevLett.93.261101} {\bibfield
  {journal} {\bibinfo  {journal} {Phys. Rev. Lett.}\ }\textbf {\bibinfo
  {volume} {93}},\ \bibinfo {pages} {261101} (\bibinfo {year} {2004})},\
  \Eprint {http://arxiv.org/abs/gr-qc/0411113} {arXiv:gr-qc/0411113 [gr-qc]}
  \BibitemShut {NoStop}%
\bibitem [{\citenamefont {Hofmann}\ and\ \citenamefont
  {Muller}(2018)}]{Hofmann:2018myc}%
  \BibitemOpen
  \bibfield  {author} {\bibinfo {author} {\bibfnamefont {F.}~\bibnamefont
  {Hofmann}}\ and\ \bibinfo {author} {\bibfnamefont {J.}~\bibnamefont
  {Muller}},\ }\href {\doibase 10.1088/1361-6382/aa8f7a} {\bibfield  {journal}
  {\bibinfo  {journal} {Class. Quant. Grav.}\ }\textbf {\bibinfo {volume}
  {35}},\ \bibinfo {pages} {035015} (\bibinfo {year} {2018})}\BibitemShut
  {NoStop}%
\bibitem [{\citenamefont {{Genova}}\ \emph {et~al.}(2018)\citenamefont
  {{Genova}}, \citenamefont {{Mazarico}}, \citenamefont {{Goossens}},
  \citenamefont {{Lemoine}}, \citenamefont {{Neumann}}, \citenamefont
  {{Smith}},\ and\ \citenamefont {{Zuber}}}]{Genova:2018NatComm}%
  \BibitemOpen
  \bibfield  {author} {\bibinfo {author} {\bibfnamefont {A.}~\bibnamefont
  {{Genova}}}, \bibinfo {author} {\bibfnamefont {E.}~\bibnamefont
  {{Mazarico}}}, \bibinfo {author} {\bibfnamefont {S.}~\bibnamefont
  {{Goossens}}}, \bibinfo {author} {\bibfnamefont {F.~G.}\ \bibnamefont
  {{Lemoine}}}, \bibinfo {author} {\bibfnamefont {G.~A.}\ \bibnamefont
  {{Neumann}}}, \bibinfo {author} {\bibfnamefont {D.~E.}\ \bibnamefont
  {{Smith}}}, \ and\ \bibinfo {author} {\bibfnamefont {M.~T.}\ \bibnamefont
  {{Zuber}}},\ }\href {\doibase 10.1038/s41467-017-02558-1} {\bibfield
  {journal} {\bibinfo  {journal} {Nature Communications}\ }\textbf {\bibinfo
  {volume} {9}},\ \bibinfo {eid} {289} (\bibinfo {year} {2018})}\BibitemShut
  {NoStop}%
\bibitem [{\citenamefont {Accetta}\ \emph {et~al.}(1990)\citenamefont
  {Accetta}, \citenamefont {Krauss},\ and\ \citenamefont
  {Romanelli}}]{Accetta:1990au}%
  \BibitemOpen
  \bibfield  {author} {\bibinfo {author} {\bibfnamefont {F.~S.}\ \bibnamefont
  {Accetta}}, \bibinfo {author} {\bibfnamefont {L.~M.}\ \bibnamefont {Krauss}},
  \ and\ \bibinfo {author} {\bibfnamefont {P.}~\bibnamefont {Romanelli}},\
  }\href {\doibase 10.1016/0370-2693(90)90029-6} {\bibfield  {journal}
  {\bibinfo  {journal} {Phys. Lett.}\ }\textbf {\bibinfo {volume} {B248}},\
  \bibinfo {pages} {146} (\bibinfo {year} {1990})}\BibitemShut {NoStop}%
\bibitem [{\citenamefont {Uzan}(2011)}]{Uzan:2010pm}%
  \BibitemOpen
  \bibfield  {author} {\bibinfo {author} {\bibfnamefont {J.-P.}\ \bibnamefont
  {Uzan}},\ }\href {\doibase 10.12942/lrr-2011-2} {\bibfield  {journal}
  {\bibinfo  {journal} {Living Rev. Rel.}\ }\textbf {\bibinfo {volume} {14}},\
  \bibinfo {pages} {2} (\bibinfo {year} {2011})},\ \Eprint
  {http://arxiv.org/abs/1009.5514} {arXiv:1009.5514 [astro-ph.CO]} \BibitemShut
  {NoStop}%
\bibitem [{\citenamefont {Zahn}\ and\ \citenamefont
  {Zaldarriaga}(2003)}]{Zahn:2002rr}%
  \BibitemOpen
  \bibfield  {author} {\bibinfo {author} {\bibfnamefont {O.}~\bibnamefont
  {Zahn}}\ and\ \bibinfo {author} {\bibfnamefont {M.}~\bibnamefont
  {Zaldarriaga}},\ }\href {\doibase 10.1103/PhysRevD.67.063002} {\bibfield
  {journal} {\bibinfo  {journal} {Phys. Rev.}\ }\textbf {\bibinfo {volume}
  {D67}},\ \bibinfo {pages} {063002} (\bibinfo {year} {2003})},\ \Eprint
  {http://arxiv.org/abs/astro-ph/0212360} {arXiv:astro-ph/0212360 [astro-ph]}
  \BibitemShut {NoStop}%
\bibitem [{\citenamefont {Umezu}\ \emph {et~al.}(2005)\citenamefont {Umezu},
  \citenamefont {Ichiki},\ and\ \citenamefont {Yahiro}}]{Umezu:2005ee}%
  \BibitemOpen
  \bibfield  {author} {\bibinfo {author} {\bibfnamefont {K.-i.}\ \bibnamefont
  {Umezu}}, \bibinfo {author} {\bibfnamefont {K.}~\bibnamefont {Ichiki}}, \
  and\ \bibinfo {author} {\bibfnamefont {M.}~\bibnamefont {Yahiro}},\ }\href
  {\doibase 10.1103/PhysRevD.72.044010} {\bibfield  {journal} {\bibinfo
  {journal} {Phys. Rev.}\ }\textbf {\bibinfo {volume} {D72}},\ \bibinfo {pages}
  {044010} (\bibinfo {year} {2005})},\ \Eprint
  {http://arxiv.org/abs/astro-ph/0503578} {arXiv:astro-ph/0503578 [astro-ph]}
  \BibitemShut {NoStop}%
\bibitem [{\citenamefont {Galli}\ \emph {et~al.}(2009)\citenamefont {Galli},
  \citenamefont {Melchiorri}, \citenamefont {Smoot},\ and\ \citenamefont
  {Zahn}}]{Galli:2009pr}%
  \BibitemOpen
  \bibfield  {author} {\bibinfo {author} {\bibfnamefont {S.}~\bibnamefont
  {Galli}}, \bibinfo {author} {\bibfnamefont {A.}~\bibnamefont {Melchiorri}},
  \bibinfo {author} {\bibfnamefont {G.~F.}\ \bibnamefont {Smoot}}, \ and\
  \bibinfo {author} {\bibfnamefont {O.}~\bibnamefont {Zahn}},\ }\href {\doibase
  10.1103/PhysRevD.80.023508} {\bibfield  {journal} {\bibinfo  {journal} {Phys.
  Rev.}\ }\textbf {\bibinfo {volume} {D80}},\ \bibinfo {pages} {023508}
  (\bibinfo {year} {2009})},\ \Eprint {http://arxiv.org/abs/0905.1808}
  {arXiv:0905.1808 [astro-ph.CO]} \BibitemShut {NoStop}%
\bibitem [{\citenamefont {Nagata}\ \emph {et~al.}(2002)\citenamefont {Nagata},
  \citenamefont {Chiba},\ and\ \citenamefont {Sugiyama}}]{Nagata:2002tm}%
  \BibitemOpen
  \bibfield  {author} {\bibinfo {author} {\bibfnamefont {R.}~\bibnamefont
  {Nagata}}, \bibinfo {author} {\bibfnamefont {T.}~\bibnamefont {Chiba}}, \
  and\ \bibinfo {author} {\bibfnamefont {N.}~\bibnamefont {Sugiyama}},\ }\href
  {\doibase 10.1103/PhysRevD.66.103510} {\bibfield  {journal} {\bibinfo
  {journal} {Phys. Rev.}\ }\textbf {\bibinfo {volume} {D66}},\ \bibinfo {pages}
  {103510} (\bibinfo {year} {2002})},\ \Eprint
  {http://arxiv.org/abs/astro-ph/0209140} {arXiv:astro-ph/0209140 [astro-ph]}
  \BibitemShut {NoStop}%
\bibitem [{\citenamefont {Ooba}\ \emph {et~al.}(2016)\citenamefont {Ooba},
  \citenamefont {Ichiki}, \citenamefont {Chiba},\ and\ \citenamefont
  {Sugiyama}}]{Ooba:2016PRD}%
  \BibitemOpen
  \bibfield  {author} {\bibinfo {author} {\bibfnamefont {J.}~\bibnamefont
  {Ooba}}, \bibinfo {author} {\bibfnamefont {K.}~\bibnamefont {Ichiki}},
  \bibinfo {author} {\bibfnamefont {T.}~\bibnamefont {Chiba}}, \ and\ \bibinfo
  {author} {\bibfnamefont {N.}~\bibnamefont {Sugiyama}},\ }\href {\doibase
  10.1103/PhysRevD.93.122002} {\bibfield  {journal} {\bibinfo  {journal} {Phys.
  Rev.}\ }\textbf {\bibinfo {volume} {D93}},\ \bibinfo {pages} {122002}
  (\bibinfo {year} {2016})},\ \Eprint {http://arxiv.org/abs/1602.00809}
  {arXiv:1602.00809 [astro-ph.CO]} \BibitemShut {NoStop}%
\bibitem [{\citenamefont {Denissenya}\ and\ \citenamefont
  {Linder}(2018)}]{Denissenya:2018mqs}%
  \BibitemOpen
  \bibfield  {author} {\bibinfo {author} {\bibfnamefont {M.}~\bibnamefont
  {Denissenya}}\ and\ \bibinfo {author} {\bibfnamefont {E.~V.}\ \bibnamefont
  {Linder}},\ }\href {\doibase 10.1088/1475-7516/2018/11/010} {\bibfield
  {journal} {\bibinfo  {journal} {JCAP}\ }\textbf {\bibinfo {volume} {1811}},\
  \bibinfo {pages} {010} (\bibinfo {year} {2018})},\ \Eprint
  {http://arxiv.org/abs/1808.00013} {arXiv:1808.00013 [astro-ph.CO]}
  \BibitemShut {NoStop}%
\bibitem [{\citenamefont {Amendola}\ \emph
  {et~al.}(2018{\natexlab{b}})\citenamefont {Amendola} \emph
  {et~al.}}]{Amendola:2016saw}%
  \BibitemOpen
  \bibfield  {author} {\bibinfo {author} {\bibfnamefont {L.}~\bibnamefont
  {Amendola}} \emph {et~al.},\ }\href {\doibase 10.1007/s41114-017-0010-3}
  {\bibfield  {journal} {\bibinfo  {journal} {Living Rev. Rel.}\ }\textbf
  {\bibinfo {volume} {21}},\ \bibinfo {pages} {2} (\bibinfo {year}
  {2018}{\natexlab{b}})},\ \Eprint {http://arxiv.org/abs/1606.00180}
  {arXiv:1606.00180 [astro-ph.CO]} \BibitemShut {NoStop}%
\bibitem [{\citenamefont {Abate}\ \emph {et~al.}(2012)\citenamefont {Abate}
  \emph {et~al.}}]{Abate:2012za}%
  \BibitemOpen
  \bibfield  {author} {\bibinfo {author} {\bibfnamefont {A.}~\bibnamefont
  {Abate}} \emph {et~al.} (\bibinfo {collaboration} {LSST Dark Energy
  Science}),\ }\href@noop {} {\  (\bibinfo {year} {2012})},\ \Eprint
  {http://arxiv.org/abs/1211.0310} {arXiv:1211.0310 [astro-ph.CO]} \BibitemShut
  {NoStop}%
\bibitem [{\citenamefont {Bacon}\ \emph {et~al.}(2018)\citenamefont {Bacon}
  \emph {et~al.}}]{Bacon:2018dui}%
  \BibitemOpen
  \bibfield  {author} {\bibinfo {author} {\bibfnamefont {D.~J.}\ \bibnamefont
  {Bacon}} \emph {et~al.} (\bibinfo {collaboration} {SKA}),\ }\href@noop {}
  {\bibfield  {journal} {\bibinfo  {journal} {Submitted to: Publ. Astron. Soc.
  Austral.}\ } (\bibinfo {year} {2018})},\ \Eprint
  {http://arxiv.org/abs/1811.02743} {arXiv:1811.02743 [astro-ph.CO]}
  \BibitemShut {NoStop}%
\bibitem [{\citenamefont {Bull}\ \emph {et~al.}(2018)\citenamefont {Bull} \emph
  {et~al.}}]{Bull:2018lat}%
  \BibitemOpen
  \bibfield  {author} {\bibinfo {author} {\bibfnamefont {P.}~\bibnamefont
  {Bull}} \emph {et~al.},\ }\href@noop {} {\  (\bibinfo {year} {2018})},\
  \Eprint {http://arxiv.org/abs/1810.02680} {arXiv:1810.02680 [astro-ph.CO]}
  \BibitemShut {NoStop}%
\bibitem [{\citenamefont {{Bellini}}\ and\ \citenamefont
  {{Sawicki}}(2014)}]{Bellini2014JCAP}%
  \BibitemOpen
  \bibfield  {author} {\bibinfo {author} {\bibfnamefont {E.}~\bibnamefont
  {{Bellini}}}\ and\ \bibinfo {author} {\bibfnamefont {I.}~\bibnamefont
  {{Sawicki}}},\ }\href {\doibase 10.1088/1475-7516/2014/07/050} {\bibfield
  {journal} {\bibinfo  {journal} {JCAP}\ }\textbf {\bibinfo {volume} {7}},\
  \bibinfo {eid} {050} (\bibinfo {year} {2014})},\ \Eprint
  {http://arxiv.org/abs/1404.3713} {arXiv:1404.3713} \BibitemShut {NoStop}%
\end{thebibliography}%

\end{document}